\documentclass[aps,prx,twocolumn,footinbib,showpacs,floatfix]{revtex4}

\usepackage{amssymb}
\usepackage[pdftex]{graphicx}	
	\DeclareGraphicsExtensions{.pdf,.png,.jpg}
\usepackage[applemac]{inputenc}
\usepackage{amsmath,amssymb,amsthm}
\usepackage{mathbbol,bbm}
\usepackage{graphicx,float}
\usepackage[final]{showkeys}
\usepackage{bm,dsfont}
\usepackage{booktabs}
\usepackage{hyperref}
\usepackage{color}
\usepackage{array}

\newcommand*{\epsfscale}[1]{\def\epsfsize##1##2{#1##1}}
\newcommand{\bi}[1]{\boldsymbol{#1}}
\renewcommand{\c}[1]{\mathcal{#1}}
\newcommand{\g}[1]{\mathfrak{#1}}
\newcommand{\s}[1]{\mathsf{#1}}
\newcommand{\calA}{\mathcal{A}}
\newcommand{\calB}{\mathcal{B}}
\newcommand{\calC}{\mathcal{C}}
\newcommand{\calD}{\mathcal{D}}
\newcommand{\calE}{\mathcal{E}}
\newcommand{\calF}{\mathcal{F}}
\newcommand{\calG}{\mathcal{G}}
\newcommand{\calH}{\mathcal{H}}
\newcommand{\calI}{\mathcal{I}}
\newcommand{\calK}{\mathcal{K}}
\newcommand{\calL}{\mathcal{L}}
\newcommand{\calM}{\mathcal{M}}
\newcommand{\calN}{\mathcal{N}}
\newcommand{\calP}{\mathcal{P}}
\newcommand{\calR}{\mathcal{R}}
\newcommand{\calS}{\mathcal{S}}
\newcommand{\calT}{\mathcal{T}}
\newcommand{\calU}{\mathcal{U}}
\newcommand{\calX}{\mathcal{X}}
\newcommand{\calZ}{\mathcal{Z}}
\newcolumntype{P}[1]{>{\centering\arraybackslash}p{#1}}

\def\ket#1{\left|#1\right>}
\def\bra#1{\left<#1\right|}
\def\Tr{ {\rm{Tr }}}
\def\C{{\mathbb{C}}} \def\F{{\mathbb{F}}}
\def\N{{\mathbb{N}}} \def\Q{{\mathbb{Q}}}
\def\R{{\mathbb{R}}} \def\Z{{\mathbb{Z}}}
\def\cH{{\mathcal H}}
\def\bfsigma{\boldsymbol{\sigma}}
\def\bfmu{\boldsymbol{\mu}}
\def\dd{\mathord{\rm d}} \def\Det{\mathop{\rm Det}}
\def\dist{\mathop{\rm dist}} \def\ee{\mathord{\rm e}}
\def\End{\mathord{\rm End}} \def\ev{\mathord{\rm ev}}
\def\id{\mathord{\rm id}} \def\ii{\mathord{\rm i}}
\def\min{\mathord{\rm min}} \def\mod{\mathord{\rm mod}}
\def\prob{\mathord{\rm prob}}
\def\tr{\mathop{\rm Tr}}
\def\half{\textstyle\frac{1}{2}}
\def\third{\textstyle\frac{1}{3}}
\def\fourth{\textstyle\frac{1}{4}}

\begin{document}
\title{A bilayer Double Semion Model with Symmetry-Enriched Topological Order }

\author{L. Ortiz  and M.A. Martin-Delgado}
\affiliation{Departamento de F\'{\i}sica Te\'orica I, Universidad Complutense, 28040 Madrid, Spain}

\vspace{-3.5cm}

\begin{abstract}
We construct a new model of two-dimensional quantum spin systems that combines intrinsic topological orders and a global symmetry called flavour symmetry. It is referred as the bilayer Doubled Semion model (bDS) and is an instance of
symmetry-enriched topological order. A honeycomb bilayer lattice is introduced to combine  a Double Semion Topolgical Order with a global spin-flavour symmetry to get the fractionalization of its quasiparticles. The bDS model exhibits non-trival braiding self-statistics of excitations and its dual model constitutes a Symmetry-Protected Topological Order with novel edge states. This dual model gives rise to a bilayer Non-Trivial Paramagnet that is invariant under the flavour symmetry and the well-known spin flip symmetry. 
\end{abstract}


\maketitle

\section{Introduction}\label{introduction}

\noindent
 Topological effects have a long history in condensed matter physics and
in strongly correlated systems in particular. The quantum Hall effects 
(integer and fractional) and the Haldane phase are two of the most outstanding
examples \cite{TKNN,Haldane_83,wen90,wen92,blokwen90,wenbook04}. 
 Yet, topological effects have acquired a more prominent role in the last
years when the notions of topological orders have emerged and pervaded many
previous seemingly unrelated areas of physics.

\noindent
Topological orders (TOs) are quantum phases of matter that exhibit ground state degeneracy
when the system is on a lattice with non-trivial topology, like a torus. Their excitations are
anyons  \cite{LM77, Wilczek82, ASW85, Nayak_08} that may exist both in the bulk and at the boundaries of the system. These properties
are robust against arbitrary local perturbations at zero temperature. Other features of TOs are the existence of a gap in the spectrum 
and long-range entanglement. A key ingredient in the construction
of these TOs is a gauge group of symmetry $\mathcal{G}$. The elements of this group act locally
on a set of $k$ spins leaving the Hamiltonian invariant. Thus, in concrete models, the Hamiltonian is
composed of $k$-local operators. 
 
\noindent
In the last decade, another class of topological orders has appeared with
the discovery of topological insulators and superconductors \cite{Haldane_88,kane_mele05,bernevig_zhang06,kane_mele05b,moore_balents07,fu_kane_mele07,rmp1,rmp2,Moore,LibroBernevig,
Kitaev01,Volovik_99,Read_00,Ivanov_01,zhang1,vishwanath1,dauphin1}. These are topological phases where the fermion character of the degrees of freedom plays
a fundamental role due to the conservation of fermion parity. More recently, bosonic versions of these phases have been discovered and
they are now understood
as symmetry-protected topological orders (SPTOs) \cite{spto1,Levin_Gu,Mesaro_Ran,spto3}. These are quantum phases of matter
with a gap to excitations in the bulk that have trivial anyonic statistics.
Whereas at the boundary,
they may exhibit edge/surface states with unusual properties like being gapless and may have
anyonic statistics. These properties are robust under local perturbations that respect a
global symmetry group $\mathbb{G}$, but not for arbitrary ones as in the TOs. The global group
$\mathbb{G}$ may act as an internal group of symmetries on the spins of the system, i.e. 
like an on-site symmetry, or may act as an spatial symmetry group of the lattice \cite{Essin_Hermele}. 
SPTOs are short-range entanglement phases since they do not have TO in the bulk.

\noindent
Given these two relevant classes of topological phases it is a natural question whether 
it is possible to combine them into a single quantum phase of matter.
The idea of unifying two classes of symmetries is not purely ascetic.
We know that when a unification happens to be possible, it comes with benefits
in a better understanding of physical phenomena underneath and with rewards in the form
of novel effects not present in the initial theories separately. 

\noindent
In the case of a topological order, the additional global symmetry produces the fractionalisation
of the topological charge. The paradigmatic example is the fractional quantum Hall effect (FQHE) \cite{fqhe1,fqhe2,fqhe3}. Namely, the Laughlin state with filling factor $\nu=1/m$ is an example of TO with global symmetry $\mathbb{G}=U(1)$ representing
charge conservation. The quasiparticle excitations are Abelian anyons and moreover, they have fractional
charge with respect to the electron charge: $q=e/m$, $m$ being the filling factor. Another modern example of TO is the Doubled Semion (DS) model
with spin degrees of freedom that has a gauge group $\mathcal{G}=\mathbb{Z}_2$. In general, the fractionalisation class of an anyon describes its characteristic type of topological charge fractionalisation.

\noindent
Thus, the combination of intrinsic topological orders TOs with global symmetries produces new topological
phases of matter that are now called symmetry-enriched topological orders (SETOs). They have been recently studied in several
remarkable works. Their aim is to classify novel SET phases \cite{Mesaro_Ran,Jiang_Ran}, or finding new mechanisms to produce them \cite{Levin_Gu,Hermele}, or constructing explicit realizations of these phases \cite{Essin_Hermele,Song_Hermele} etc.

\noindent
A recent fundamental discovery by Levin and Gu \cite{Levin_Gu} introduces a new example of SPTO with global spin flip symmetry. The non-trivial braiding statistics of the excitations implies the existence of protected edge modes while bosonic
or fermionic statistics yield no edge modes. There is  a duality
transformation between spin models and string models. Namely, the Kitaev and the DS models are referred to as string models since
their anyon excitations are carried  by string operators. The excitations are attached at the endpoints of these strings and consequently they appear in pairs.

\noindent
The string-flux mechanism proposed by Hermele \cite{Hermele} is another explicit instance of how fractionalization of the topological
charge occurs when a global symmetry is preserved: an anyon of a topological ordered phase with global symmetry may carry  fractionalized quantum numbers with respect to the original TO without that global symmetry. This phenomenon occurs when the string attached to the anyon sweeps over a background pattern of fluxes in the ground state of the TO model.

\noindent
Although  these two different methods presented above act on different kind of systems, i.e. Levin and Gu introduced a new type of paramagnet that is a SPTO and Hermele worked out a general mechanism to obtain charge fractionalization, they bear some similarities.  The Non-Trivial paramagnet derived in the Levin-Gu method has  plaquette operators with  opposite sign to the operators for the Trivial case. It can be interpreted as a pattern of fluxes as demanded by the
string-flux mechanism of Hermele.

\noindent
At this point, a fundamental question arises: how to make compatible the duality method of Levin-Gu and the string-flux mechanism of Hermele. Both of them provide us with very useful explicit techniques to fractionalize the topological charge in spin models with TO.
However, to do so we face two main obstacles that look unsurmountable at first sight:

\noindent  { \bf a)} On the one hand, the Hermele model is constructed with a square lattice in a multilayer structure that forms a quasi-three-dimensional
 model, but the square lattice cannot support a DS model. (see Appendix \ref{app:square_lattice} and \cite{Fiona_Burnell}).
 
\noindent  { \bf b)} On the other hand, the Levin-Gu method is realized on a single  hexagonal layer, but we would need a  multilayer realization of that construction.  This is problematic since the necessary coordination condition (3) is incompatible with a multilayer structure of honeycomb layers.
 
 \noindent
 Interestingly enough, we can rephrase this compatibility problem between these two fractionalization methods as a compatibility condition
 between global symmetries. The key point is to realize that the Levin-Gu method deals with a spin-flip symmetry,
 e.g. $\mathbb{G}=\mathbb{Z}^{fs}_2$, explicitly shown in the spin model introduced in Section \ref{edge_states}, while the Hermele method is about a spin-flavour symmetry among lattice layers, e.g. $\mathbb{G}=\mathbb{Z}^{fv}_2$. This spin-favour symmetry is present explicitly in the string model presented in Eq.(\ref{eq: hamiltonian})
  
\noindent
We hereby summarise briefly some of our main results: 

\noindent i/ We have constructed a bilayer Doubled Semion (bDS) model that has
intrinsic topological orders of type $\mathcal{G}=\mathbb{Z}_2$ and is invariant
under the global symmetry group $\mathbb{G}=\mathbb{Z}^{fv}_2$
corresponding to  spin-flavour transformations.

\noindent ii/ Two different types of fractionalized topological charges appear due to the group $\mathbb{Z}^{fs}_2$. These two kinds of charge  fractionalizations are due to the fact that there are two types of  charge-like excitations in the system.

\noindent iii/ We calculate the dual model of the TO associated bilayer Double Semion model. The dual model constitutes a new bilayer Non-Trivial Paramagnet (bNTP). Similarly as in \cite{Levin_Gu}, we also get a bilayer Trivial Paramagnet, which corresponds to the dual model of the model introduced by Hermele. 

 \noindent iv/ Within this dual model, which is now a SPT phase we construct the edge states that appear in the bilayer lattice when boundaries are present. These edge modes are invariant under both the flavour symmetry and the spin symmetry.

\noindent v/ As a bonus, we also construct the edge states corresponding to the dual model for the original Hermele model in a 
bilayer squared lattice that serve for comparison with the bDS model.

\noindent
It is worthwhile to mention that bilayer models of different kinds have been recently proposed in the context 
of topological orders, both intrinsic and with symmetry protection \cite{cross_arx,stacked,yoshida0,yoshida2,yoshida1}. They deal with topological orders of color code type \cite{tcc1,tcc2,tcc3,tcc4, terhal1,brell1,experimental}
for topological quantum computation.

\noindent
This paper is organized as follows: 
in Sec.\ref{preliminars} we review the models needed to build the new bDS model, in particular the Doubled Semion model and the string-flux mechanism. In Sec.\ref{model} we construct an exactly solvable Hamiltonian representing a DS model on a bilayer
hexagonal lattice and compatible with the global symmetry of exchanging the two layers in the lattice, the $\mathbb{Z}^{fv}_2$.  Explicit formulas for the ground states are also given. In Sec.\ref{edge_states} we report the construction of edge modes in the dual model of the bDS TO. To do so, we construct dual spin models to 
the string bilayer DS model and the Hermele model. Sec.\ref{conclusions} is devoted to conclusions and outlook.
In order to facilitate the reading and exposition of our results, we present the proofs of our main results and their technicalities  in independent appendices.

\section{MODELS WITH DIFFERENT $\mathbb{Z}_2$ TOPOLOGICAL ORDERS}\label{preliminars}

\noindent
This section is devoted to  explain the two main concepts we employ to build the bilayer Doubled Semion model presented in Section \ref{model}.  As mentioned in the introduction, the quantum topological phase that we construct is an instance of  symmetry-enriched topological order (SETO) since it presents fractionalization of the topological charges of certain intrinsic topological orders. This kind of phases combines two elements: a topological order associated to a local gauge group symmetry denoted as $\mathcal{G}$  and a global symmetry
group denoted as $\mathbb{G}$. In the following we shall review both elements and introduce the appropriate notation. 

\subsection{Doubled Semion model as topological order} \label{ssec:double_semion}

\noindent
The well-known Kitaev model (toric code) is the simplest example of an exactly solvable topologically ordered model \cite{Kitaev01}. This model is characterised by the gauge group $\mathcal{G}=\mathbb{Z}_2$. 
Less familiar is the Doubled Semion (DS) model \cite{Levin_Wen2}, a non-trivial modification of the Kitaev model in two dimensions. The DS model constitutes a different topological phase of matter but with the same gauge group, $\mathcal{G}=\mathbb{Z}_2$ \cite{Levin_Wen,Fiona_Burnell,Freedman1,pollmnn1,pollmnn2,orus1,yao1,pramod1}.

\noindent 
A question naturally arises: are there more distinct topological phases of matter with $\mathbb{Z}_2$ as their gauge group? The classification of topological phases presented in  \cite{Mesaro_Ran} shows that the Kitaev and the DS model are all the possible bosonic/spin gapped quantum phases with $\mathbb{Z}_2$ gauge group. This result is derived from group cohomology classification of SPTOs \cite{spto1}.

\begin{figure}
\includegraphics[width=0.7\linewidth]{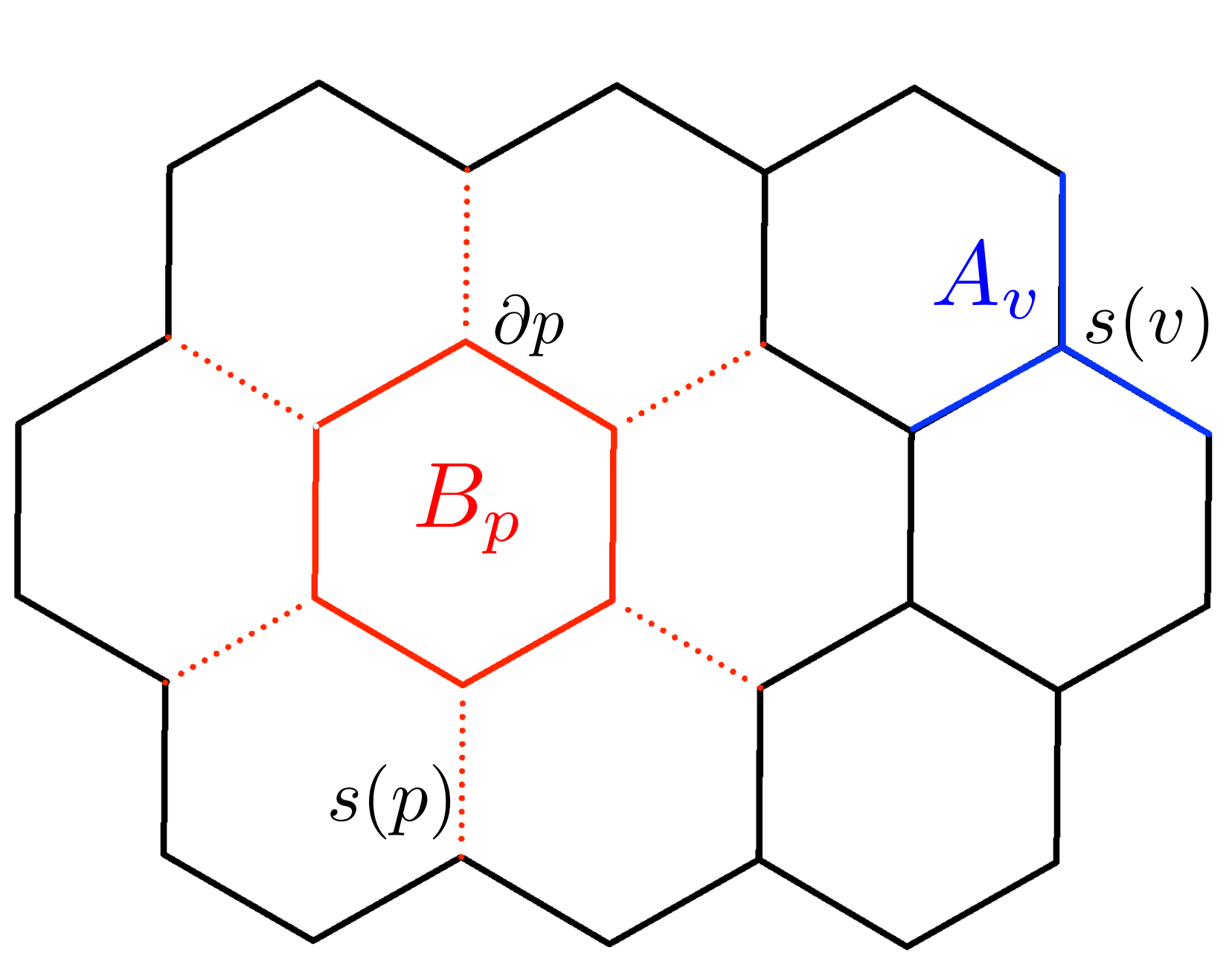}
\caption{\label{fig:ds_model} Double Semion model in a 2D hexagonal lattice. In the picture, a vertex and plaquette operator are shown (defined in Eq.\eqref{eq: vertex_plaquettes_ds}). The links in $s(v)$ are in solid blue lines. Red solid lines represent the links in $\partial p $ and the set of links in $s(p)$ is in dotted red lines.}
\end{figure}

\noindent
 We shall need the explicit construction of the DS model in order to construct the bDS model.
The DS model  is defined on an hexagonal lattice. There is a spin 1/2 located at each link of the lattice. The Hamiltonian is built out of  two types of $k$-local operators: $A_v$ vertex ($k=3$) and $B_p$ plaquette ($k=12$) operators. They are defined as follows:
 \begin{align}\label{eq: vertex_plaquettes_ds}
 &A_v:=\prod_{i\in{s(v)}}\sigma^z_i, \hspace{1cm }B_p:=\prod_{i\in \partial p}\sigma^x_i\prod_{j\in s(p)}\textrm{i}^{(1-\sigma^z_j)/2},\\\nonumber
 \end{align}
 and $s(v)$ is the set of three links entering the vertex $v$. The six links delimiting a plaquette in the lattice are denoted by  $\partial p$ and $s(p)$ are the six links radiating from the plaquette $p$.  Fig.\ref{fig:ds_model} shows a picture of a vertex and a plaquette operator for the DS model. 
Having defined the plaquette and vertex operators, the hamiltonian of the DS model is written as:\begin{align}\label{DS_hamiltonian}
&H=-\sum_v A_v +\sum_{p}B_p.
\end{align}
This hamiltonian per se is not an exactly solvable model in the whole Hilbert space of states since plaquette operators only commute within an invariant subspace. This subspace is called the \textit{zero-flux} subspace and is defined as  those  states that fulfil the constraint :
 \begin{align}\label{zero-flux-rule}
 &A_v=\prod_{i\in{s(v)}}\sigma^z_i =+1.
 \end{align}
In Appendix \ref{app:zero_flux_rule}, we explicitly show that the DS hamiltonian plus this restriction turns out to be exactly solvable. Moreover, despite the fact that plaquette operators defined in Eq.(\ref{eq: vertex_plaquettes_ds}) have imaginary phases, the hamiltonian is real in the invariant subspace where it is exactly sovable. Consequently it is also hermitian  (see Appendix \ref{app:zero_flux_rule}).
\vspace{1mm}

\noindent
The ground state $\ket{\psi_0}$ of  the DS hamiltonian  fulfils the following conditions: 
 \begin{align}\label{eq: gs_conditions}
 A_v \ket{\psi_0} = + \ket{\psi_0} \; \forall v, \hspace{1cm} B_p \ket{\psi_0}=-\ket{\psi_0} \; \forall  p.
 \end{align}
 The explicit form of the ground state is obtained in the following way:  1/ select a vacuum state being an eigenvector of the vertex operators, i.e. $|0\rangle^{\otimes N}$ (where $|0\rangle$ is the +1 eigenvector of $\sigma^z_l$);  2/use the plaquette operators to build projectors and apply them onto the vacuum state. The resulting (unnormalised) state is:
\begin{align}\label{eq: gs}
 \ket{\psi_0}=\prod_{p\in B_p}(1-B_p)|0\rangle^{\otimes N}.
\end{align} 
It is straightforward to check that this state fulfills the lowest energy conditions in Eq.\eqref{eq: gs_conditions} since $ B_p^2=1$ within the invariant subspace (see Appendix \ref{app:zero_flux_rule}). Expanding the product in Eq.\eqref{eq: gs}, one can see that the ground state is a superposition of closed loops configurations or closed strings. Each bit of these strings is represented by a filled link of the lattice where there is $|1\rangle$, whereas links with values $|0\rangle$ are left empty. Due to the condition  $B_p=-1$ for the ground state, the coefficients of this superposition of closed loops alternate sign. Taking this fact into account, we can write the ground state in a different way: 
\begin{align}
\ket{\psi_0}=\sum_x (-1)^{\text{number of loops}}|x\rangle,
\end{align}
where $\ket{x}$ denotes all possible closed string configurations compatible with the lattice geometry. Here the number of loops corresponds
to the different closed string configurations contributing to the ground state \cite{Levin_Wen,Fiona_Burnell}.
\begin{figure}
\includegraphics[width=0.79\linewidth]{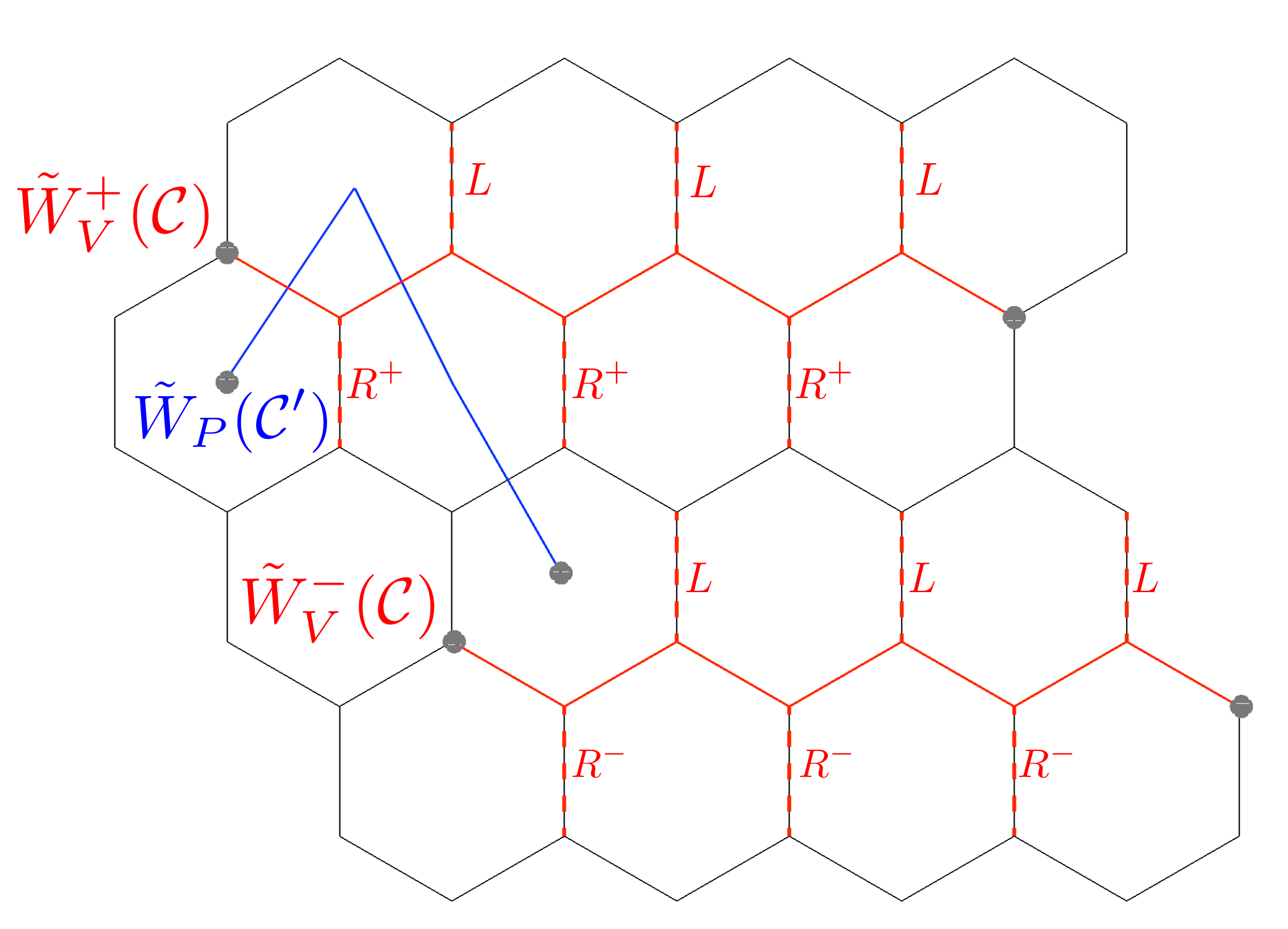}
\caption{\label{fig:string_statistics} A picture showing the three possible  types of string operators in the DS model. The explicit expressions are in Eq.\eqref{eq: string-operators} and Eq.\eqref{eq: string-operators-p}. Grey circles denote vertices or plaquettes where the ground-state conditions in \eqref{eq: gs}, are violated. A plaquette-type string operator is associated to a blue solid line $\mathcal{C}'$ crossing links in the lattice. It represents a path in the dual lattice that connects the two plaquettes  where quasiparticles reside. The two possible chiralities of vertex-type operators are represented by red paths $\mathcal{C}$
connecting vertices in the direct lattice. The paths $\mathcal{C}$ are also endowed with dotted red lines corresponding to $L=(-1)^{\frac{1}{4}(1-\sigma^z_i)(1+\sigma^z_j)}$ and $R^+=(+\textrm{i})^{(1-\sigma^z_l)/2}$ or $R^-=(-\textrm{i})^{(1-\sigma^z_l)/2}$ respectively.}
\end{figure}

\noindent
Excitations in the DS model are created by violating the ground state conditions in Eq.(\ref{eq: gs_conditions}). These violations always involve a pair of quasiparticles or defects. The operators that create these defects  are called open \textit{string operators} and act on the links of a path $\mathcal{C}$ ($\mathcal{C}'$) joining the pair of quasiparticles in the direct lattice (dual lattice). There are two types of string operators:  vertex-type string operators $\tilde{W}^{\pm}_V(\mathcal{C})$, which come in two chiralities, and plaquette-type string operators $\tilde{W}_P(\mathcal{C}')$. The expressions for these operators are: 
\begin{align}\label{eq: string-operators}
&\tilde{W}^{\pm}_V(\mathcal{C})= \prod_{i\in \mathcal{C}}\sigma^x_{i}\prod_{k\in L}(-1)^{\frac{1}{4}(1-\sigma^z_i)(1+\sigma^z_j)}\prod_{l \in R}(\pm\textrm{i})^{(1-\sigma^z_j)/2},\\[2mm]
&\tilde{W}_P(\mathcal{C}')=\prod_{i\in \mathcal{C}'}\sigma_i^z, \label{eq: string-operators-p}
\end{align}
where $L$ and $R$ denote the links on the left or right of the path $\mathcal{C}$, when an orientation is chosen, as shown in Fig.\ref{fig:string_statistics}. $L$-links has two edges attached, $i$ and $j$, where $i$ is before in $\mathcal{C}$. The sign $(\pm \textrm{i})$ in $R$-links indicates the quirality of the string operator.

\noindent
 The quasiparticles created by vertex-type string operators  are called \textit{semions}. Semions have  different self-statistics from the well-know fermions and bosons. Namely, if we interchange two semions, the wave function of the system acquires a phase of $\pi/2$ (instead of $0$ or $\pi$ as in the case of bosons or fermions, respectively).  On the contrary, plaquette excitations create  quasiparticles that are bosons (trivial self-statistics). However, plaquette excitations have semionic mutual statistics with respect to the vertex excitations.  Self-statistics and mutual statistics of all quasiparticles in the DS model are summarised in Table \ref{tab:statistics}.
\begin{table}[t]
\centering
\begin{tabular}{|P{2 cm}|P{2 cm}|P{2 cm}|P{2 cm}|}
\toprule
\hline
\hline
\multicolumn{4}{c}{Self and mutual  statistics of excitations in the DS model} \\
\hline
\cmidrule(r){1-2}
& \ \ $\tilde{W}^+_V$ \ \ & \ \ $\tilde{W}^-_V$ \ \ & \  $\tilde{W}_P$ \\
\hline
\midrule
\ \ $\tilde{W}^+_V$\ \ & anticommute & commute & anticommute \\
\hline
$\tilde{W}^-_V$ & commute & anticommute  & anticommute \\
\hline
$\tilde{W}_P$  & anticommute & anticommute & commute \\
\hline
\hline
\bottomrule
\end{tabular}
\caption{This table shows the self and mutual statistical of the string operators in the DS model.  The three columns represent the vertex string operators, which come in two different chiralities, and the plaquette string operators. The string operators create two quasiparticles. If they anticommute the wave function of each quasiparticle acquires a $\pm i$ phase, giving rise to semions. Then, we can observe that vertex excitations are semions because their string operators anticommute among themselves. Moreover, there are semionic mutual statistics among plaquette and vertex string operators.  }
\label{tab:statistics}
\end{table}

\begin{figure}
\includegraphics[width=0.5\linewidth]{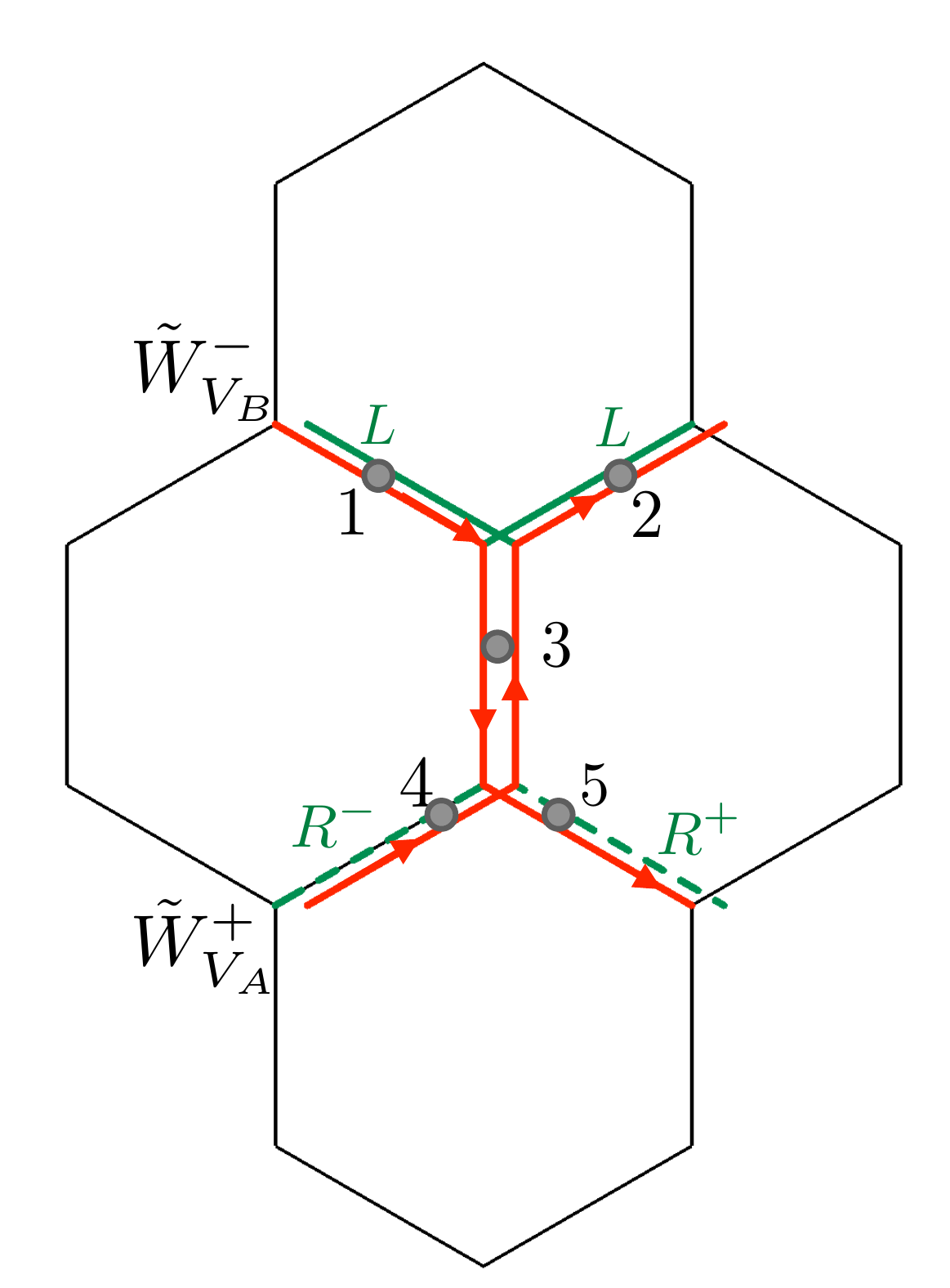}
\caption{\label{fig:two_vertex_strings} A detailed description of two vertex operators that cross their paths. Red solid lines indicate the edges where $\sigma^x$ is acting. The strings are orientated, as Eq. (\ref{eq: string-operators}) requires. The edges attached to the left(right) are plotted in green solid (dashed) lines. Grey circles denote the spins which play a role in the crossing. They are labeled with numbers in order to easily compute the calculations in Eq.\eqref{eq:commutator_strings}.}
\end{figure}

\noindent
We can determine the results presented in Table \ref{tab:statistics}  by calculating the commutator of two string operators. The definition of the commutator used throughout this work is: $[A,B]=AB-BA$.
As a matter of example, we give the details of the commutator for two vertex-type strings, which are shown in Fig.\ref{fig:two_vertex_strings}.
The explicit expressions of these strings  are:
\begin{align}
W^+_{V_A}=\sigma^x_4\sigma^x_3\sigma^x_2 (-1)^{\frac{1}{4}(1-\sigma^z_3)(1+\sigma^z_2)}(+\textrm{i})^{(1-\sigma^z_5)/2} \nonumber,\\
W^-_{V_B}=\sigma^x_1\sigma^x_3\sigma^x_5(-1)^{\frac{1}{4}(1-\sigma^z_1)(1+\sigma^z_3)}(-\textrm{i})^{(1-\sigma^z_4)/2}\nonumber.
\end{align}
Using the commutation relations of Pauli matrices, we calculate the commutator between these two vertex string operators:
\begin{align}\label{eq:commutator_strings}
[W^+_{V_A},W^-_{V_B}]= \sigma^x_1 \sigma^x_2  \sigma^x_4 \sigma^x_5 \;e^{i\frac{\pi}{4}(-\sigma^z_1+\sigma^x_2)}\times\\
\times\;2\textrm{i}\sin\left (\frac{\pi}{4}(2\sigma^z_3+\sigma^z_2\sigma^z_3+\sigma^z_5-\sigma^z_1\sigma^z_3+\sigma^z_4)\right)\nonumber
\end{align}

\noindent
 In order to obtain the commutator we consider all the possible configurations compatible with  the zero-flux rule. It turns out that the commutator is zero for all cases. Consequently, we have already shown that the vertex string operators with different quiralities commutate. As a consequence, the mutual statistic between two semions is bosonic. Similarly, we can reproduce the entire Table \ref{tab:statistics}.
\begin{table}[t]
\centering
\begin{tabular}{|P{1.2 cm}|P{1.2 cm}|P{1.2 cm}|P{1.2 cm}|P{1.2 cm}||P{1.2 cm}| }
\toprule
\hline
\hline
\multicolumn{6}{c}{configurations compatible with zero-flux rule} \\
\hline
\cmidrule(r){1-2}
 $\sigma^z_1$ \ \ & \ \ $\sigma^z_2$ \ \ & \  $\sigma^z_3$\ \ & \  $\sigma^z_4$\ \ & \  $\sigma^z_5$\ \ & \  $ X$ \\
\hline
\midrule
\ \ +\ \ & + & + & + & + & $\pi$\\
\hline
\ \ +\ \ & + & + & - & - & $0$\\
\hline
\ \ +\ \ & - & - & + & - & $0$\\
\hline
\ \ +\ \ & - & - & - & + & $0$\\
\hline
\ \ +\ \ & - & - & +& - & $0$\\
\hline
\ \ -\ \ & - & + & - & - & $0$\\
\hline
\ \ -\ \ & - & + & + & + & $\pi$\\
\hline
\ \ -\ \ & + & - & - & + & $-\pi$\\
\hline
\ \ -\ \ & + & - & + & - & $-\pi$\\
\hline
\hline
\bottomrule
\end{tabular}
\caption{The configurations compatible with zero-flux rule are shown in this table.  The first five columns show the possible eigenvalues of the operators $\sigma^z_l$ for the five spins represented in Fig.\ref{fig:two_vertex_strings}. The last column is devoted to the quantity $X=\frac{\pi}{4}(2\sigma^z_3+\sigma^z_2\sigma^z_3+\sigma^z_5-\sigma^z_1\sigma^z_3+\sigma^z_4)$, which is the argument of the sine that appears in Eq. (\ref{eq:commutator_strings}). As it follows from the Table, the commutator is zero for all possible configurations. }
\label{tab:commutator}
\end{table}

\noindent
So far, we have reviewed the main special features of the DS model that will be needed to construct a bilayer DS model with generalized SPTO. Furthermore,
in order to construct that new model we need also to introduce another concept, namely,  the string-flux mechanism for topological charge fractionalization.

 \subsection{String-flux mechanism }\label{string flux mechanism}
\begin{figure}
\includegraphics[width=\linewidth]{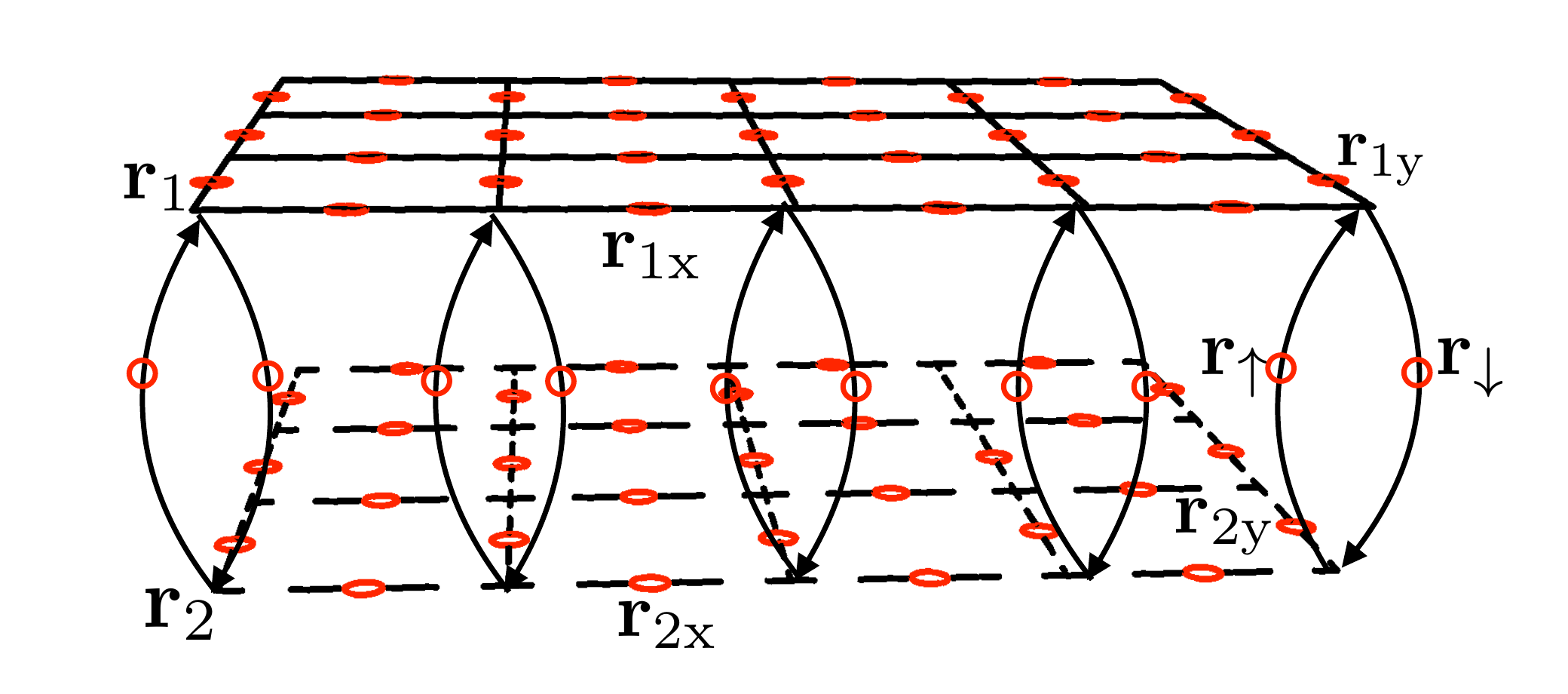}
\caption{\label{fig:hermele_lattice} A picture showing the lattice and notation used in the Hermele model \cite{Hermele}. Red circles represent spins. Lattice vertices are denoted by $\mathbf{r}_i$ with $i=1,2$, corresponding to the two layers. Horizontal links are denoted as $\mathbf{r}_{ix}$ and $\mathbf{r}_{iy}$, depending on which spatial direction they point to. Vertical links are represented by $\mathbf{r}_\uparrow$ and $\mathbf{r}_\downarrow$. Each vertex on the top layer has 4 horizontal
links and 2 vertical links, and similarly for each bottom-layer vertex (see also Fig.\ref{fig:z2_flavour}).}
\end{figure}

\begin{figure}
\includegraphics[width=0.99\linewidth]{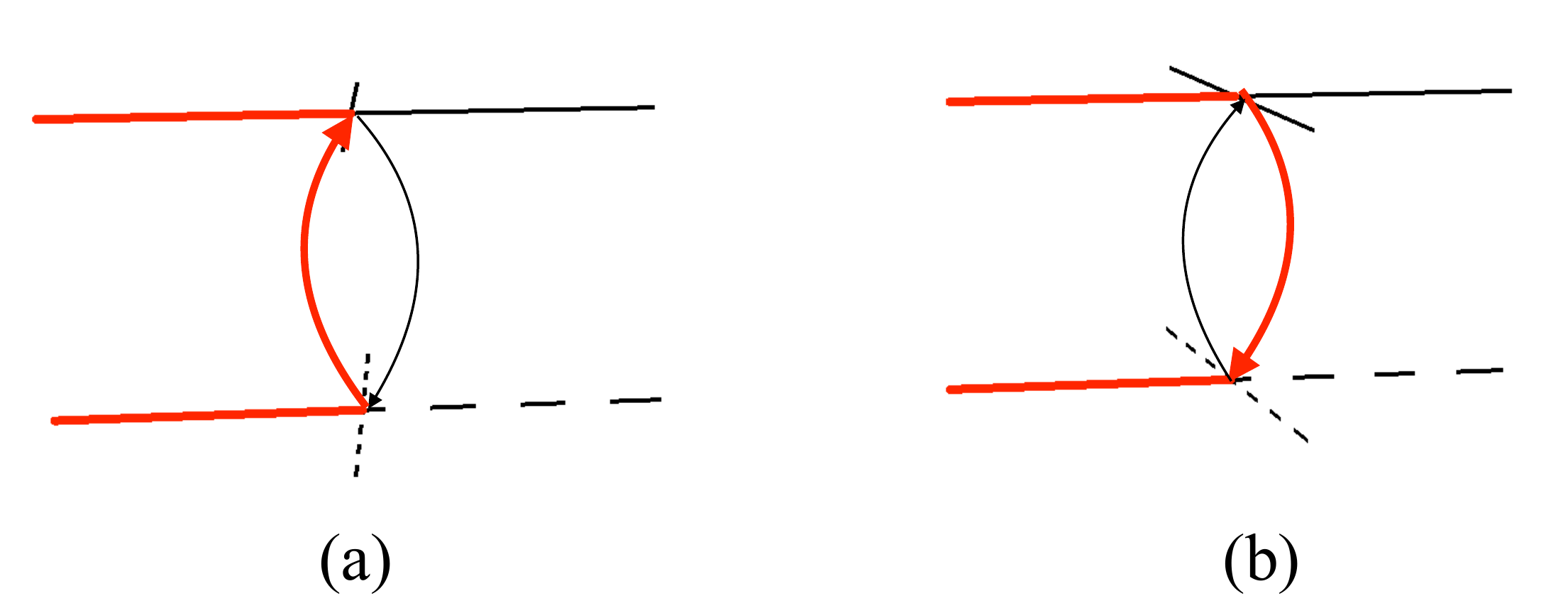}
\caption{\label{fig: string_flux} Two strings that differ by a $B^{\textrm{I}}_{\mathbf{r}}$ plaquette are represented in the figure with red solid lines. The wave function of the two strings in (a) and (b) differs on a phase determined by $K$, which is the coupling constant associated to plaquettes of type I in the hamiltonian in Eq.(\ref{eq: hermele_hamiltonian}). The string feels a $\mathbb{Z}_2$ flux through $B^{\textrm{I}}_{\mathbf{r}}$ when $K=-1$. }
\end{figure}

\begin{figure}
\includegraphics[width=0.8\linewidth]{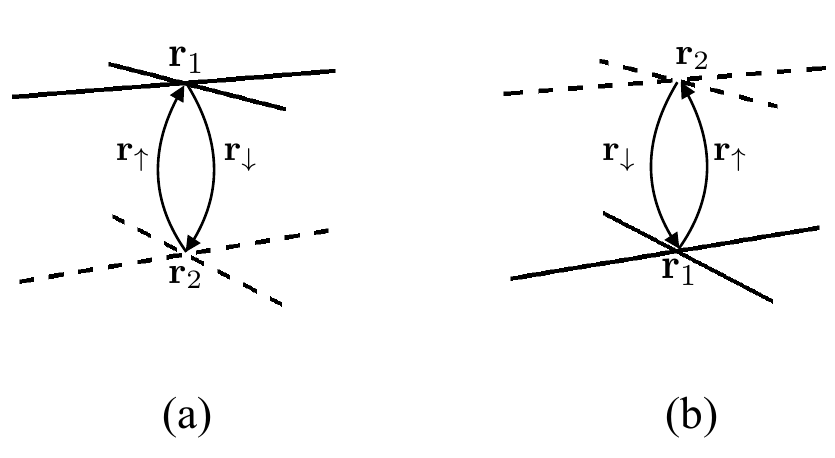}
\caption{\label{fig:z2_flavour}(a) and (b) are related by the action of the $\mathbb{Z}^{fv}_2$ flavour symmetry on the Hermele lattice in Fig.\ref{fig:hermele_lattice}. This symmetry changes layer 1 to layer 2, and viceversa. Vertical links joining the two hexagonal layers are also exchanged, i.e. $\mathbf{r}_\uparrow$ turns into $\mathbf{r}_\downarrow$ and viceversa.  }
\end{figure}

\noindent
 The \textit{string flux mechanism} is a recently proposed technique by Hermele \cite{Hermele} that allows us to produce
 the fractionalization of topological orders with a finite global symmetry group, $\mathbb{G}$. The idea relies in the interconnection between
 non-trivial braiding statistics of string operators creating quasiparticle excitations on one hand, and charge fractionalization of
 a topological order with a global symmetry group $\mathbb{G}$ on another. The mechanism is rather general, but we describe here the simplest case that we shall be using for the bilayer DS model in Sec.\ref{model} along with appropriate notation. As a new result, we shall also  obtain the edge states of the Hermele model in Sec.\ref{edge_states}.
The Hermele model  \cite{Hermele} is a version of the toric code defined on a bilayer square lattice (see Fig.\ref{fig:hermele_lattice}). A spin-$1/2$ or qubit is located on each horizontal link of the square layers. The top and bottom layers are joined by two vertical links per site. There is also one spin on each vertical link,  as it is shown in Fig.\ref{fig:hermele_lattice}.
The model depends on a coupling constant  $K$ that only takes on two values, $K=\pm1$. Both values represent a topological order, but charge fractionalisation only occurs  for $K=-1$. The hamiltonian is also defined by means of vertex and plaquette operators, explicitly: 
\begin{align}\label{eq: hermele_hamiltonian}
H=&-\sum_{\mathbf{r}}\sum_{i=1,2}A_{\mathbf{r}i}-\sum_{\mathbf{r}}\sum_{i=1,2}B^{\textrm{II}}_{\textbf{r}i}\nonumber\\
&-\sum_{\mathbf{r}}\sum_{\alpha=x,y}[B^{\textrm{III}}_{\textbf{r}\alpha\uparrow}+B^{\textrm{III}}_{\textbf{r}\alpha\downarrow}]-K\sum_{\mathbf{r}}B^{\textrm{I}}_\textbf{r},
\end{align}
where vertex and plaquette operators are defined as follows: 
\begin{align}
A_{\mathbf{r}i}:=\prod_{l\sim \mathbf{r}i}\sigma^z_l,
\end{align}
where the product runs over the six links meeting at the vertex $\mathbf{r}i$ (see Fig.\ref{fig:hermele_lattice}).
As for the plaquette operators, they come into three classes \cite{change_notation}:
\begin{align}
B^\textrm{I}_{\textbf{r}}&:=\sigma^x_{\mathbf{r}\uparrow}\sigma^x_{\mathbf{r}\downarrow},\\
B^\textrm{II}_{\textbf{r}i}&:=\sigma^x_{\mathbf{r}ix}\sigma^x_{\mathbf{r}+\hat{e}_x,iy}\sigma^x_{\mathbf{r}+\hat{e}_y,ix}\sigma^x_{\mathbf{r}iy},\\
B^\textrm{III}_{\textbf{r}\alpha\uparrow}&:=\sigma^x_{\mathbf{r}\uparrow}\sigma^x_{\mathbf{r}1\alpha}\sigma^x_{\mathbf{r}2\alpha}\sigma^x_{\mathbf{r}+\hat{e}_\alpha\uparrow},\\
B^\textrm{III}_{\textbf{r}\alpha\downarrow}&:=\sigma^x_{\mathbf{r}\downarrow}\sigma^x_{\mathbf{r}1\alpha}\sigma^x_{\mathbf{r}2\alpha}\sigma^x_{\mathbf{r}+\hat{e}_\alpha\downarrow}.\nonumber
\end{align}
The hamiltonian is exactly solvable since vertex and plaquette operators always share an even number of spins. Thus vertex and plaquette operators form a set of commuting observables. The explicit expression for the ground state can be constructed applying the same technique as for \eqref{eq: gs}: 
\begin{align}\label{eq:gs_hermele}
|\psi_0\rangle=&\prod_{p\in B^{\textrm{I}}_{\mathbf{r}}}(1+K  B^{\textrm{I}}_{\mathbf{r}})
\prod_{p\in B^{\textrm{III}}_{\mathbf{r}\alpha\uparrow}}(1+ B^{\textrm{III}}_{\mathbf{r}\alpha\uparrow})\\
&\prod_{p\in B^{\textrm{III}}_{\mathbf{r}\alpha\downarrow}}(1+ B^{\textrm{III}}_{\mathbf{r}\alpha\downarrow})\prod_{p\in B^{\textrm{II}}_{\mathbf{r}i}}(1+ B^{\textrm{II}}_{\mathbf{r}i})|0\rangle^{\otimes N}.\nonumber
\end{align}
One can readily check that this state is the lowest energy state of the Hermele hamiltonian in Eq.\eqref{eq: hermele_hamiltonian}. The corresponding ground state  is a superposition of all closed string configurations compatible with the bilayer lattice. For $K=1$, all coefficients of those closed configurations have positive sign. Loop configurations with $K=-1$ alternate in sign depending on the parity of the number of plaquettes of type $\textrm{I}$ contained in a given configuration. Remarkably, for $K=-1$, two string configurations that differ by a plaquette $B^\textrm{I}_{\textbf{r}}$ have opposite signs, as it is shown in Fig.\ref{fig: string_flux}.
These ground-state factors in the case of $K=-1$ are directly responsible for the fractionalization of the $\mathbb{Z}_2$ topological charge. To show this, we  have to take into account the global symmetry that is present in the hamiltonian in Eq.(\ref{eq: hermele_hamiltonian}). An important feature of both the hamiltonian and the ground state is their invariance under what may be called a  $\mathbb{Z}^{fv}_2$ \textit{flavour} symmetry. 
This symmetry acts exchanging the two layers, and also exchanging the two vertical spins joining the layers, i.e. $\mathbf{r}\uparrow$ and $\mathbf{r}\downarrow$ as it is shown in Fig.\ref{fig:z2_flavour}.
Consequently, the action of this symmetry is defined as:
\begin{subequations}\label{eq: flavour symmetry}
\begin{equation}\label{eq:symmetry layer}
U_F\sigma^{\mu}_{\mathbf{r}1\alpha}U_F^{-1}:=\sigma^{\mu}_{\mathbf{r}2\alpha},
\end{equation}
\begin{equation}\label{eq:symmetry frontal}
U_F\sigma^{\mu}_{\mathbf{r}\uparrow}U_F^{-1}:=\sigma^{\mu}_{\mathbf{r}\downarrow},
\end{equation}
\end{subequations}
where $U_F$ is the unitary operator which implements the $\mathbb{Z}^{fv}_2$ flavour symmetry and satisfies the condition $U_F^2=1$. The superindex $\mu$ of $\sigma^{\mu}_{\mathbf{r}i\alpha}$ runs over $x,y$ and $z$ to include the action of $\mathbb{Z}^{fv}_2$ flavour symmetry on all types of Pauli matrices. 

\noindent
The invariance under this $\mathbb{Z}^{fv}_2$ flavour symmetry has important consequences on the system, namely, the fractionalization of the charge.
 This phenomenon occurs when a global symmetry is introduced in an intrinsic topological order phase. The action of the Global Symmetry ($\mathbb{G}$) on the ground and excited states of a quantum system is given by  irreducible representations of this precise $\mathbb{G}$. However, there is a special feature of topological orders which gives rise to the fractionalization mentioned before: excitations are created in pairs and as a consequence, a state with a single excitation is not physical. Therefore, it is not surprising to have a phase ambiguity when we specify the action of the $\mathbb{G}$ on a single excitation. This ambiguity is captured mathematically by projective representations of the $\mathbb{G}$. In contrast to the irreducible representations, the projective representations have an extra degree of freedom. If we multiply two projective representations, we obtain a phase factor,  $\omega(g_1,g_2)$, where $g_1$ and $g_2$ are two elements of the group of the $\mathbb{G}$. The set of equivalent classes of $\omega(g_1,g_2)$ is isomorphic to the cohomology group $H^2(\mathbb{G},\mathcal{G})$, where $\mathcal{G}$  is the Gauge Group specifying the topological order. The set denoted as $H^2(\mathbb{G}, \mathcal{G})$ will give us the different  fractionalization classes for an specific global symmetry in an specific topological order \cite{Mesaro_Ran}.

\noindent
For the concrete example of the model in Eq.(\ref{eq: hermele_hamiltonian}), the cohomology group is $H^2(\mathbb{Z}^{fv}_2, \mathbb{Z}_2)=\mathbb{Z}_2$. It implies that  two possible fractionalization are present in the system.
To specify it, we  write the action of the $\mathbb{Z}^{fv}_2$ flavour symmetry on each single quasiparticle excitation or defect explicitly. As we have explained in Section \ref{ssec:double_semion}, these excitations are located  at the endpoints of the open strings. Let us study  the action of $U_F$ ($U_F^2=1$) on a single vertex-type violation (quasiparticle) as an example. We call this operator $U^e_F$ and its local action on the corresponding quasiparticle excitation  is defined as (a detailed derivation of this result can be found in \cite{Hermele} ):
\begin{align}
(U^e_F)^2=K.
\end{align}
From the above equation, we can conclude that the action of $\mathbb{Z}^{fv}_2$ flavour symmetry on a single $e$-particle is determined by $K$. Equivalently, we can state that $\omega (U_F, U_F)=K$. The constant $K$ takes the values $K=\pm1$ and these two values give the two possible fractionalisation classes. The physical interpretation of this fact is the following:   the coupling constant $K=-1$ creates a pattern of fluxes on the ground state. When we move a single excitations applying the symmetry operator $U^e_F$, it interacts with the background fluxes and acquires a non-trivial phase. 
As a result, we conclude that there is a trivial fractionalization for $K=+1$ and a non-trivial fractionalization for $K=-1$. 

\section{The Bilayer Doubled Semion Model}\label{model}

\subsection{Defining Properties}
\noindent
In the previous section we have introduced the necessary elements to construct a new model with symmetry-enriched 
topological order (SETO): the bilayer Doubled Semion model. This model aims at the unification of the global flavour symmetry of the Hermele model with the Topological Order of the DS model. As a result, we will achieve the fractionalization of the underlying topological orders .
 
\noindent
We seek for a model involving two layers of the Doubled Semion model having the following two important properties:
i/ the model must be exactly solvable; ii/ the model must be invariant under the global flavour symmetry $U_F$ explained in Eq.\eqref{eq: flavour symmetry}.

\begin{figure}
\includegraphics[width=\linewidth]{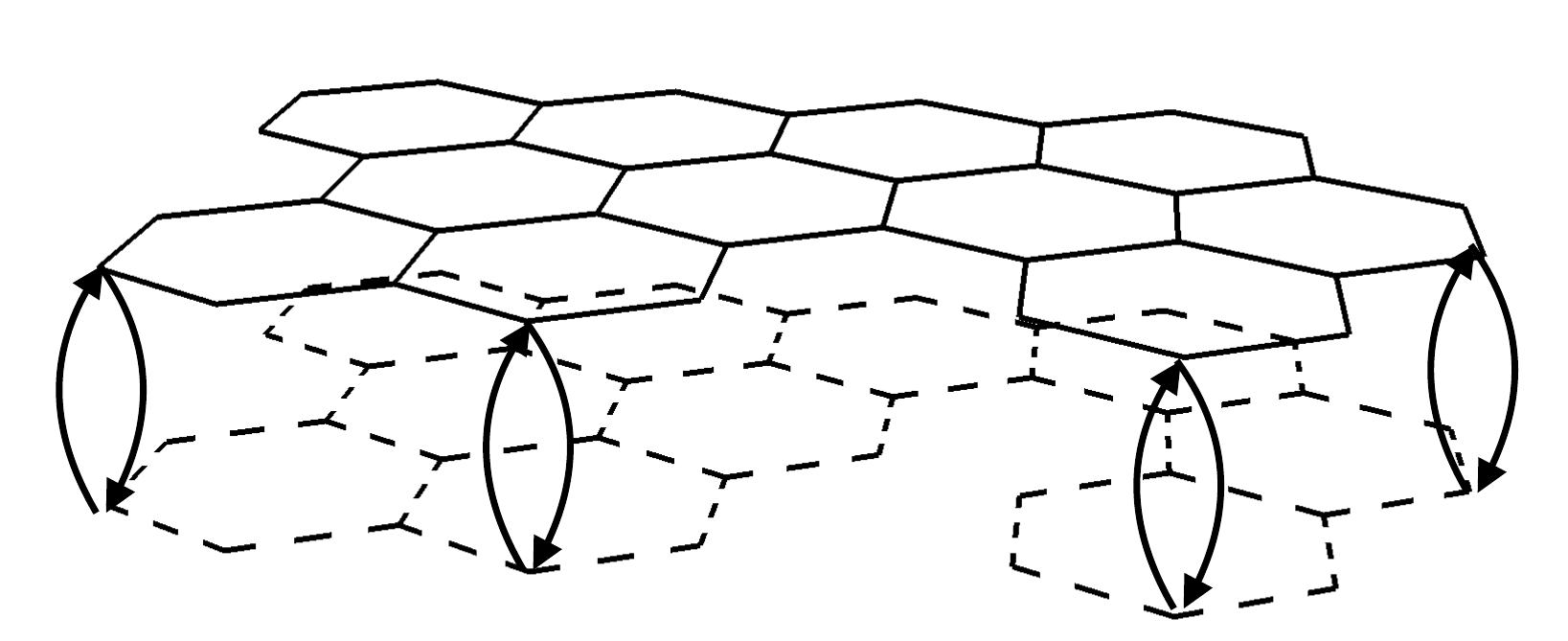}
\caption{\label{fig:model} Bilayer hexagonal lattice needed to construct the bilayer Double Semion model. The degrees of freedom are spins residing  on the links, both horizontal and vertical. Vertical links connect the two layers by two distinct links with one spin on each. For clarity, only some of the vertical links are shown.}
\end{figure}

\begin{figure}
\includegraphics[width=0.60\linewidth]{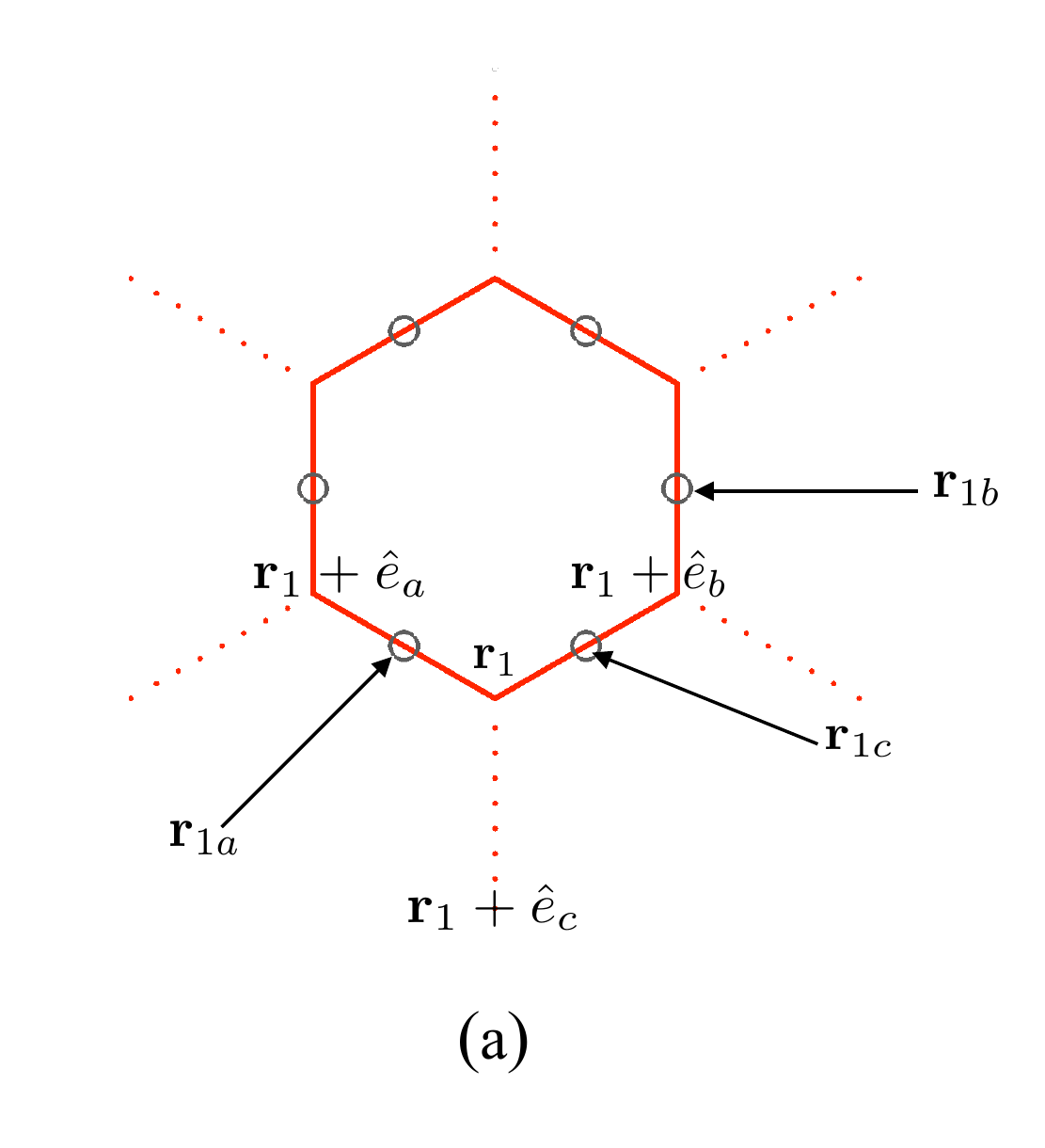}
\includegraphics[width=0.70\linewidth]{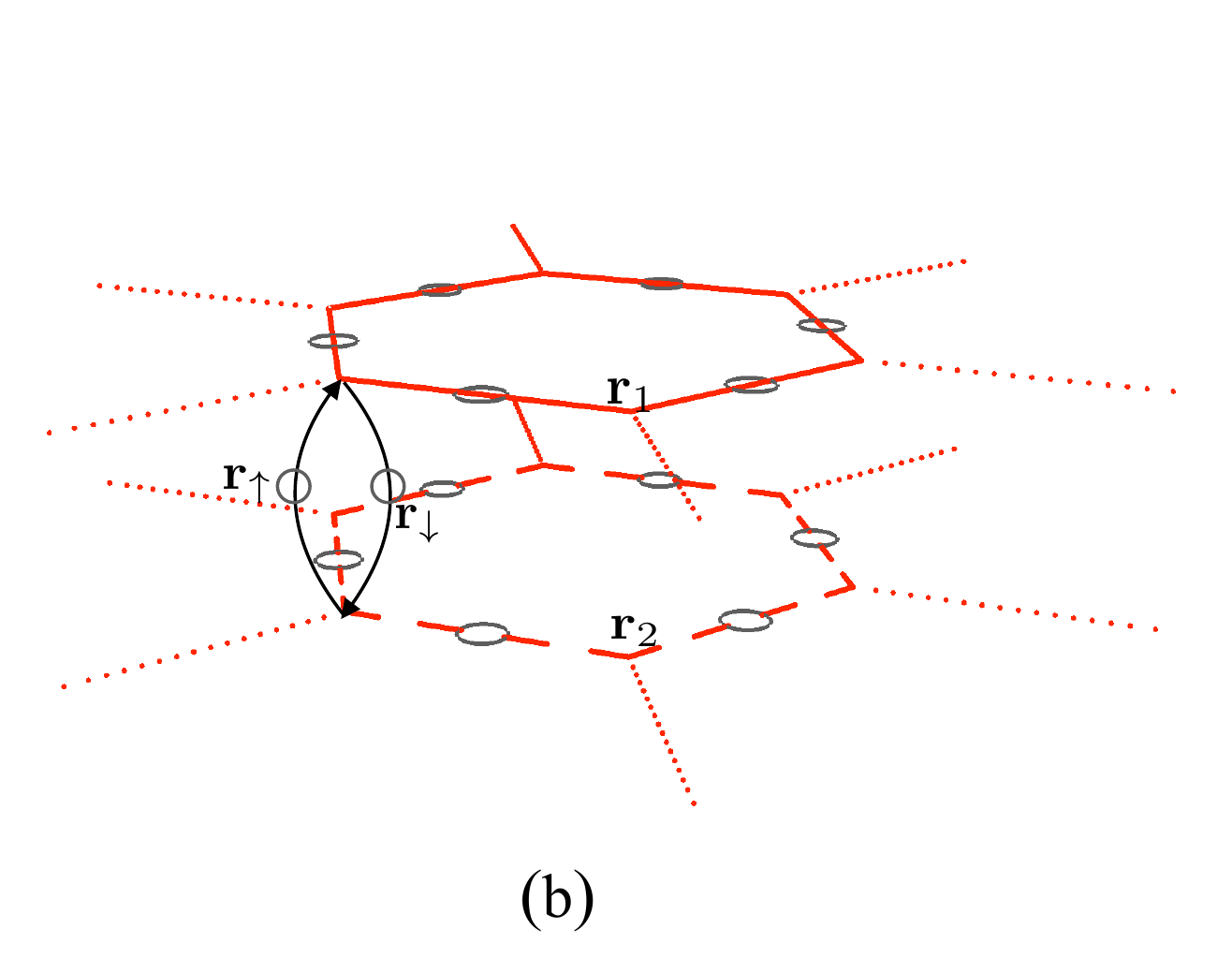}
\caption{\label{fig:notation}Details of a representative hexagonal plaquette in the bilayer lattice showing the labelling of vertices and links. Spins are located on the links and are represented by grey circumferences. The three posible directions on each hexagonal layer are denoted as $a$, $b$ and $c$, respectively. (a) A plaquette that belongs to one hexagonal layer is shown indicating the notation for the vertices. Links are labeled appropriately with their positions. Red solid lines represent links in $\partial p$, the boundary of the plaquette $p$, and dotted red lines denote the set $s(p)$ of outgoing links from the plaquette. (b) Vertical links joining the two layers are shown ($\mathbf{r}_\uparrow$ and $\mathbf{r}_\downarrow$). As we can see from the figure, there is a spin (grey circumference) on each vertical link.)}
\end{figure}

\noindent
The model we seek for is defined on an hexagonal bilayer lattice shown in Fig.\ref{fig:model}. On each link of each  hexagonal layer we place a spin-1/2.
There is also  a spin-1/2 spins on each vertical link joining the two layers.
Some notation is introduced to label the links and the vertices on the lattice, as well as the Pauli matrices $\sigma^{\mu}_l$ acting on each spin at a given link $l$  (see Fig.\ref{fig:notation}). 
Each vertex in an hexagonal cell is denoted with $\mathbf{r}$. The vectors $\hat{e}_\alpha$ (with $\alpha=a,b,c$) are unit vectors in the three possible directions indicated in Fig.\ref{fig:notation}. The vertices in the upper and lower layers are denoted by $\mathbf{r}_1$ and $\mathbf{r}_2$, respectively. Links within each hexagonal lattice are labeled by $\mathbf{r}_{1\alpha}$  and $\mathbf{r}_{2\alpha}$ being $\alpha=a,b,c$. Finally, we denote  by $\mathbf{r}_\uparrow$ and $\mathbf{r}_\downarrow$ the two vertical links joining the hexagonal layers. 

\noindent
The model combines two different topological orders. First, we place the DS model on the hexagonal layers. Then, a version of Kitaev model is placed on the links joining the two layers. This Kitaev-like hamiltonian depends on a coupling constant, $K=\pm1$, that gives rise to charge fractionalization as we will see later on. We must demand that the set of vertex and plaquette operators of the bilayer DS model commute among themselves in order to construct an exactly solvable hamiltonian. However, we know in advance that this cannot be true as such since plaquette operators of DS type do not commute among themselves. Moreover, these plaquettes are now embedded into the bilayer model with additional interactions. Thus, we need some kind of invariant subspace condition in which commutativity is preserved. 
 As the bilayer DS model combines DS topological order on the layers and Kitaev topological order on the interlayers, it is natural to impose a mixed invariant subspace condition, namely: to preserve the DS zero-flux rule on the layers while leaving the interlayer couplings free of this condition. These considerations lead to define the invariant subspace for our bilayer DS model:
\begin{subequations}\label{eq:zero_flux_bDS}
\begin{equation}\label{eq: zero_flux_rule_cover}
A^{\text{cover}}_{\mathbf{r}i}\ket\psi=+\ket\psi,
\end{equation}
\begin{equation}\label{eq: zero_flux_rule_frontal}
A^{\text{frontal}}_{\mathbf{r}}\ket\psi=\pm\ket\psi,
\end{equation}
\end{subequations}
where $i=1,2$ stands for the upper and lower layers, respectively and
 \begin{align}\label{cover_frontal}
A_{\mathbf{r}i}:=A^{\text{cover}}_{\mathbf{r}i}\otimes A^{\text{frontal}}_{\mathbf{r}}.
\end{align}
We split the vertex operator in two different parts. The first part, $A^{\text{cover}}_{\mathbf{r}i}$, corresponds to a purely vertex operator
on the layers $i=1,2$, either top or bottom, while $A^{\text{frontal}}_{\mathbf{r}}$ is defined on the links of the interlayers. Explicit forms of these
components will be given in the next Subsection, \ref{ssec:model_hamiltonian}. The conditions in Eq.\eqref{eq:zero_flux_bDS} have the following interpretation: we impose the zero-flux rule only on the hexagonal layers, where the DS model is implemented. In the invariant subspace where Eq.(\ref{eq: zero_flux_rule_cover}) is fulfilled, the DS model on the hexagonal lattice is an exactly solvable model. Eq.(\ref{eq: zero_flux_rule_frontal}) shows that for the Kitaev model that we implement on the interlayer spins, the zero-flux condition is not necessary to be exactly solvable. Consequently, no restriction is imposed on these spins. Despite being a natural choice of conditions for the new vertex operators, it is far from trivial to find out a set of vertex and plaquette operators on the bilayer lattice such that they commute among themselves and fulfil these new invariant-subspace conditions. In the following, we construct such a set of operators leaving for Appendix \ref{app:commutativity} the proof of their commutativity. 

\subsection{Model Hamiltonian}\label{ssec:model_hamiltonian}

\noindent
Taking into account the considerations explained in the previous Section we now specify the expressions for vertex and plaquette operators. These operators are local and this property makes the system gauge invariant. The gauge group characterising the topological order. underlying our model is $\mathcal{G}=\mathbb{Z}_2$. Then, we formulate our hamiltonian as a sum of every plaquette and vertex operator on the spins of the system. Using these building blocks we ensure that our hamiltonian is also invariant under the gauge symmetry $\mathcal{G}=\mathbb{Z}_2$.
\begin{figure}
\includegraphics[width=\linewidth]{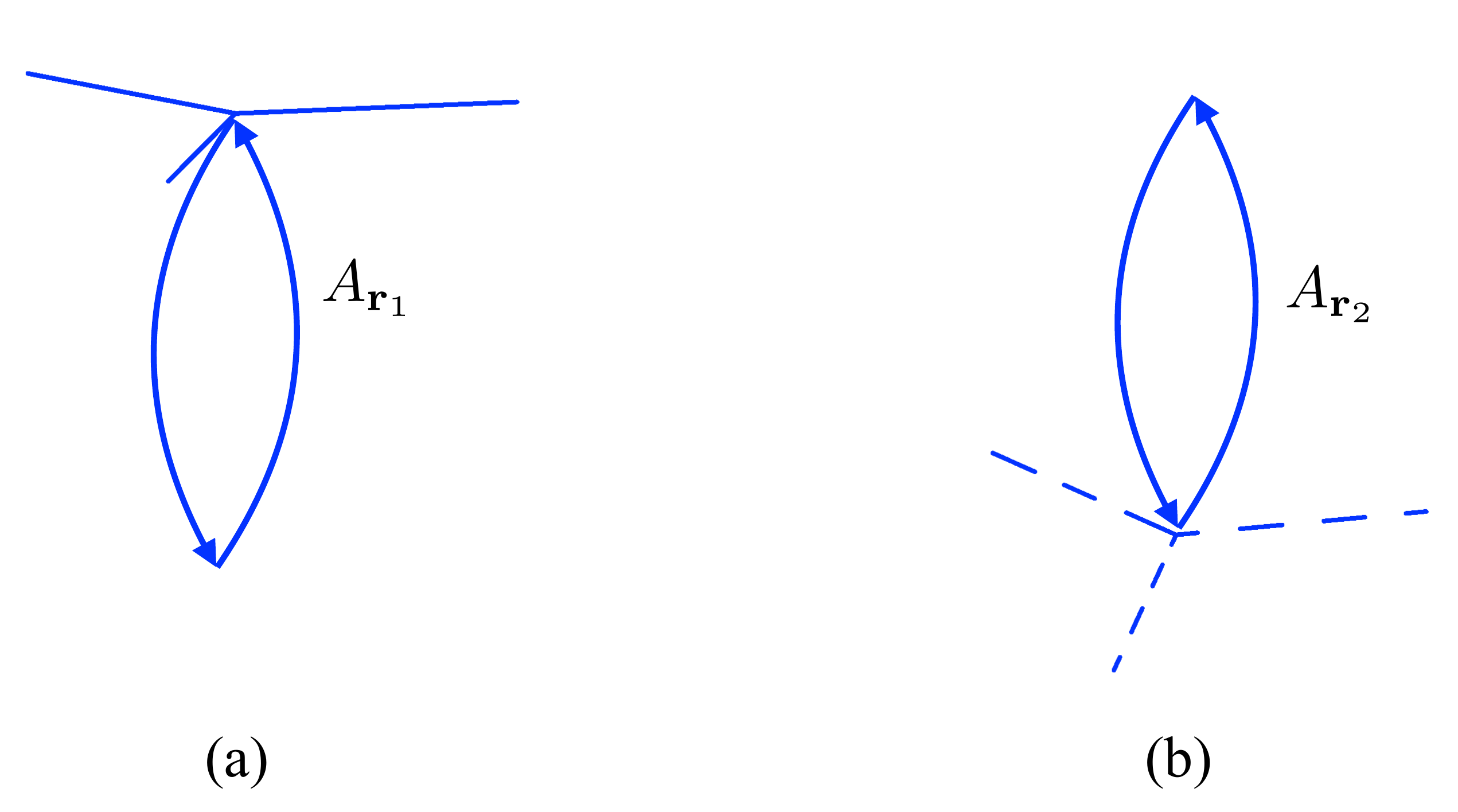}
\includegraphics[width=0.85\linewidth]{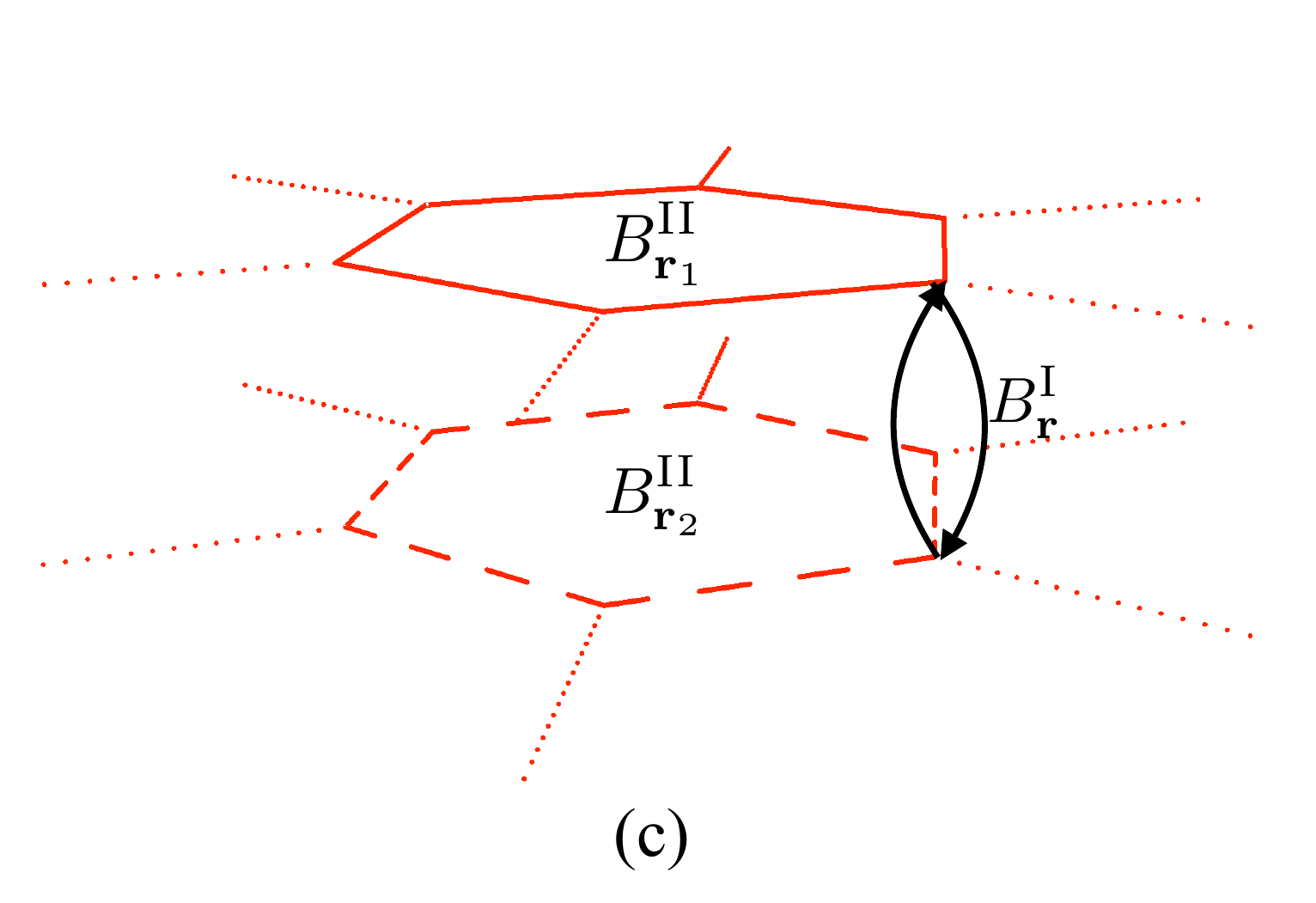}
\includegraphics[width=0.95\linewidth]{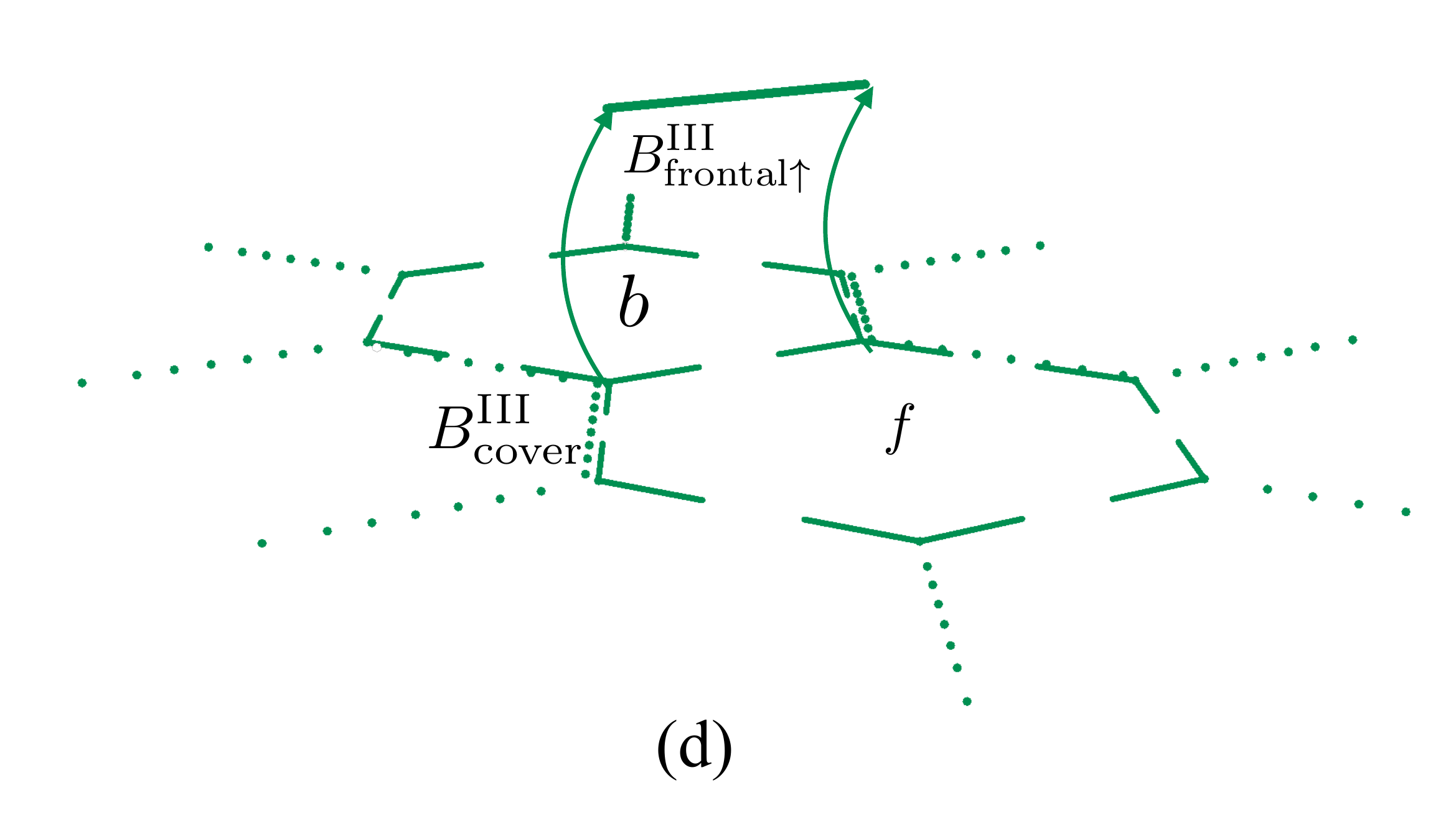}
\caption{\label{fig:vertex_plaquettes}  Examples of vertex an plaquette operators are shown. (a) Blue solid lines represent a vertex operator located on the upper layer. (b) Similar vertex operator for the bottom layer. Dashed lines are used to distinguish the lower layer from the upper layer. (c) Plaquettes of type $\textrm{I}$ are pictured in black solid lines. In red, two examples of type II plaquettes $B_{\mathbf{r}i}^{\textrm{II}}$ are shown. (d) An example of a plaquette  of type ${\textrm{III}}$ is illustrated in green lines, namely $B_{\mathbf{r}i\alpha \uparrow}^{\textrm{III}}$. The two parts introduced in Eq.(\ref{eq:plaquette_III}) are also indicated in the figure, together with backward and  forward plaquettes denoted as $b$ and $f$. }
\end{figure}

\noindent
We first study vertex operators. These operators act not only on the layers but also on the vertical spins joining the two layers.  These two distinguished parts are the already mentioned factors $A^{\text{cover}}_{\mathbf{r}i}$ and $A^{\text{frontal}}_{\mathbf{r}}$. Therefore, vertex operators are defined as (see Fig.\ref{fig:vertex_plaquettes} (a) and (b)):
\begin{align}\label{eq: vertex}
A_{\mathbf{r}i}:=A^{\text{cover}}_{\mathbf{r}i}\otimes A^{\text{frontal}}_{\mathbf{r}}= \prod_{\mathbf{r}i\alpha\in s(v)}\sigma^z_{\mathbf{r}i\alpha}\;\sigma^z_{\mathbf{r}\uparrow}\sigma^z_{\mathbf{r}\downarrow},
\end{align}
where the first subindex $\mathbf{r}$ locates the vertex on the hexagonal layer and the second subindex $i=1,2$, sets on which layer the vertex operator is defined.  The first product runs over $s(v)$,  which is the set of three edges radiating from every vertex $v$ on the hexagonal layer. The second part of the operator denotes a product of the two spins  joining the layers. Thus, the vertex operators implement a spin interaction coupling spins within the same layer and also spins on the vertical links.

\noindent
We move on to specify the plaquette operators. Three types of plaquette operators are defined, namely: $B_{\mathbf{r}}^{\textrm{I}}$, that acts just on the two vertical links  joining the hexagonal layers; $B_{\mathbf{r}i}^{\textrm{II}}$, which is defined on the upper or lower layer  and $B_{\mathbf{r}i\alpha\uparrow}^{\textrm{III}}$ ($B_{\mathbf{r}i\alpha\downarrow}^{\textrm{III}}$) which couples spins both on the layers and on the interlayer links. 
Their explicit definitions are given by: 
\begin{align}
B_{\mathbf{r}}^{\textrm{I}}:=\sigma_{\mathbf{r}\uparrow}^x\sigma_{\mathbf{r}\downarrow}^x.
\end{align}

\noindent 
These plaquette operators (see Fig.\ref{fig:vertex_plaquettes} (c))  consist  of the two links $\mathbf{r}\uparrow$ and $\mathbf{r}\downarrow$ on each vertex $\mathbf{r}$. Since they act only on two spins, we refer to them as degenerate plaquettes. These are the plaquette operators which have a coupling constant $K=\pm1$ in the Kitaev-like Hamiltonian. This factor $K$ introduces a non trivial fractionalization in the charge of the spin on the interlayer links when $K=-1$.
\noindent 
Plaquette operators of type $\textrm{II}$  are the well-known DS-model plaquettes  on the hexagonal lattice. They act on the two layers independently. Fig.\ref{fig:vertex_plaquettes} (c) shows two examples in red lines: 
\begin{align}
&B_{\mathbf{r}i}^{\textrm{II}}:=\prod_{\mathbf{r}i \alpha\in \partial p}\sigma_{\mathbf{r}i\alpha}^x\prod_{\mathbf{r}'i \alpha'\in s(p)}i^{(1-\sigma_{\mathbf{r'}i\alpha'}^z)/2}.
\end{align}
 The first product runs  over $\partial p$, i.e., the set of links in the boundary of plaquette $p$. The second product is over $s(p)$, the six edges radiating from the vertices in  $\partial p$ (dotted  red lines in Fig.\ref{fig:vertex_plaquettes} (c)). This second product includes a phase which depends on the spin in $s(v)$. Due to these extra phases in the plaquettes, semions will  appear in the system. We point out that the hamiltonian is hermitian despite the factors $\textrm{i}$ (See Appendix \ref{app:zero_flux_rule}).

\begin{figure*}
\includegraphics[width=0.8\linewidth]{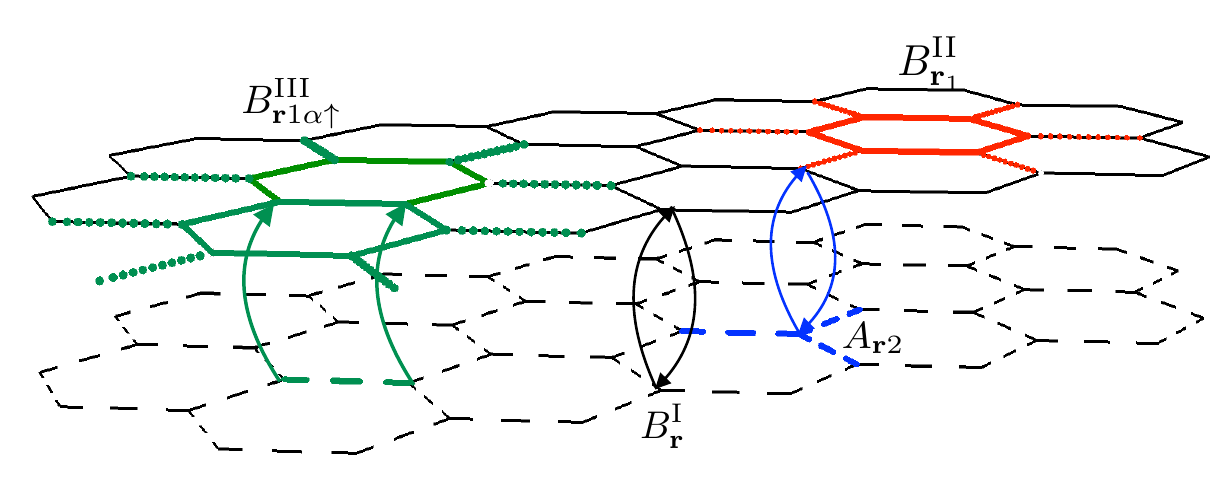}
\caption{\label{fig:lattice_plaquettes}
General view of the lattice structure associated to the construction of the bilayer DS model. A vertex operator on the lower hexagonal
layer is shown $A_{\mathbf{r}2}$ (blue). A plaquette operator of type I, $B^{\textrm{I}}_{\mathbf{r}}$, appears connecting upper and lower layers  (black solid). A plaquette
operator of type II, $B_{\mathbf{r}i}^{\textrm{II}}$, is shown in red. The novel plaquette of type III corresponding to the
upper layer $B_{\mathbf{r}1\alpha\uparrow}^{\textrm{III}}$ is shown with its cover and frontal components (green lines). }
\end{figure*}

\noindent 
Finally, we describe the structure of the plaquette operator of type $\textrm{III}$. It acts on the frontal links as well as  on the hexagonal layers. Therefore, it couples the two topological orders present in the model: the Kitaev- like on the interlayer links and the DS model implemented on both hexagonal layers.  To clarify these two different parts, the operator is  written as a  tensor product of two factors (see Fig.\ref{fig:vertex_plaquettes} (d)): 
\begin{align}\label{eq:plaquette_III}
B_{\mathbf{r}i\alpha \uparrow}^{\textrm{III}}:=B_{i,\text{cover}}^{\textrm{III}}\otimes B_{\text{frontal}\uparrow}^{\textrm{III}},\\
B_{\mathbf{r}i\alpha \downarrow}^{\textrm{III}}:=B_{i,\text{cover}}^{\textrm{III}}\otimes B_{\text{frontal}\downarrow}^{\textrm{III}}.\nonumber
\end{align}
As it is shown in the above equation, plaquettes of type $\textrm{III}$ come in two flavours: $B_{\mathbf{r}i\alpha \uparrow}^{\textrm{III}}$ and $B_{\mathbf{r}i\alpha \downarrow}^{\textrm{III}}$. The differences among them are in the frontal part, i.e. $\mathbf{r}\uparrow$ or $\mathbf{r}\downarrow$  interlayer links are included.
The cover part consists of two hexagonal plaquettes endowed with external legs, similar to two plaquettes of type $\textrm{II}$. The two cover plaquette operators are labeled backward ($b$) and forward ($f$) (see Fig.\ref{fig:vertex_plaquettes} (d)):
\begin{align}\label{eq:cover_b_f}
B_{i,\text{cover}}^{\textrm{III}}:=B_{\mathbf{r}i,b}^{\textrm{II}}B_{\mathbf{r}i,f}^{\textrm{II}}.
\end{align}
 The subindex $i=1,2$ in $B_{\mathbf{r}i\alpha \uparrow}^{\textrm{III}}$ shows on which layer the cover operator acts. To specify the $B_{\mathbf{r}i\alpha \uparrow}^{\textrm{III}}$ operator the subindex $\alpha$ is needed. $\alpha=a,b,c$ runs over the three spatial direction on the hexagonal layer.  As we have already mentioned, the frontal part includes the spins  $\mathbf{r}_\uparrow$ (or $\mathbf{r}_\downarrow$) of two neighbouring vertices ($\mathbf{r}$ and $\mathbf{r+\hat{e}_\alpha}$) and one link of each layer: 
 \begin{align}\label{eq:frontal_op}
B_{\text{frontal}\uparrow}^{\textrm{III}}:= \sigma^{x}_{\mathbf{r}\uparrow} \sigma^{x}_{\mathbf{r}1\alpha}\sigma^{x}_{\mathbf{r}2\alpha} \sigma^{x}_{\mathbf{r+\hat{e}_\alpha}\uparrow}, \\
B_{\text{frontal}\downarrow}^{\textrm{III}}:= \sigma^{x}_{\mathbf{r}\downarrow} \sigma^{x}_{\mathbf{r}1\alpha}\sigma^{x}_{\mathbf{r}2\alpha} \sigma^{x}_{\mathbf{r+\hat{e}_\alpha}\downarrow}.\nonumber
\end{align}
In Fig.\ref{fig:lattice_plaquettes} a general view of the bilayer hexagonal lattice is pictured along with the previously introduced vertex and plaquette
operators needed to construct the bilayer DS model.
In Appendix \ref{app:commutativity} we show that the set of vertex and plaquette operators thus constructed commute. Then, we can built an exactly solvable model out of these operators.

\noindent
Now, with this set of vertex and plaquette operators just defined,  we can introduce the hamiltonian of our bilayer DS model as follows: 
\begin{align}\label{eq: hamiltonian}
&\mathcal{H}_{bDS}:=-\sum_{\mathbf{r}}\sum_{i=1,2} A_{\mathbf{r}i}- K\sum_{\mathbf{r}}B^{\textrm{I}}_{\mathbf{r}}
+\sum_{\mathbf{r}}\sum_{i=1,2}B_{\mathbf{r}i}^{\textrm{II}}
\\\nonumber
&-\sum_{\mathbf{r}}\sum_{i=1,2}\sum_{\alpha= a,b,c}(B_{\mathbf{r}i\alpha\uparrow}^{\textrm{III}}+B_{\mathbf{r}i\alpha\downarrow}^{\textrm{III}}).\end{align}
The above hamiltonian fulfils the two properties we were searching for, namely, it is exactly solvable and invariant under the global
flavour symmetry that exchanges the upper and lower layers. The latter property is shown in
the next subsection. This model also combines the DS model and the Kitaev-like model, creating two different kinds of fractionalization. The excitations created by a violation of the ground state condition for the plaquette operators on the layers corresponds to semions. Within the interlayer space, violations of degenerate plaquette operators create an e-charge. The former and the latter type of excitations coexist with the bilayer structure and have different charge fractionalization.

\begin{table}[t]
\begin{tabular}{|P{2.5 cm}|P{1 cm}|}
\toprule
\hline
\hline
\multicolumn{2}{c}{Two different fractionalization classes} \\
\hline
& \ \ $\mathbb{Z}^{fv}_2$ \\
\hline
$[\omega]_{K}$\ \ &\hspace{6mm} K \\
\hline
\hspace{1.5mm}$[\omega]_{DS}$ &\hspace{4mm}${-K}$\\
\hline
\bottomrule
\end{tabular}
\caption{This table shows the fractionalization classes for a modified-Kitaev model, $[\omega_K]$ and the DS model, $[\omega_{DS}]$ . As we can see from the table, these fractionalization classes are due to  the flavour symmetry ($\mathbb{Z}^{fv}_2$) -- which is explained in Eq.\eqref{eq: flavour symmetry} --.  The equivalence class $[\omega]=1$ corresponds to a trivial fractionalization and a non-trivial fractionalization is obtained otherwise. $K=\pm1$  is the coupling constant in Eq.\eqref{eq: hermele_hamiltonian}
.  }
\label{tab:fractionalization class}
\end{table}

\subsection{Ground state and $\mathbb{Z}_2$ Flavour Symmetry }\label{ssec:gs_flavour_sym}

Now that we understand how the hamiltonian of the bilayer DS model is built and under which conditions is an exactly solvable model, we move on to calculate the ground state of it. This will play a central role in the construction of protected edge states in Section \ref{edge_states}. We must also 
check that the ground state is invariant under the global $\mathbb{Z}^{fv}_2$ on-site flavour symmetry introduced in Eq.\eqref{eq: flavour symmetry}. Using the fact that the eigenvalues of all vertex and plaquette operators in Eq.(\ref{eq: hamiltonian}) are $\pm1$, we can easily obtain the lowest-energy state conditions. 
Thus, any state satisfying:
\begin{align}\label{eq: gs_condition}
A_{\mathbf{r}i}&=+1, \hspace{5mm} B_{\mathbf{r}}^{\textrm{I}}=K,
\\
B_{\mathbf{r}i}^{\textrm{II}}=-1,   \hspace{5mm} &B_{\mathbf{r}i\alpha \uparrow}^{\textrm{III}}=+1, 
 \hspace{5mm} B_{\mathbf{r}i\alpha \downarrow}^{\textrm{III}} =+1, \nonumber
\end{align}
is a ground state of the system. From the above equation, we remark that one of the conditions depends on the coupling constant $K$, i.e. the energy of the ground state remains the same for the two values of K, $K=\pm1$, because the ground state condition accordingly changes. Every ground state is an equal superposition of closed strings generated by combinations of plaquette operators compatible with the lattice structure. However, the coefficients of this superposition  for $K=+1$ are different from those for $K=-1$. Similarly to Eq.\eqref{eq: gs} and Eq.\eqref{eq:gs_hermele}, we can write explicitly the ground states by means of the plaquette operator projectors. Thus, we consider the vacuum state corresponding to the vertex operators, i.e. $\ket{0}^{\otimes N}$. Then, the plaquette projectors  act on this vacuum giving rise to a loop gas configuration. Models which have this characteristic loop gas ground state are called \textit{string} models.

\begin{figure}
\includegraphics[width=\linewidth]{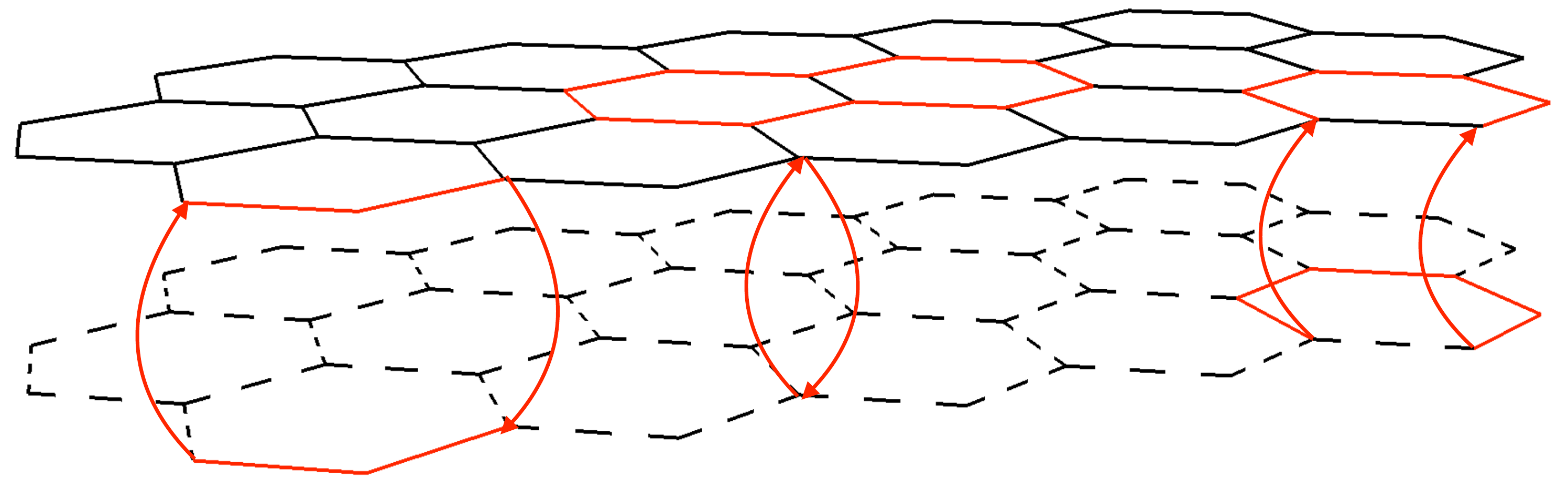}
\caption{\label{fig:loop_gas}. In this figure, a sample of a possible ground state for the bilayer DS model is shown (Eq.\eqref{eq: gs: no_simplication}). The loops in this model are made by strings that are spread all over the  system.  Links in red represent the spins where a product of plaquette operators is acting. Links where the projectors do not act are shown in black.}
\end{figure}

\noindent
The complete expression for the ground state of the bilayer DS model is the following:
\begin{align}\label{eq: gs: no_simplication}
|\psi_0\rangle=\prod_{p\in B^{\textrm{I}}_{\mathbf{r}}}(1+K  B^{\textrm{I}}_{\mathbf{r}})
\prod_{p\in B^{\textrm{II}}_{\mathbf{r}i}}(1- B^{\textrm{II}}_{\mathbf{r}i})\\
\prod_{p\in B^{\textrm{III}}_{\mathbf{r}i\alpha\uparrow}}(1+ B^{\textrm{III}}_{\mathbf{r}i\alpha\uparrow})\prod_{p\in B^{\textrm{III}}_{\mathbf{r}i\alpha\downarrow}}(1+ B^{\textrm{III}}_{\mathbf{r}i\alpha\downarrow})|0\rangle^{\otimes N}\nonumber
\end{align}
This expression has the desired  structure to fulfil conditions in Eq.\eqref{eq: gs_condition}. 
 A schematic picture of a possible loop gas ground state is plotted in Fig.\ref{fig:loop_gas}.

\noindent
We have seen in Section \ref{string flux mechanism} that the global symmetry plays an important role in the charge fractionalisation of a topological order. Now we move on to show that the hamiltonian and the ground state of the bilayer DS model are invariant under  $\mathbb{Z}^{fv}_2$ flavour symmetry.

\noindent
From Eq.(\ref{eq: flavour symmetry}), we know how the $\mathbb{Z}^{fv}_2$ flavour symmetry acts on Pauli operators. Taking into account this information, it is possible to show that the hamiltonian is invariant under this symmetry. Let us analyse each term of the hamiltonian in Eq.(\ref{eq: hamiltonian}) and prove that each one of them is invariant under  $\mathbb{Z}^{fv}_2$ flavour symmetry. Firstly, we analyse the behaviour of the vertex operators under $\mathbb{Z}^{fv}_2$  symmetry:
\begin{align}
U_FA_{\mathbf{r}1}U_F^{-1}=U_F\prod_{\mathbf{r}1\in s(v)}\sigma^z_{\mathbf{r}1}\prod_{\mathbf{r}\in \{\uparrow, \downarrow\}}\sigma^z_{\mathbf{r}}\hspace{2mm}U_F^{-1}\nonumber,
\end{align}
where $U_F$ is the unitary operator that implement the $\mathbb{Z}^{fv}_2$ flavour symmetry on Pauli operators. Now, we can work out how the operator $U_F$ acts on the two products separately:
\begin{align}\label{eq: vertex_flavour_symmetry}
U_F\prod_{\mathbf{r}1\in s(v)}\sigma^z_{\mathbf{r}1}\hspace{1mm}U_F^{-1}=\prod_{\mathbf{r}2\in s(v)}\sigma^z_{\mathbf{r}2}\nonumber,\\
U_F\hspace{2mm}\sigma^z_{\mathbf{r}\uparrow}\sigma^z_{\mathbf{r}\downarrow}\hspace{2mm}U_F^{-1}=\sigma^z_{\mathbf{r}\downarrow}\sigma^z_{\mathbf{r}\uparrow}=\sigma^z_{\mathbf{r}\uparrow}\sigma^z_{\mathbf{r}\downarrow}\nonumber.
\end{align}
\newline
As we can see from the above results, the first part of the vertex operator is not invariant under $\mathbb{Z}^{fv}_2$ flavour symmetry, i.e. $A_{\mathbf{r}1}$ transforms into $A_{\mathbf{r}2}$ and viceversa. However, if we take into account the sum over the two layers $i=1,2$ in the first term in Eq.(\ref{eq: hamiltonian}), this term is indeed invariant under the flavour symmetry.  

\noindent
Now let us  consider the plaquette operator $B_{\mathbf{r}}^{\textrm{I}}$ under the action of the flavour symmetry:
\begin{align}
U_F B_{\mathbf{r}}^{\textrm{I}}U_F^{-1}=U_F \sigma_{\mathbf{r}\uparrow}^x\sigma_{\mathbf{r}\downarrow}^xU_F^{-1}=\sigma_{\mathbf{r}\downarrow}^x\sigma_{\mathbf{r}\uparrow}^x=B_{\mathbf{r}}^{\textrm{I}}.
\end{align}
Thus, $B_{\mathbf{r}}^{\textrm{I}}$ is itself invariant under $\mathbb{Z}^{fv}_2$ flavour symmetry. This is the reason why we only sum over all possible $\mathbf{r}$ in the lattice and no other sum has to be taken into consideration.

\noindent 
Next, we check the same property for plaquettes of type $\textrm{II}$:
\begin{align}
U_F B_{\mathbf{r}1}^{\textrm{II}}&U_F^{-1} =U_F \prod_{\mathbf{r}1 \alpha\in \partial p}\sigma_{\mathbf{r}1\alpha}^x\prod_{\mathbf{r}'1 \alpha'\in s(p)}\textrm{i}^{(1-\sigma_{\mathbf{r}1\alpha}^z)/2} \hspace{1mm}U_F^{-1}\nonumber\\
&=\prod_{\mathbf{r}2 \alpha\in \partial p}\sigma_{\mathbf{r}2\alpha}^x\prod_{\mathbf{r}'2 \alpha'\in s(p)}\textrm{i}^{(1-\sigma_{\mathbf{r}2\alpha}^z)/2}=B_{\mathbf{r}2}^{\textrm{II}}.\nonumber
\end{align}
\newline
The above equation shows that the plaquettes of type $\textrm{II}$ are not invariant under $\mathbb{Z}^{fv}_2$ flavour symmetry as such. As we did previously for vertex operators, considering the sum over $i=1,2$ in the hamiltonian, the third term in Eq.\eqref{eq: hamiltonian} remains invariant term because $ B_{\mathbf{r}1}^{\textrm{II}}$ turns into $B_{\mathbf{r}2}^{\textrm{II}}$ and viceversa.

\noindent
Finally, let us address the last term of the Hamiltonian in Eq. (\ref{eq: hamiltonian}). As for the previous plaquette operators, $B^{\textrm{III}}_{\mathbf{r}\alpha\uparrow}$ and $B^{\textrm{III}}_{\mathbf{r}\alpha\downarrow}$ are not invariant under $\mathbb{Z}^{fv}_2$ flavour symmetry by themselves. 
Explicitly, by analysing the cover part and the frontal part separately, we find:
\begin{align}
&U_F \hspace{1mm} B_{\text{1,cover}}^{\textrm{III}}U_F^{-1}=U_F \hspace{1mm}B_{\mathbf{r}1,b}^{\textrm{II}}B_{\mathbf{r}1,f}^{\textrm{II}}U_F^{-1}=B_{\mathbf{r}2,b}^{\textrm{II}}B_{\mathbf{r}2,f}^{\textrm{II}}\nonumber\\
&= B_{\text{2,cover}}^{\textrm{III}}\nonumber\\[4mm]
&U_F B_{\text{frontal}\uparrow}^{\textrm{III}} U_F^{-1}= U_F \sigma^{x}_{\mathbf{r}\uparrow} \sigma^{x}_{\mathbf{r}1\alpha}\sigma^{x}_{\mathbf{r}2\alpha} \sigma^{x}_{\mathbf{r+\hat{e}_\alpha}\uparrow} U_F^{-1}\nonumber\\
&= \sigma^{x}_{\mathbf{r}\downarrow}\sigma^{x}_{\mathbf{r}2\alpha}\sigma^{x}_{\mathbf{r}1\alpha} \sigma^{x}_{\mathbf{r+\hat{e}_\alpha}\downarrow}= B_{\text{frontal}\downarrow}^{\textrm{III}}\nonumber
\end{align}
\begin{figure*}
\includegraphics[width=0.49\linewidth]{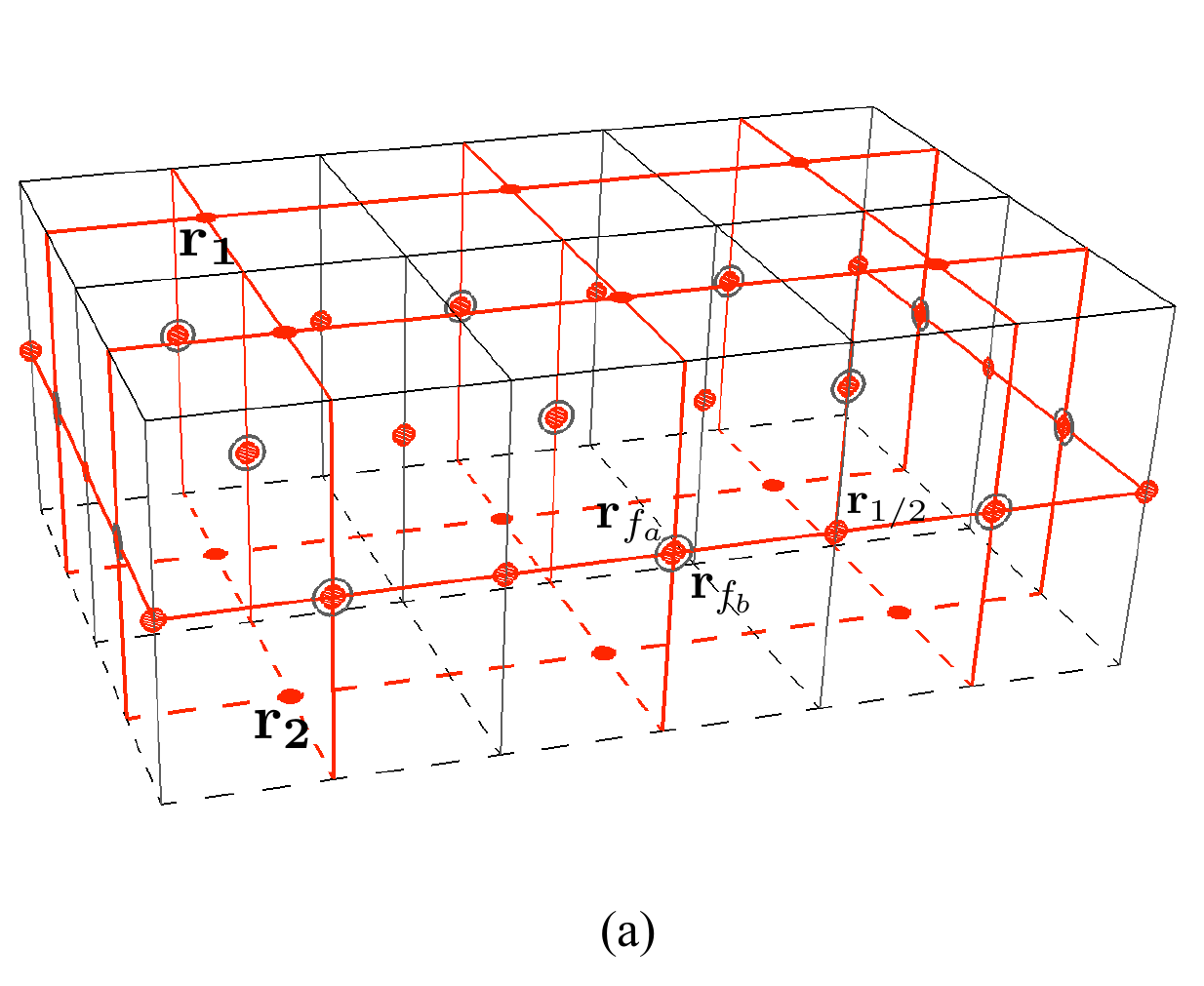}
\includegraphics[width=0.49\linewidth]{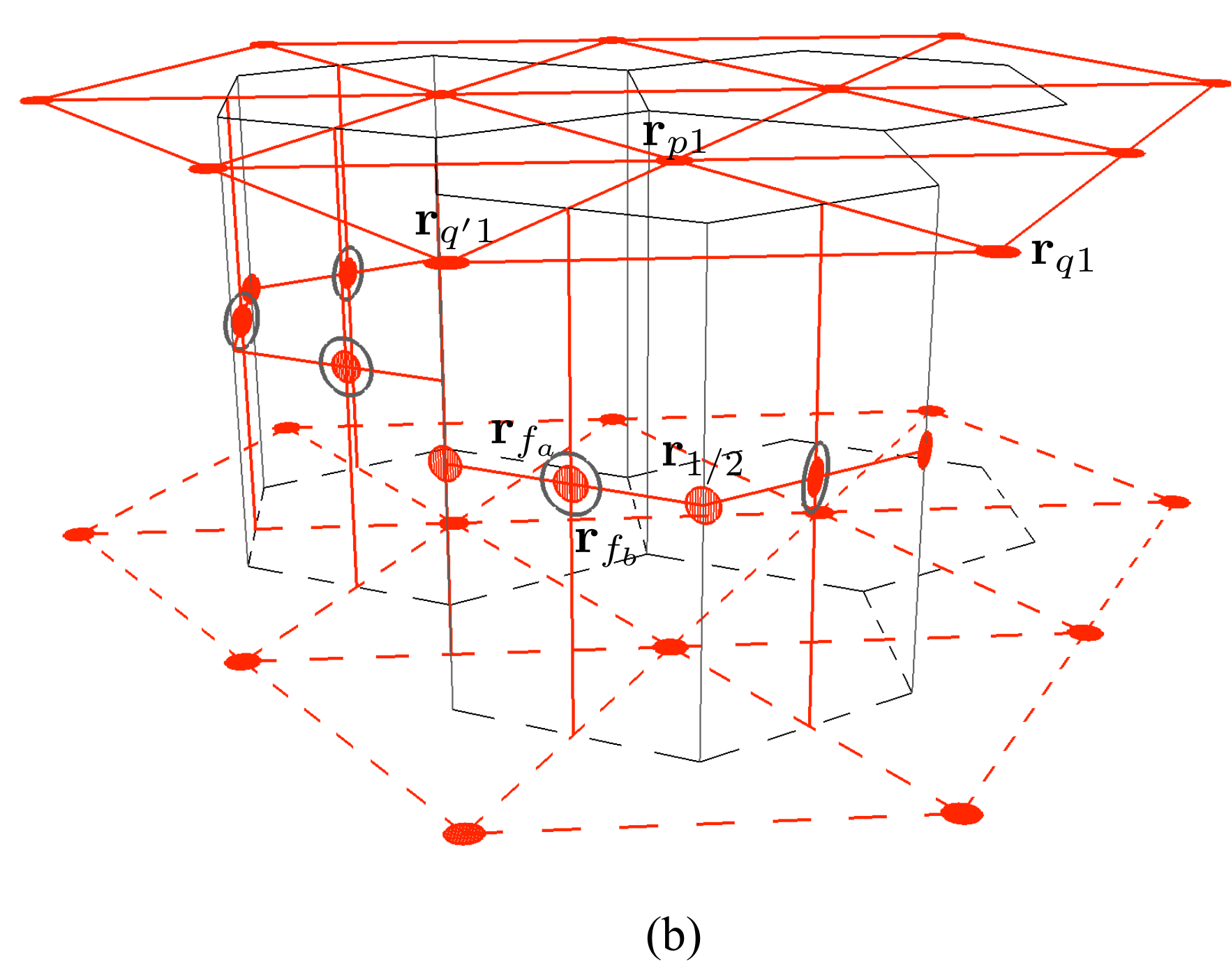}
\caption{\label{fig:dual_lattice_hermele_our_model} (a)  The dual lattice of the Hermele bilayer model presented in \cite{Hermele}.  (b) The dual lattice of the bilayer DS model presented in Section \ref{model}. Red lines correspond to the dual lattice and black is for the direct lattice. Dashed lines are used to distinguish the lower from the upper layer. Red circles presents the spins in the dual lattice.  Concentric red and grey circles means that there are two spins in the very same place (this is a consequence of the plaquettes of type II on the string model). Notation: spins on the layers are named by $\mathbf{r_1}$ (upper) and $\mathbf{r_2}$ (lower), respectively. Two vertices located on the same place are called $\mathbf{r}_{f_a}$ and $\mathbf{r}_{f_b}$. The vertices denoted as $\mathbf{r}_{1/2}$ come from the plaquettes of type I. In (b), the subindices $p$, $q$ and $q'$ indicate the vertices of a triangle used to define the hamiltonian. The spin hamiltonians that describe the interactions on these two lattices are in Eq.\eqref{eq: spin_model_hermele} and Eq.\eqref{eq: spin_model_our_model} respectively. We called as bilayer Trivial Paramagnet and bilayer Non-Trivial Paramagnet.} 
\end{figure*}

\noindent
Previous expressions prove that this plaquette operator is not invariant under the  $\mathbb{Z}^{fv}_2$  symmetry. However, there are two sums in the hamiltonian that run over $i=1,2$ and over $\uparrow$ and $\downarrow$. Since $B_{\text{frontal}\uparrow}^{\textrm{III}}$ transforms into $B_{\text{frontal}\downarrow}^{\textrm{III}}$ and $B_{1,\text{cover}}^{\textrm{III}}$  into $B_{\text{2,cover}}^{\textrm{III}}$, the fourth term in Eq.\eqref{eq: hamiltonian} remains invariant under $\mathbb{Z}^{fv}_2$ flavour symmetry.

\noindent
Taking into account the previous results we can prove that the hamiltonian is invariant under $\mathbb{Z}^{fv}_2$ :
\begin{equation}
U_F\mathcal{H}_{bDS}U^{-1}_F=\mathcal{H}_{bDS}.
\end{equation}
We can also prove that the ground state of the bilayer DS model (Eq.\eqref{eq: hamiltonian}) is also invariant under the flavour symmetry:
\begin{align}
U_F \ket{\psi_0} = \ket{\psi_0}.
\end{align}
This follows straightforwardly from the transformation laws of the different plaquette operators shown above when applied to
Eq.\eqref{eq: gs: no_simplication}. Consequently, the $\mathbb{Z}^{fv}_2$ flavour symmetry is not spontaneously broken in the bilayer DS model. 
As a result of all these calculations, we have proved that the hamiltonian and the ground state are invariant under the global $\mathbb{Z}_2$ flavour symmetry.

\noindent
To conclude, we want to remark that the model thus constructed succeeds  in combining the DS model and the string-flux mechanism. Therefore we have open strings from the DS model, which create semions and bosons on the hexagonal layers. Additionally, we also have open strings carrying  fractionalized charge at their endpoints due to the string-flux mechanism on the links joining the two layers. As we pointed out in Table \ref{tab:fractionalization class}, there are two types of fractionalization, corresponding with the two types of vertex excitations.

\section{Edge states for the bilayer Doubled Semion model}\label{edge_states}

\noindent
In the previous Section, a microscopic model that realises a SETO, called the bilayer Double Semion ($bDS$) model, is constructed. Interestingly enough, the string model we have already introduced is a gauged model. To obtain the spin we are searching for through this Section, we turned off the gauging symmetry. As mentioned in the Introduction, these kind of topological phases have a global symmetry on top of an intrinsic topological order. For the $bDS$ model, the global symmetry is $\mathbb{Z}^{fv}_2$ and the topological order is a combination of the DS model and a Kitaev-like model.This SETO phase can be mapped to a spin model that exhibits  edge states, a signature of SPTOs. This Section is devoted to build the edge states present in this dual model that we call Bilayer Non-Trivial Paramagnet (bNTP). Moreover,  the model introduced by Hermele in \cite{Hermele}, called  the bilayer Kitaev ($bK$) model through this Section, can also be mapped to a different spin model, the Bilayer Trivial Paramagnet (bTP) and its edge states  are shown for the first time. 

\noindent 
To construct the edge states for both models, the method introduced by Levin and Gu in \cite{Levin_Gu} is employed. According to it, the non-trivial self-statistics of the open string operators describing the excitations are responsible for the presence of edge states. This occurs when the model has a global symmetry group and it is placed
on a lattice with boundaries. This construction is based upon a duality between models with spin degrees of freedom (spin models, which are SPTO ) and models with string configurations (string models,  which are TO). Following this terminology, the models with topological order presented thus far are string models, like the Kitaev model and the DS model. Thus, in order to apply the Levin-Gu method, we shall need to construct the spin
model for the $bK$ and the $bDS$ model. Spin models provide a natural scheme to construct the edge states and to show them pictorially.

\noindent 
As the spin models are defined on dual lattices, we begin by constructing the dual lattice for a bilayer square lattice (corresponding to the $bK$ model) and for a bilayer hexagonal lattice (in the case of the $bDS$ model). Although both models are built on bilayer lattices, they are truly two-dimensional (2D) models as far as the dual lattice is concerned. Therefore, we apply the duality rules of 2D lattices first on the layers and then on the vertical links. The basic duality rules among lattices in 2D are: a vertex in the direct lattice becomes a plaquette in the dual lattice and viceversa. The number of links remains the same although they change orientation: a dual lattice link should be orthogonal to the equivalent one in the direct lattice. The resulting dual lattices for both the $bK$ and the $bDS$ models are shown in Fig.\ref{fig:dual_lattice_hermele_our_model} in red. The direct lattice is also pictured in the figure (black lines) to show the relation between them. 

\begin{figure}
\includegraphics[width=0.3\linewidth]{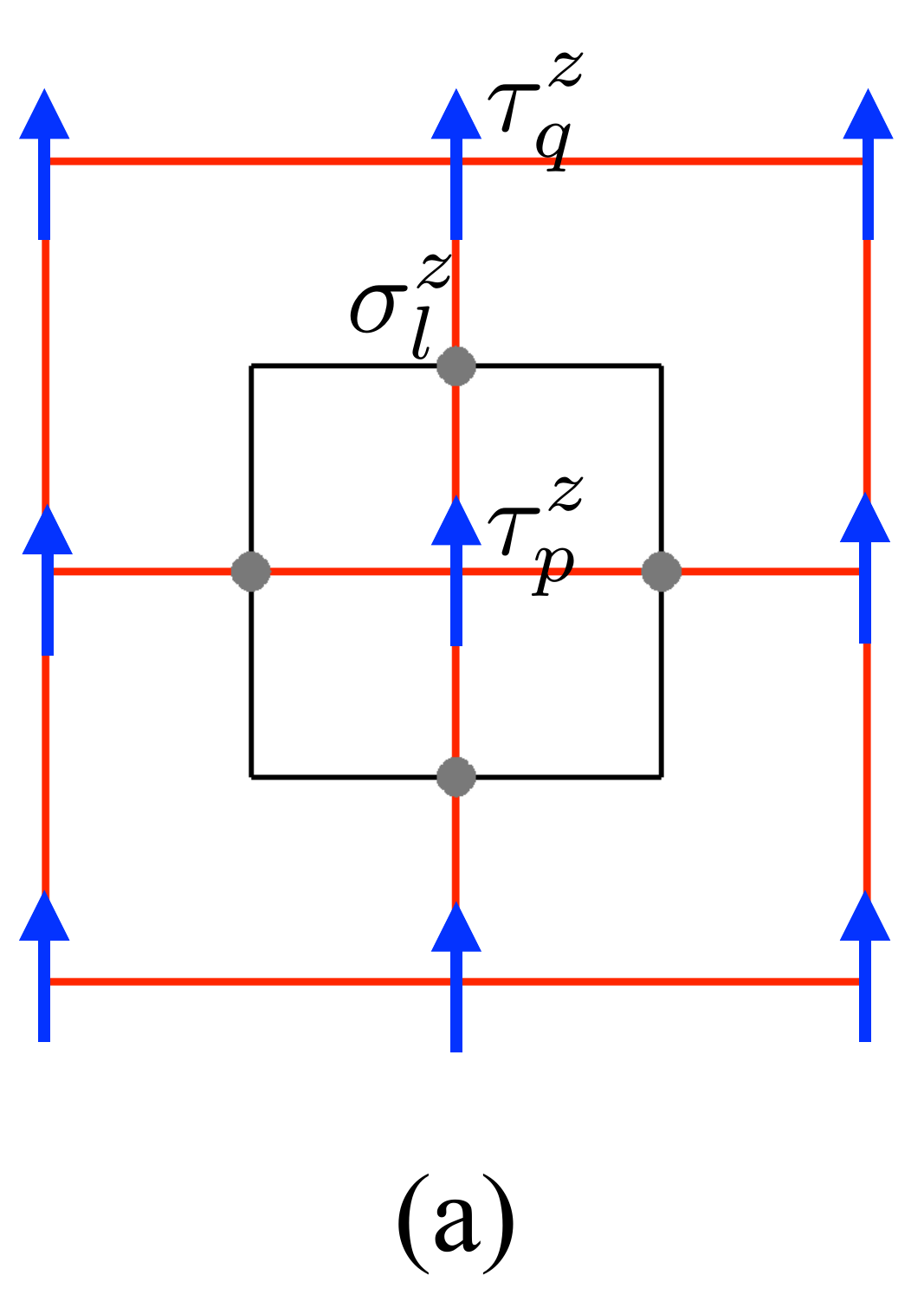}
\includegraphics[width=0.6\linewidth]{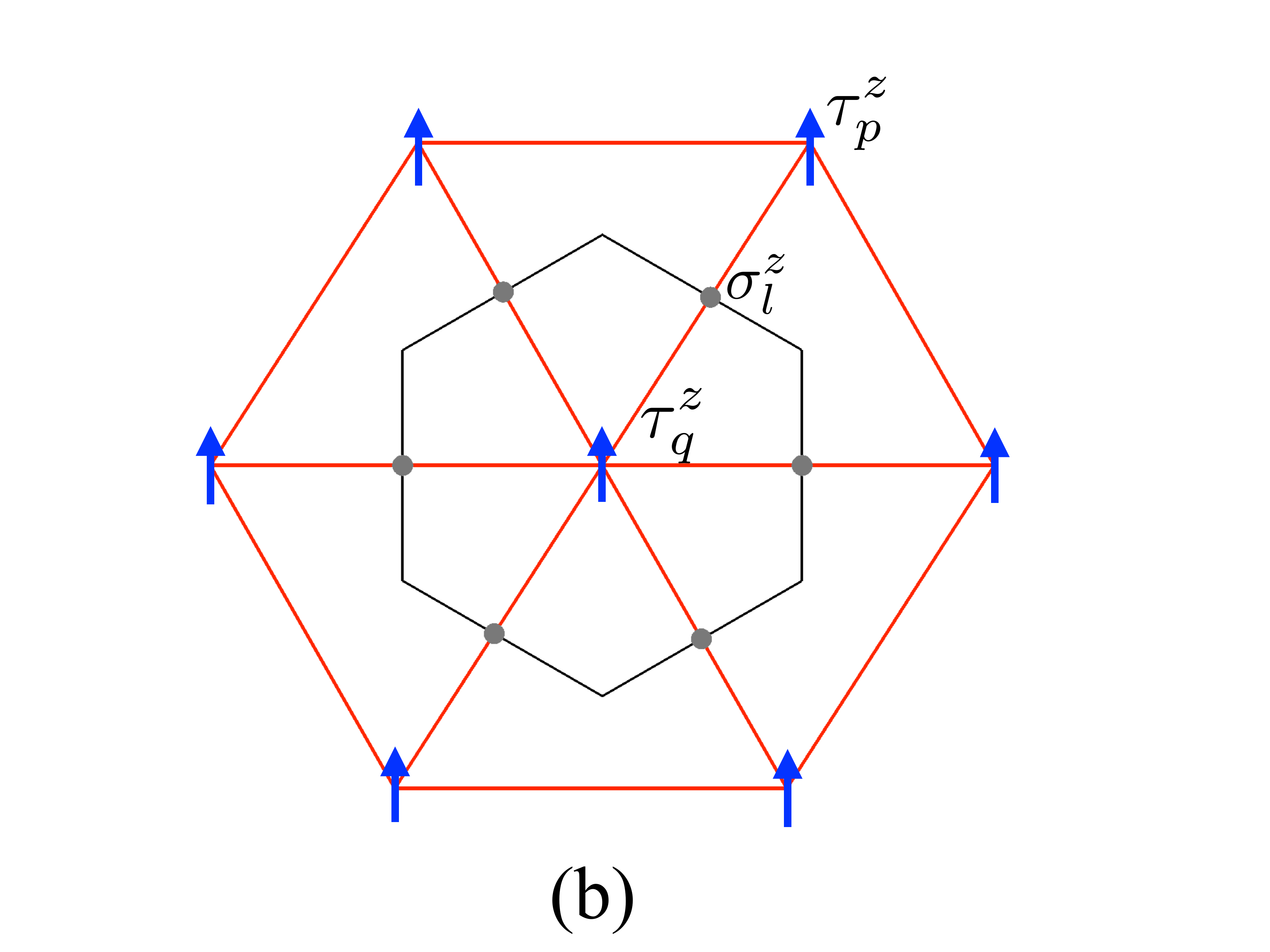}
\includegraphics[width=0.4\linewidth]{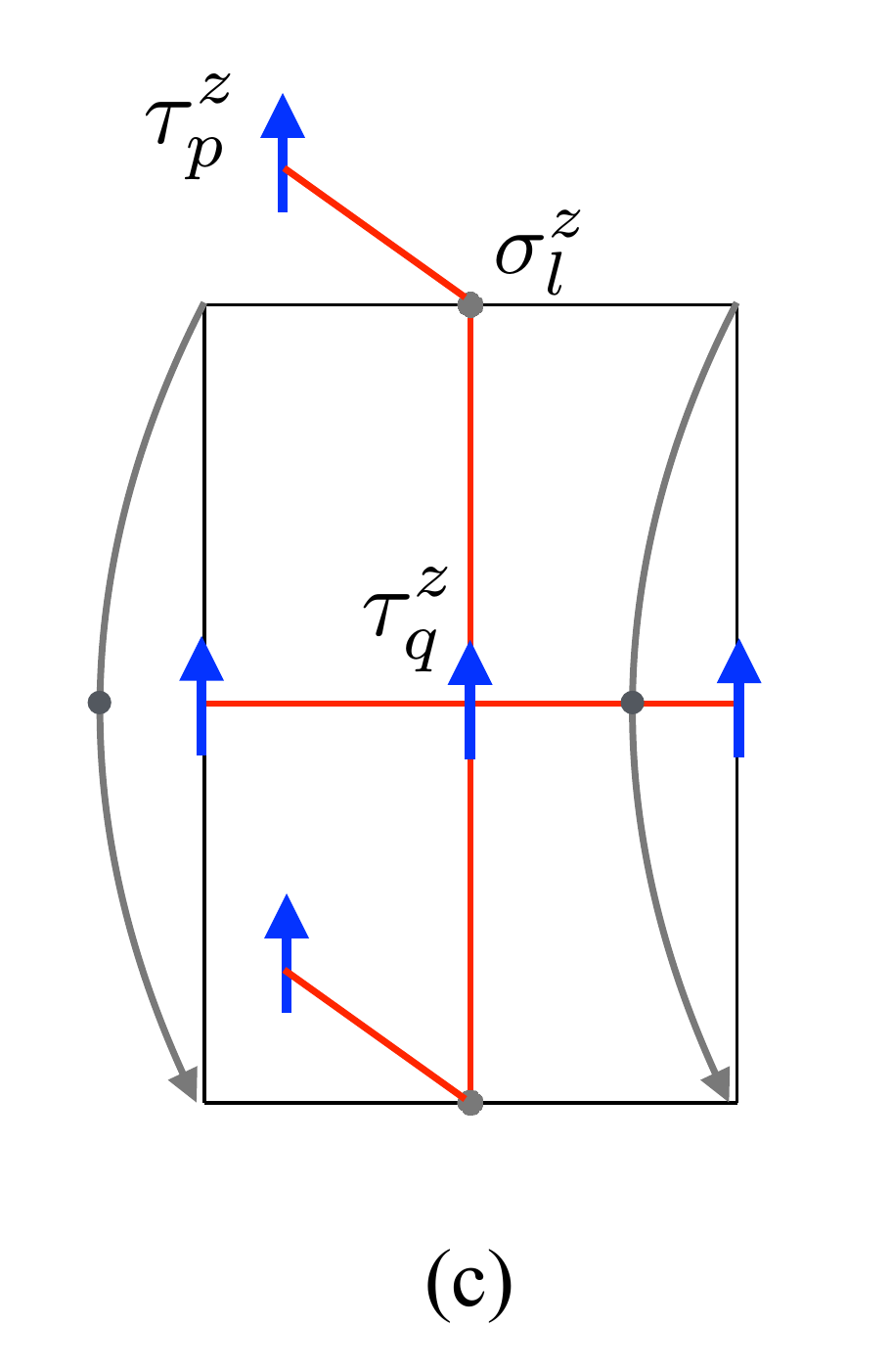}
\includegraphics[width=0.4\linewidth]{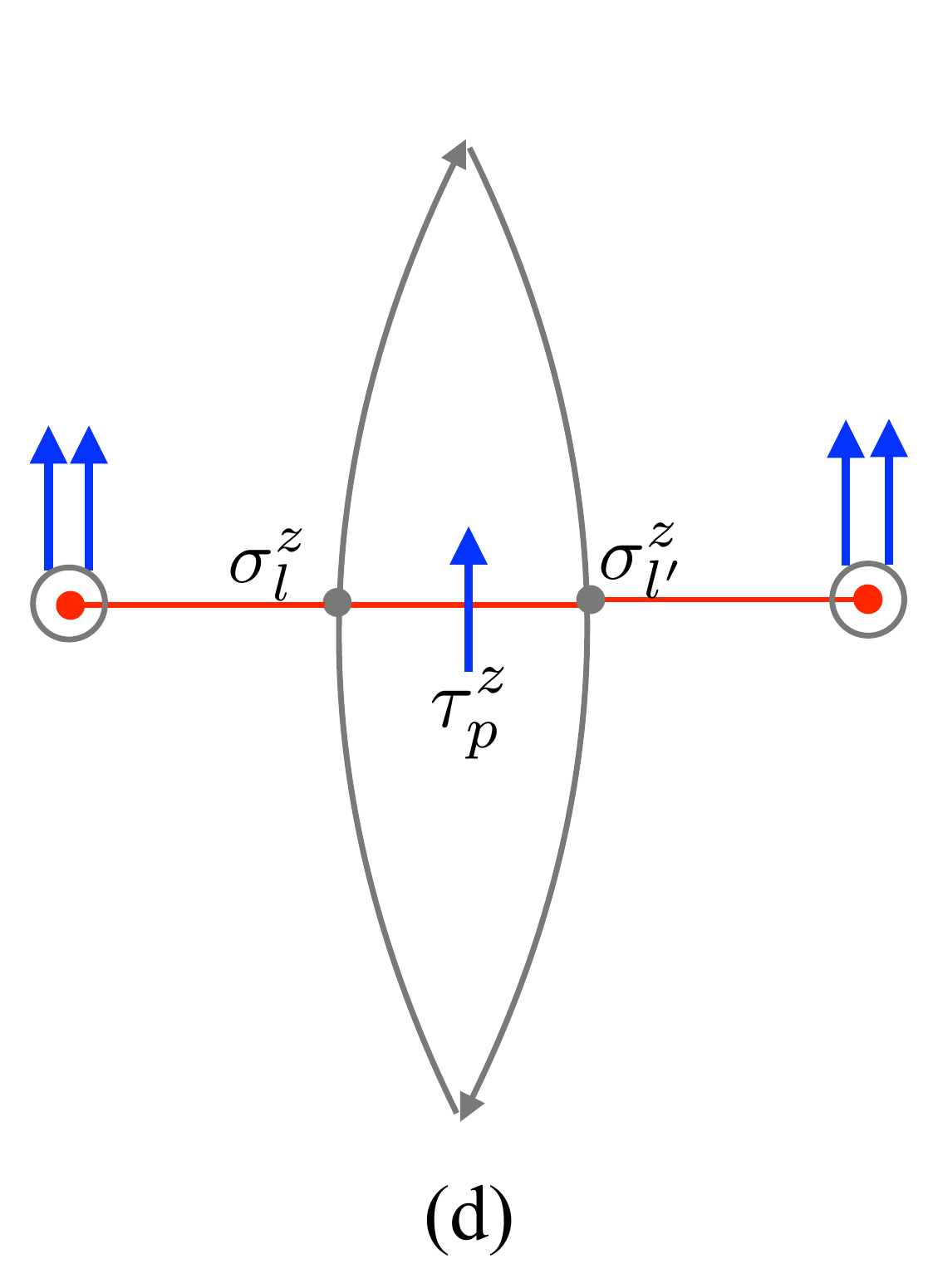}
\caption{\label{fig:duality_map}  Duality map between string and spin models for the different types of vertices in the $bTP$ and the $bNTP$ models. The duality map in Eq.\eqref{eq:dual_map} is used to relate the spins in the direct lattice (grey circles on the links) with the spins in the dual lattice (blue arrows on the vertices). (a) and (b) represent the direct (grey solid lines) and dual (red solid lines) lattices for a square and hexagonal layer, respectively. (c) shows the frontal structure for a bilayer lattice. The figure illustrates a frontal plaquette (grey solid lines) for the direct lattice that corresponds to a single vertex in the dual lattice (blue arrow at the center of the plaquette). However, we know there is another frontal plaquette on top of the one shown in the figure. This is the reason why there are double vertices (concentric red and grey circles in Fig.\ref{fig:dual_lattice_hermele_our_model} when we picture the whole frontal structure. (d) represents a degenerate plaquette (grey solid lines) in the direct lattice with its corresponding vertex (blue arrow at the center of the plaquette) in the dual lattice. The structure of the double vertices described in figure (c) is also illustrated by the two blue arrows on the same vertex on the left and on the right of the figure (d).  Frontal plaquettes have the same structure for the  bilayer square lattice as for the bilayer hexagonal lattice.}
\end{figure}

\noindent 
Next, we define a duality map that relates the spins on the string models with the ones on the spin models. For string models, spins are on the links  whereas  for  the spin models,  they  are placed  on the vertices. Every spin configuration in the spin model defines a corresponding domain wall configuration (closed string) in the string model. Formally, the correspondence is given by: 
 \begin{align}\label{eq:dual_map}
 \sigma^z_l:=\tau^z_p\tau^z_q,
 \end{align}
where $\tau^z_p$ and $\tau^z_q$ are Pauli operators acting on the vertices $p$ and $q$ of the dual lattice. The link connecting $p$ and $q$ is denoted as $l$. On this link $l$ there is a spin on the direct lattice. This duality map is shown graphically in Fig.\ref{fig:duality_map} for both models.

\noindent
Using the dual map described in Eq.\eqref{eq:dual_map}, the  spin hamiltonian for the   Bilayer Trivial Paramagnet model is:
\begin{align}\label{eq: spin_model_hermele}
\hat{H}_{bTP}=-\sum_{i=1,2}\sum_{\mathbf{r}i}\tau^x_{\mathbf{r}i}-\sum_{j=a,b}\sum_{\mathbf{r}_{f_j}}\tau^x_{\mathbf{r}_{ f_j}}+K\sum_{\mathbf{r}_{1/2}}\tau^x_{\mathbf{r}_{1/2}}.
\end{align}
The hamiltonian has three terms, corresponding to three types of vertices in the dual lattice (shown in Fig.\ref{fig:dual_lattice_hermele_our_model}):   i/ The vertices denoted as $\mathbf{r}_i$ are the vertices on the upper and the lower square layers.  The subindex $i=1,2$ runs over the two layers. ii/  The interlayer \textit{double} vertices, where there are two spins on the very same place, are denoted as $\mathbf{r}_{ f_a}$ and $\mathbf{r}_{f_b}$. The subindices $a$ (top) and $b$ (bottom) distinguish the two spins located on the same vertex. These vertices are represented by red and grey concentric circles in Fig.\ref{fig:dual_lattice_hermele_our_model}. iii/  The other kind of interlayer vertices are denoted as $\mathbf{r}_{1/2}$. They come from the plaquettes of type I in the string model. Consequently, they are associated to the coupling constant $K$. It takes on the values $K=\pm1$ .

\noindent
Similarly, the dual map in Eq.\eqref{eq:dual_map} leads to the Bilayer Non-Trivial Paramagnet model, dual to the bilayer DS model derived in Section \ref{model} :
 \begin{align}\label{eq: spin_model_our_model}
&\hat{H}_{bNTP}=+\sum_{i=1,2}\sum_{\mathbf{r_{pi}}}\tau^x_{\mathbf{r_{pi}}}\prod_{<\mathbf{r_{pi}}\mathbf{r_{qi}}\mathbf{r_{q'i}}>}\textrm{i}^{\frac{1-\tau^z_{\mathbf{r_{qi}}}\tau^z_{\mathbf{r_{q'i}}}}{2}}\\
&-\sum_{j=a,b}\sum_{\mathbf{r}_{f_{j}}}\tau^x_{\mathbf{r}_{f_{j}}}\left(\sum_{i=1,2}\prod_{\gamma}\tau^x_{\mathbf{r_{pi}}\gamma}\prod_{<\mathbf{r_{pi}\gamma}\mathbf{r_{qi}\gamma}\mathbf{r_{q'i}\gamma}>}\textrm{i}^{\frac{1-\tau^z_{\mathbf{r_{qi\gamma}}}\tau^z_{\mathbf{r_{q'i\gamma}}}}{2}}\right)\nonumber\\
&+K\sum_{\mathbf{r}}\tau^x_{\mathbf{r}_{1/2}} \;,\nonumber
\end{align}
 where $\gamma$ runs over the two vertices on the dual lattice corresponding to the backward and forward plaquettes in the direct lattice (see Fig.\ref{fig:vertex_plaquettes} (d)). The second term in the above  hamiltonian corresponds to the spin version of the plaquettes in Eq.\eqref{eq:plaquette_III}. The operator $\tau^x_{\mathbf{r}_{f_{j}}}$ is the equivalent to the frontal part in the string model (Eq.\eqref{eq:frontal_op}) and the products included in the parenthesis correspond to the cover part (Eq.\eqref{eq:cover_b_f}).  
\noindent
The main difference between the two models, $\hat{H}_{bTP}$ and $\hat{H}_{bNTP}$,  relies  on the interaction inserted on the layers. For the bNTP model, the spins on the layers are denoted as $\mathbf{r_{pi}}$, where $p$, $q$ and $q'$ are the three vertices forming a triangle in the dual lattice. The subindex $i=1,2$ runs over the two layers. The new product introduced in the hamiltonian of the bDS,  $\prod_{\mathbf{r_{pi}}\mathbf{r_{qi}}\mathbf{r_{q'i}}}\textrm{i}^{\frac{1-\tau^x_{\mathbf{r_{qi}}}\tau^x_{\mathbf{r_{q'i}}}}{2}}$,  gives rise to new complex phases. Despite of these factors, the hamiltonian turns out to be hermitian.

\begin{figure}
\includegraphics[width=\linewidth]{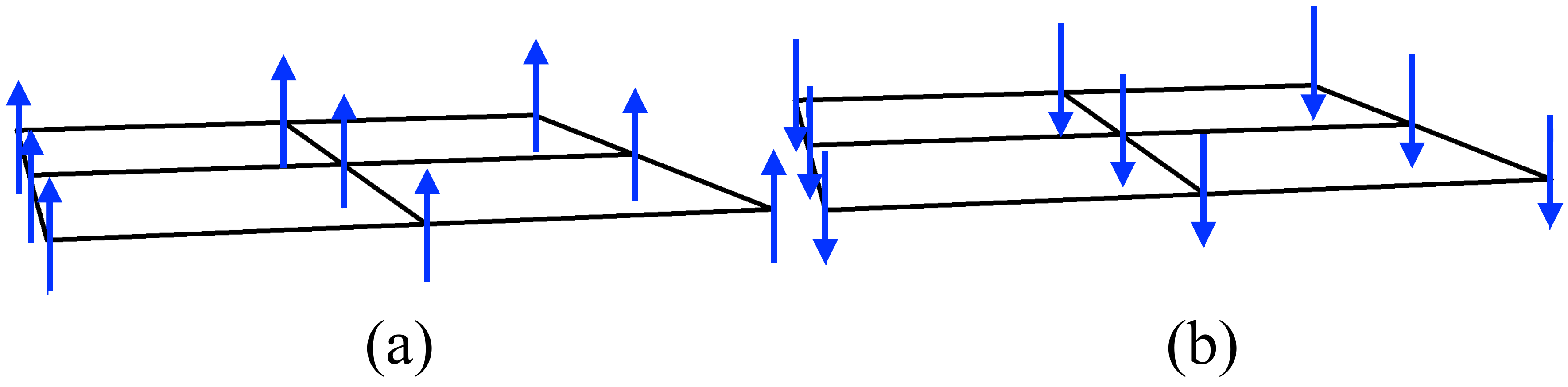}
\caption{\label{fig:z2_spin} Action of  $\mathbb{Z}^{fs}_2$ spin symmetry on a square lattice layer. It works in the same way  for triangular and frontal plaquettes.  Spins are represented by  blue arrows. The $U_S$ operator (defined in Eq.\eqref{eq:U_S_operator}) changes the orientation of the arrows, i. e. turn the spins on the whole lattice over.}
\end{figure}

\noindent
Both  spin models (Eq.(\ref{eq: spin_model_hermele}) and Eq.(\ref{eq: spin_model_our_model})) are invariant under $\mathbb{Z}_2^{fs}$ spin symmetry.
The $\mathbb{Z}_2^{fs}$  is defined by the following operator:
\begin{align}\label{eq:U_S_operator}
U_S=\prod_{\mathbf{r}} \tau^x_{\mathbf{r}}.
\end{align}
The subindex $\mathbf{r}$ runs over all types of vertices on the dual lattice. The action of this symmetry is to flip all the spins in the lattice as Fig.(\ref{fig:z2_spin}) shows.  

\noindent
This is a global symmetry, as the previously studied $\mathbb{Z}_2^{fv}$ flavour symmetry.    Remarkably , the bTP and the bNTP models are invariant under an interchange of the two layers. Therefore, both bTP and bNTP models are invariant under the symmetry $\mathbb{Z}^{fs}_2\times\mathbb{Z}^{fv}_2$. As long as they are SPTO, edge modes appear when the systems have boundaries. 

\noindent
Once it is known the symmetry structure of the hamiltonian, we are in a position to obtain the ground states for $\hat{H}_{bTP}$ and $\hat{H}_{bNTP}$ models. The hamiltonians $\hat{H}_{bTP}$ (Eq. \eqref{eq: spin_model_hermele}) and $\hat{H}_{bNTP}$ (Eq. \eqref{eq: spin_model_our_model}) are built as a sum of terms corresponding to different kind of vertices. Each term in both hamiltonians represent an interaction among the spins in the system. The ground state for each term in the hamiltonians is explained in the following. 
\begin{figure}
\includegraphics[width=0.83\linewidth]{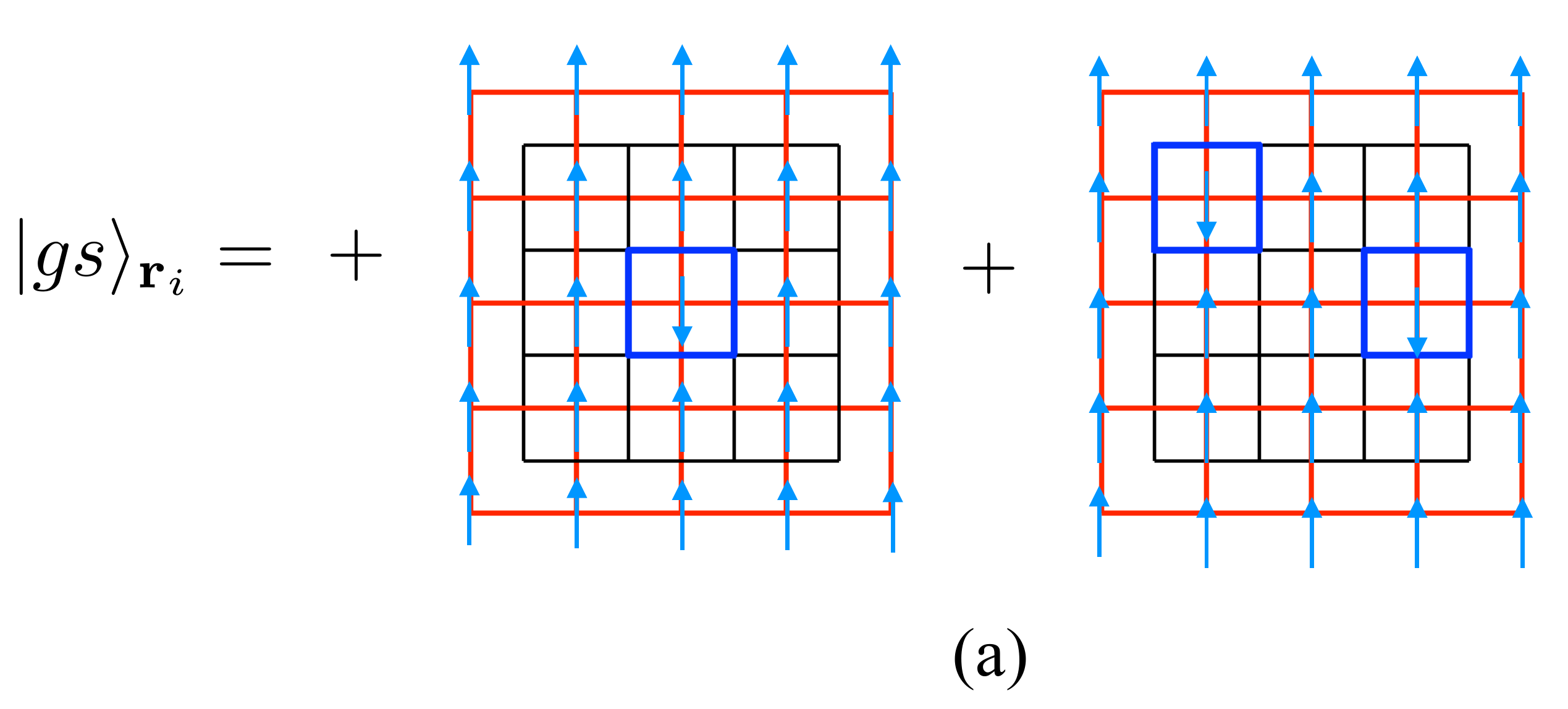}
\includegraphics[width=0.9\linewidth]{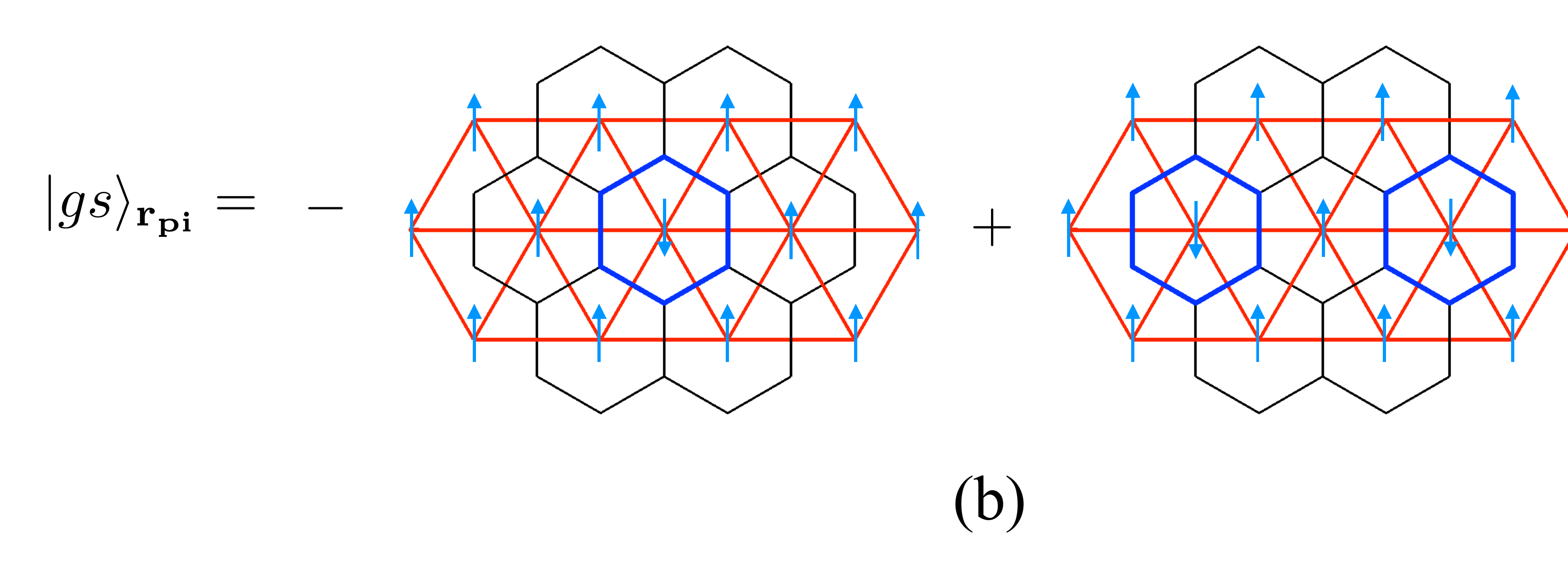}
\caption{\label{fig:ground_states1} Graphical representation of the ground states  $|gs\rangle_{\mathbf{r}_i}$  and $|gs\rangle_{\mathbf{r_{pi}}}$. (a) It is a sample of the ground state factor on the layers, for the bTP for the bilayer Kitaev model. The direct lattice (black solid lines) is square and the domain walls (dark blue solid lines) are also squares because square lattices are self-dual. At each vertex of the spin model lattice (red solid lines), there is a light blue arrow representing the spins.  (b) $|gs\rangle_{\mathbf{r_{pi}}}$ is a part of the ground state in the bNTP model (Eq.\eqref{eq:gs_frontal_denenerate_ds}).  The triangular lattice (red solid lines) hosts spins (light blue arrows) on its vertices. The domain walls corresponding to each spin configuration (dark blue solid lines) belong to the honeycomb lattice (black solid lines). It is important to point out the alternation of sign in the superposition of all spin configurations, due to the interaction placed on the layers. }
\end{figure}

\noindent

\begin{figure}
\includegraphics[width=0.75\linewidth]{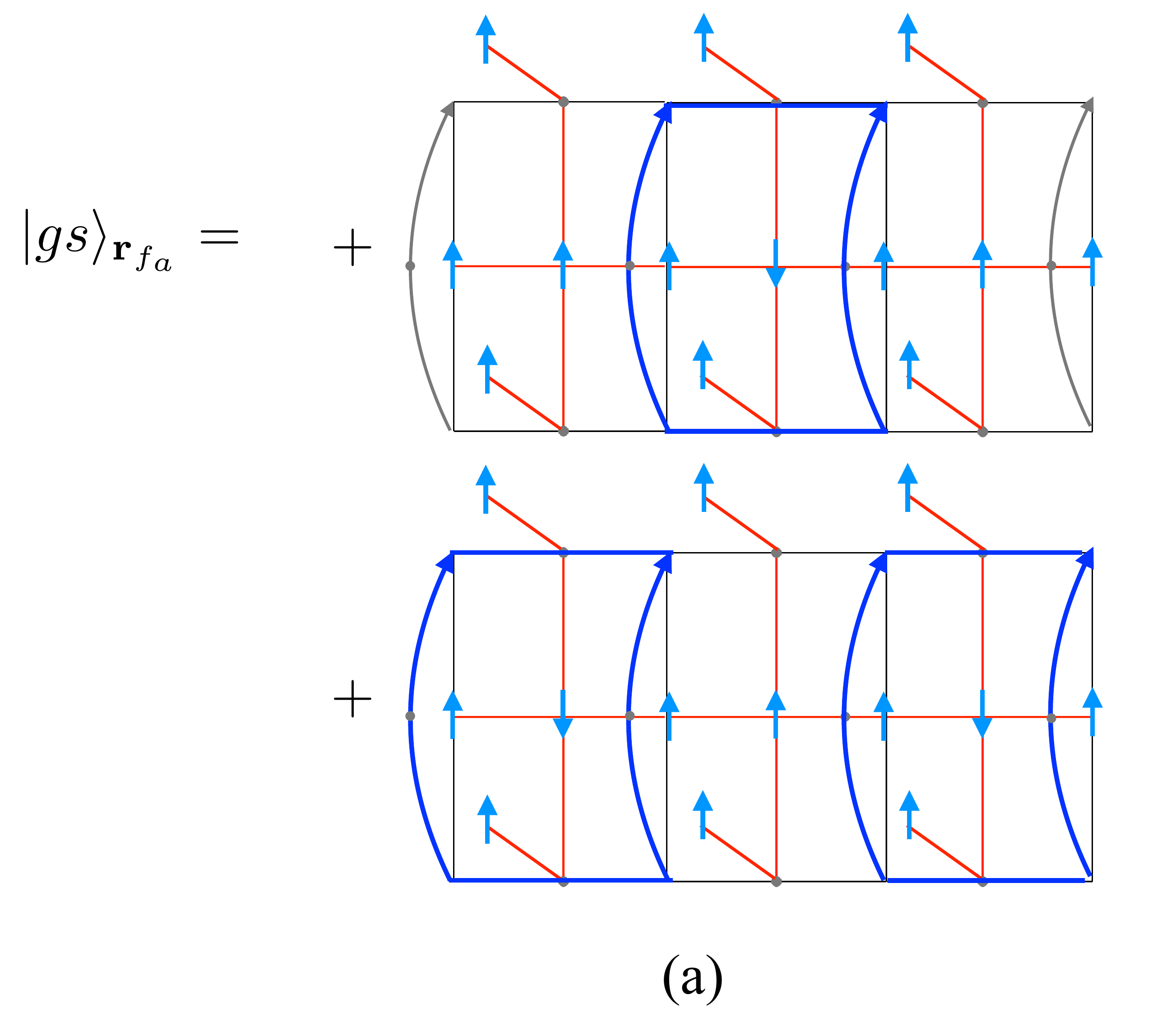}
\includegraphics[width=0.9\linewidth]{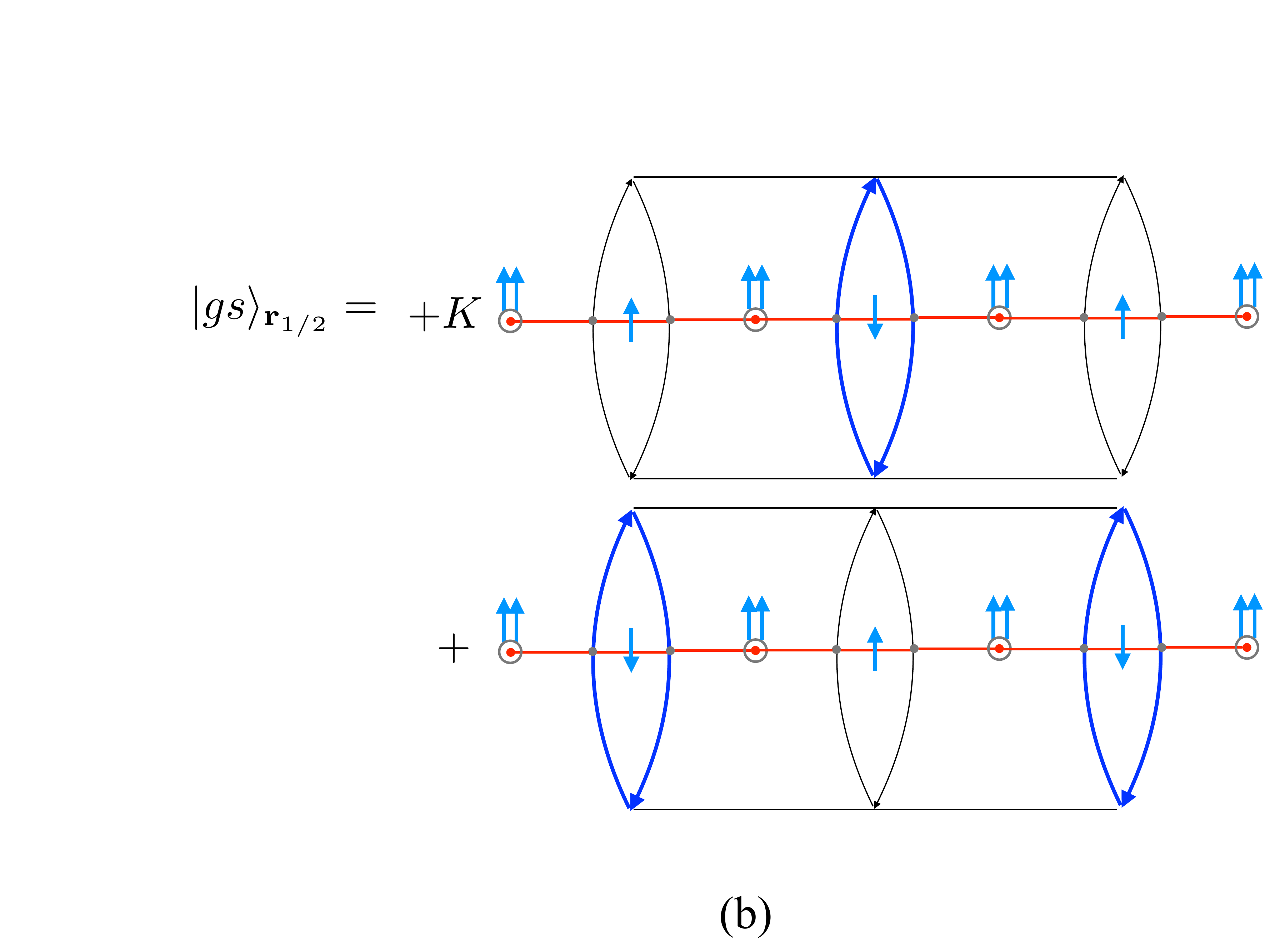}
\caption{\label{fig:ground_states2}  Pictorial representation of $|gs\rangle_{\mathbf{r}_{f_a}}$  and $|gs\rangle_{\mathbf{r}_{1/2}}$. These kind of vertices describe  the same frontal structure for both the  bTP and bNTP models. (a) is a sample of the superposition of all configurations of domain walls (dark blue squares) for $|gs\rangle_{\mathbf{r}_{f_a}}$ vertices. In this case, the coefficients of the superposition are equal. (b) The domain walls (dark blue solid lines) created when $\tau^z_{\mathbf{r}_{1/2}}=-1$ are degenerated. The figure also shows that each  spin configuration enters in the superposition with a sign $K^{\mathbf{dw_{1/2}}}$ (see Eq.\eqref{eq:gs_frontal_denenerate_ds}), where $K=\pm$  is the coupling constant in Eq.\eqref{eq: spin_model_hermele} and Eq.\eqref{eq: spin_model_our_model}  and $\mathbf{dw_{1/2}}$ is the number of domain walls corresponding to vertices $\mathbf{r}_{1/2}$.}.
\end{figure}

Regarding the bilayer Trivial Paramagnet spin model in Eq.\eqref{eq: spin_model_hermele}, there are three types of vertices: $\mathbf{r}_i,\mathbf{r}_{f_j}$ and $\mathbf{r}_{1/2}$. The ground state for $\mathbf{r}i$ vertices ($|gs\rangle_{\mathbf{r}_i}$) is obtained when $\tau^x_{\mathbf{r}i}=1$ for every spin configurations. Every spin configuration ($\tau^z_{\mathbf{r}i}=\pm1$) on the dual lattice defines a corresponding domain wall configuration on the direct lattice. Considering $|gs\rangle_{\mathbf{r}_i}$, the domain walls are square plaquettes in the direct lattice (shown in Fig. \ref{fig:ground_states1} (a)). The domain walls correspond to vertices in the dual lattice where $\tau^z_{\mathbf{r}i}=-1$.  The ground state is an equally weighted superposition of all configurations of domain walls. The ground states $|gs\rangle_{\mathbf{r}_{f_j}}$ and $|gs\rangle_{\mathbf{r}_{1/2}}$ are obtained when $\tau^x_{\mathbf{r}{f_j}}=1$ and $\tau^x_{\mathbf{r}{1/2}}=K$. The domain walls defined  by $\mathbf{r}_{f_j}$ are very similar to the ones describe for $\mathbf{r}_i$ vertices. However, there are two spins in every $\mathbf{r}_{f_j}$ vertex: $\mathbf{r}_{f_a}$ and $\mathbf{r}_{f_b}$. Fig.\ref{fig:ground_states2} (a) shows an example of a ground state spin configuration for  $|gs\rangle_{\mathbf{r}_{f_a}}$. The spin configuration for  $|gs\rangle_{\mathbf{r}_{f_b}}$ is not shown and it is independent from the previous one. Despite these geometrical differences, the ground state $|gs\rangle_{\mathbf{r}_{f_j}}$ is also an equally weighted superposition of all configurations of domain walls. 

\noindent
The ground state $|gs\rangle_{\mathbf{r}_{1/2}}$ presents geometrical and conceptual differences. First of all, the vertices $\mathbf{r}_{1/2}$ in the dual lattice come from a degenerate plaquette in the direct lattice. Then, the domain walls defined by the spin configurations are precisely plaquettes of two links. Moreover, these kind of vertices are associated with a coupling constant $K=\pm1$. This has an important consequence: the ground state is also a superposition of all spin configurations but each configuration with a factor $K^{\mathbf{dw_{1/2}}}$, which is non trivial for $K=-1$. $\mathbf{dw_{1/2}}$ is the number of degenerate domain walls for each spin configuration. This alternating factor is the reason why an alternating sign is present  in the superposition of all possible configurations of domain walls. An schematic figure is shown in Fig.\ref{fig:ground_states2} (b).

\noindent
 The mathematical expressions for each factor of the ground state of the bK model are the following:
 \begin{align}\label{eq:gs_frontal_denenerate}
&|gs\rangle_{\mathbf{r}i}=\sum_{\beta}|\beta\rangle_{\mathbf{r}i}, \hspace{1cm} |gs\rangle_{\mathbf{r}f_j}=\sum_{\beta}|\beta\rangle_{\mathbf{r}f_j},\\
& \text{and} \;\;\;|gs\rangle_{\mathbf{r}1/2}=\sum_{\beta}K^{\mathbf{dw}_{1/2}} |\beta\rangle_{\mathbf{r}1/2},\nonumber
\end{align}
where $|\beta\rangle_{x}$ represents all possible configurations $\beta=\{\uparrow,\downarrow\}$ for the three kind of vertices, $x=\mathbf{r}_i,\mathbf{r}_{f_j} $ and $\mathbf{r}_{1/2}$.

\vspace{0.2cm}
\noindent
The calculation of the ground state of the 
bilayer Non Trivial Paramagnet model is similar to the previous construction for the bTP model. In the
bNTP model, there are also three kind of vertices: $\mathbf{r_{pi}}, \mathbf{r_{f_j}}$ and $\mathbf{r_{1/2}}$. However, the interaction implemented on the layers is different.
 $\mathbf{r_{pi}}$ denotes the vertices of a triangular lattice (in red solid lines in Fig.\ref{fig:ground_states1} (b)). The domain walls  determined by $\tau^z_{\mathbf{r}_{pi}}=-1$ are plaquettes of a honeycomb lattice. The subindex $\mathbf{p}$ in $\mathbf{r_{pi}}$ denotes the vertex of a triangle formed by $<p,q,q'>$ shown in Fig.\ref{fig:dual_lattice_hermele_our_model}. The new phases attached to the vertices of this triangle (see Eq.\eqref{eq: spin_model_our_model}) give rise to an altering sign in the superposition of the spin configurations. The ground state corresponding to $\mathbf{r}_{pi}$  in the
bNTP model is a superposition of all configuration of domain walls with a factor $(-1)^{\mathbf{dw}_{\mathbf{r_{pi}}}}$, where $\mathbf{dw}_{\mathbf{r_{pi}}}$ is the number of hexagonal domain walls corresponding to vertices $\mathbf{r}_{pi}$.

\noindent
The frontal vertices in the bilayer hexagonal lattice, denoted as $\mathbf{r_{f_j}}$, give rise to an identical domain walls superposition to the one explained for the 
bTP model.  
Although the term in the hamiltonian in Eq.\eqref{eq: spin_model_our_model} related to $\mathbf{r_{f_j}}$ has a factor attached, $\left(\sum_{i=1,2}\prod_{\gamma}\tau^x_{\mathbf{r_{pi}}\gamma}\prod_{<\mathbf{r_{pi}\gamma}\mathbf{r_{qi}\gamma}\mathbf{r_{q'i}\gamma}>}\textrm{i}^{\frac{1-\tau^z_{\mathbf{r_{qi}}}\tau^z_{\mathbf{r_{q'i}}}}{2}}\right)$ , the frontal structure is the same as in the previous case (see Fig.\ref{fig:ground_states2} (a)). The mentioned factor is just a product of vertices of type $\mathbf{r_{pi}}$. As we are explaining the ground state by its factors, we include that factor in the superposition created by the vertices $\mathbf{r_{pi}}$. The last type of vertex is $\mathbf{r_{1/2}}$. These vertices are the same in both paramagnets
. Thus, each term enters with a factor $K^{\mathbf{dw_{1/2}}}$ in the superposition of all the possible spin configurations. $\mathbf{dw_{1/2}}$ is the number of  degenerate domain walls, shown in Fig.\ref{fig:ground_states2} (b). The expressions that sum up the above description are:
 \begin{align}\label{eq:gs_frontal_denenerate_ds}
&|gs\rangle_{\mathbf{r_{pi}}}=\sum_{\beta}(-1)^{\mathbf{dw}_{\mathbf{r_{pi}}}}|\beta\rangle_{\mathbf{r_{pi}}}, \hspace{0.4cm} |gs\rangle_{\mathbf{r}f_j}=\sum_{\beta}|\beta\rangle_{\mathbf{r}f_j},\\
& \text{and} \;\;\;|gs\rangle_{\mathbf{r}1/2}=\sum_{\beta}K^{\mathbf{dw}_{1/2}} |\beta\rangle_{\mathbf{r}1/2},\nonumber
\end{align}
where $|\beta\rangle_{x}$ represents all possible configurations $\beta=\{\uparrow,\downarrow\}$ for the three kind of vertices, $x=\mathbf{r_{pi}},\mathbf{r}_{f_j} $ and $\mathbf{r}_{1/2}$.

\noindent
The global structure of the ground state for both paramagnets
 is difficult to be written explicitly. However, we can say that it is a loop class built with the different domain walls we have presented separately. Fortunately, the precise general structure of the ground state is not required for constructing the edge states of the system. The building blocks of the ground state are enough to get the edge states. Their structure is obtained in terms of the boundary conditions and the type of excitations
 we are considering for each case.  

\noindent
Now that the ground states are constructed, we are in a position to study the edge modes in both the bTP and the bNTP models. 
Although the two spin hamiltonians in Eq.\eqref {eq: spin_model_hermele}  and Eq.\eqref{eq: spin_model_our_model} are invariant under the spin-flip symmetry ($\mathbb{Z}^{fs}_2$), there is a dramatic distintion between them: $\hat{H}_{bNTP}$ model has edge modes due to $\mathbb{Z}^{fs}_2$ while 
$\hat{H}_{bTP}$ model does not. These edge states are closely connected to the  non-trivial  statistic of the string operators in this model
. However, there are another kind of edge states that the two models exhibit, coming from the flavour global symmetry ($\mathbb{Z}^{fv}_2$). This symmetry gives rise to edge modes  when $K=-1$ ($K$ is the coupling constant in Eq.\eqref {eq: spin_model_hermele} and Eq. \eqref{eq: spin_model_our_model} and it takes values $K=\pm1$). The existence of the edge states when $K=-1$  is related to the alternating sign in the superpositions of domain walls in the ground state configurations.

\noindent
The existence of these modes can be verified by constructing their explicit wave functions. Concretely, a wave function can be defined for each boundary spin configuration. This wave function  depends on the spin configuration, $\beta_{\text{int}}=\{\uparrow, \downarrow\}$ , lying strictly in the interior of the system. All the domain walls that end at the boundary are closed up by assuming that there is a {\it ghost} \cite{nt:ghost} spin in the exterior of the system, pointing in the $\uparrow$ direction.

\noindent
Before constructing the edge states we have to study the possible boundaries that can be created in a bilayer square lattice (bTP model) and a bilayer hexagonal lattice (bNTP model). As both models have a bilayer structure, there are more scenarios that need to be accounted for than on a single layer model. 

\begin{figure}
\includegraphics[width=0.6\linewidth]{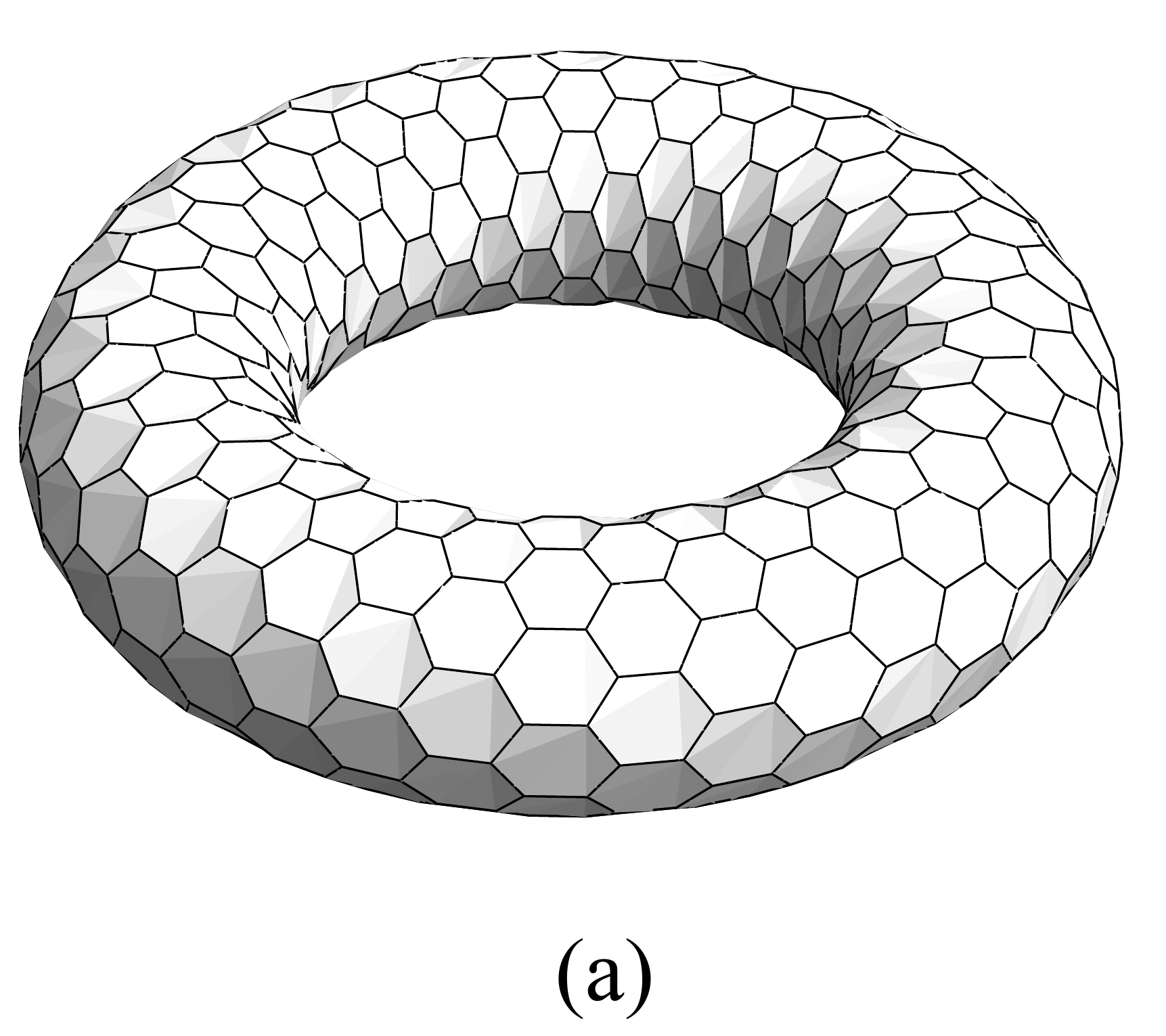}
\includegraphics[width=0.35\linewidth]{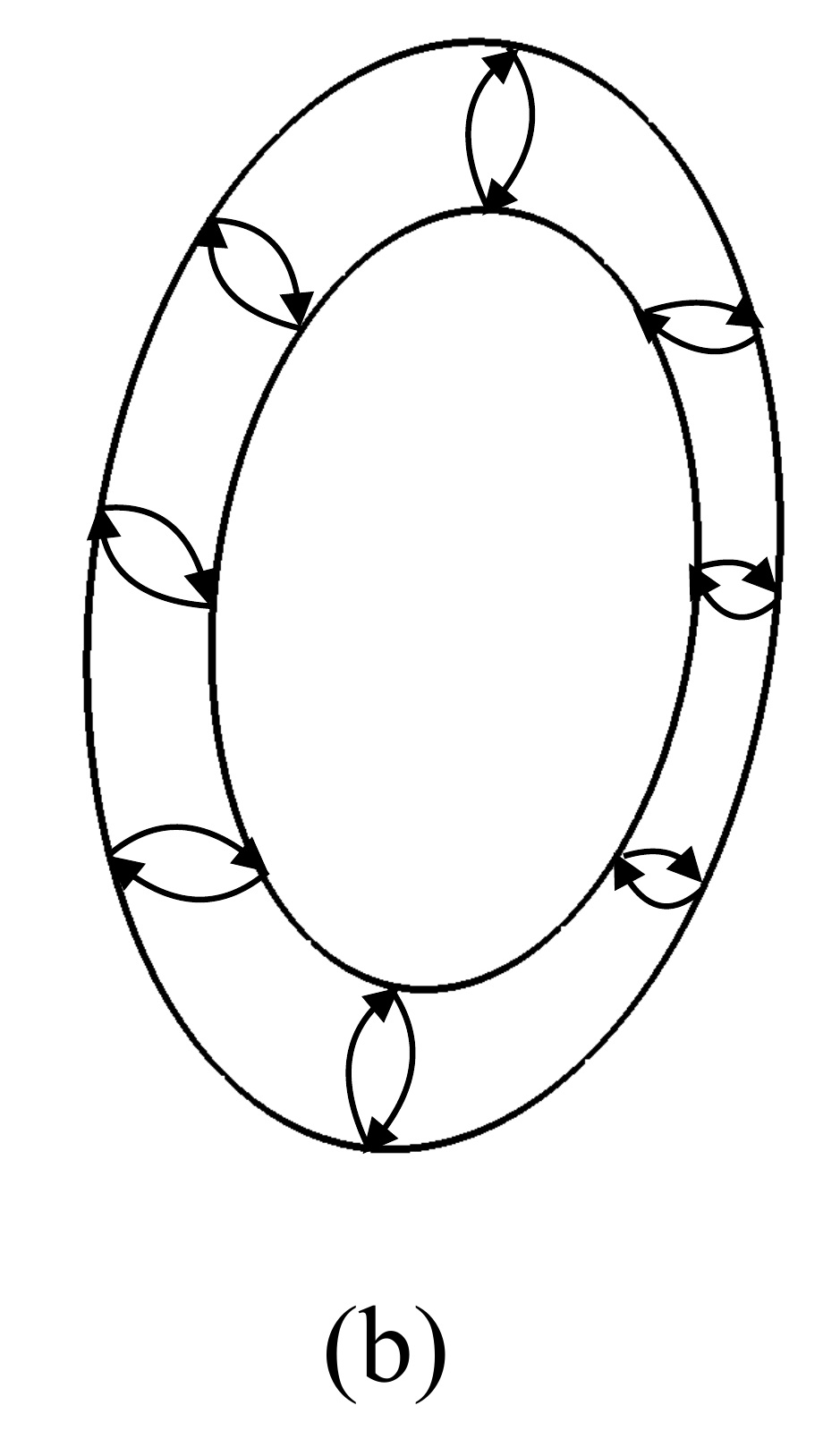}
\caption{\label{fig:torus} (a) Hexagonal bilayer lattice with periodic boundary conditions embedded into a torus. (b) Schematic pictorial view of the internal structure that has a thickness (interlayer) where degenerated domain walls are present. (b) is a transversal section of (a).  }
\end{figure}

\begin{figure}
\includegraphics[width=0.90\linewidth]{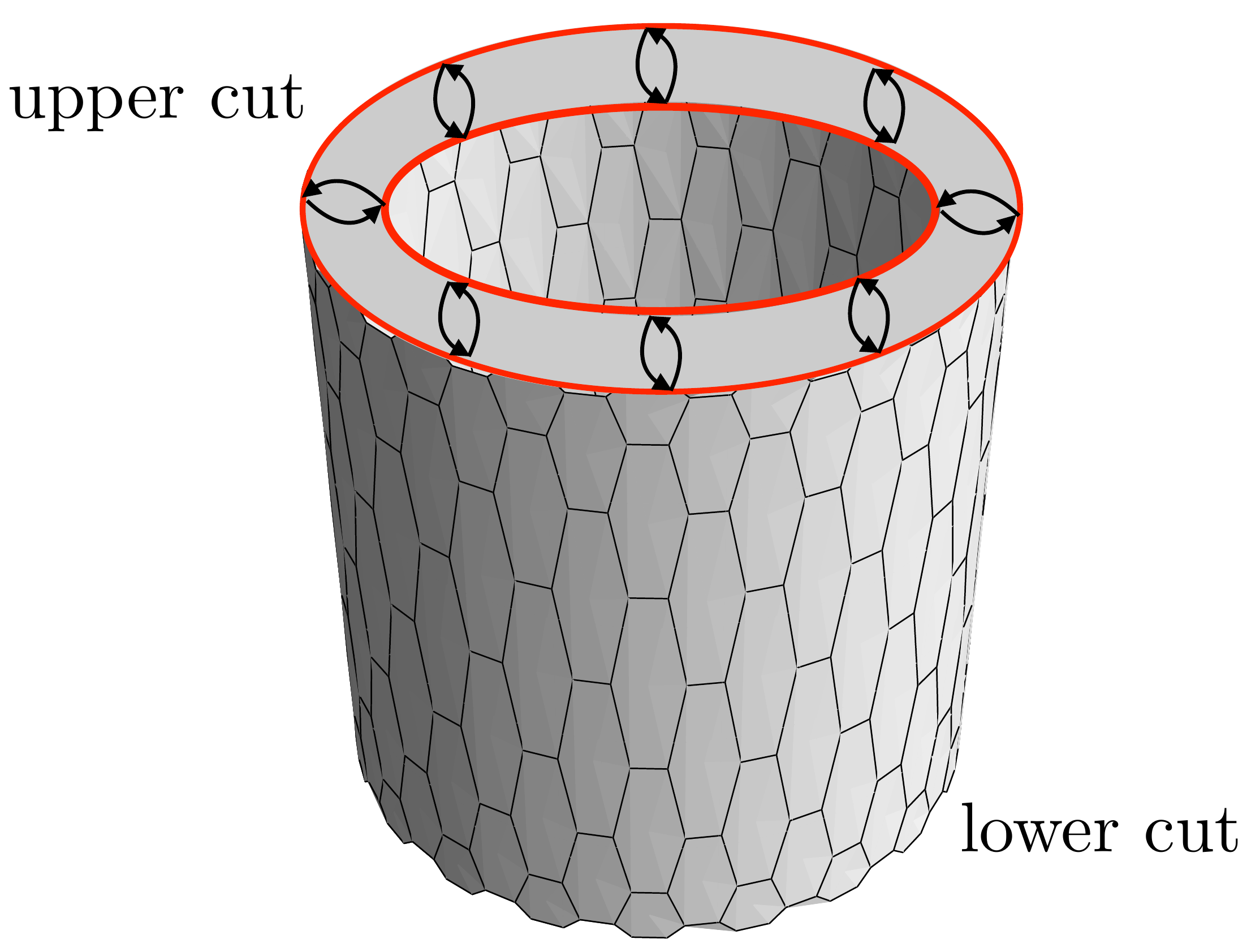}
\caption{\label{fig: Cylinder}The cylinder is the result of cutting the torus by its transversal section. In red full lines, we plot the geometrical edges that correspond to edge states on  the bNTP model. The edge models are protected by the $\mathbb{Z}^{fs}_2$ symmetry in the geometry shown in the figure. }
\end{figure}

\begin{figure*}
\includegraphics[width=0.70\linewidth]{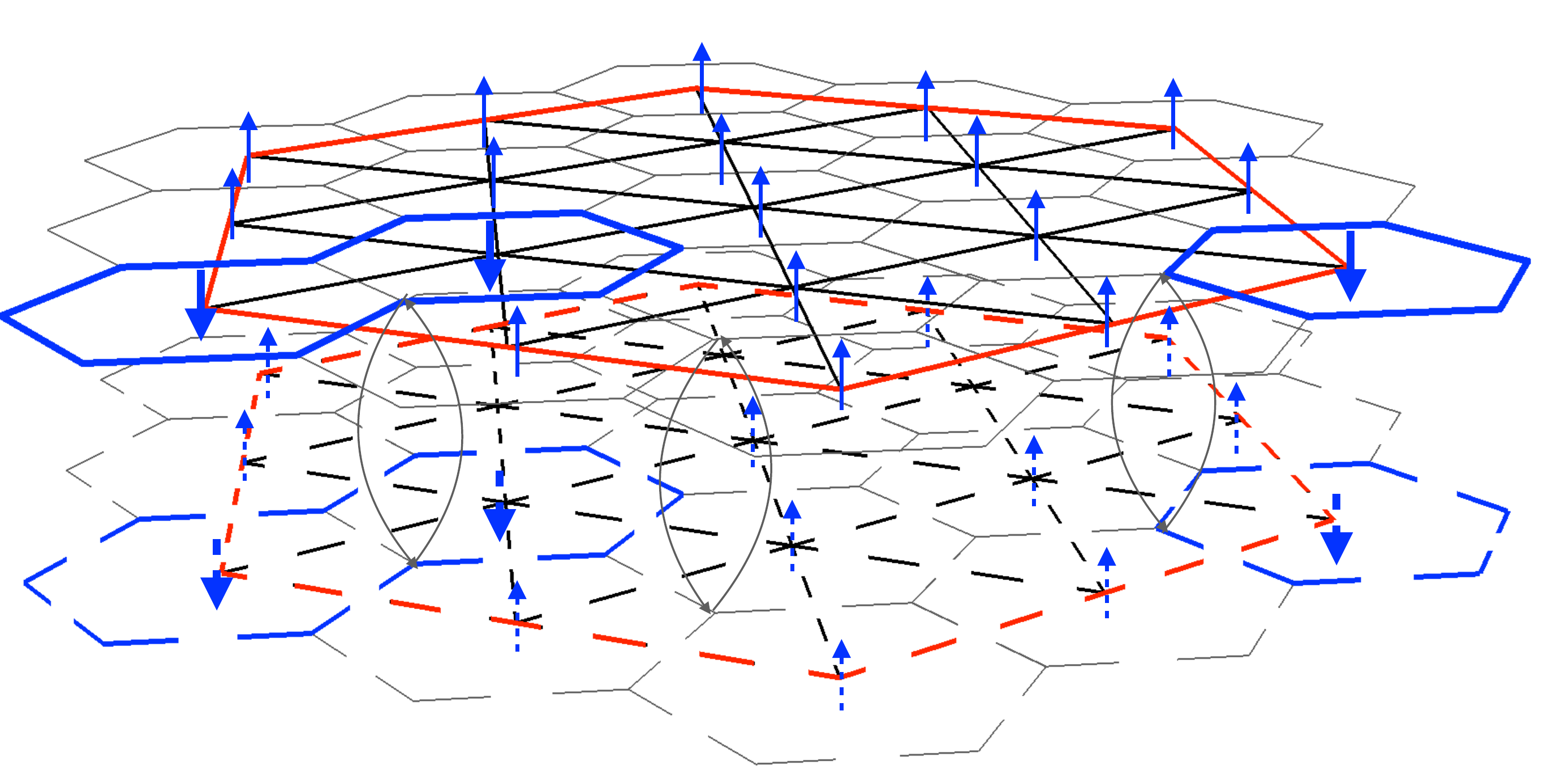}
\caption{\label{fig: boundaries_bDS}The picture shows the bilayer  Non Trivial Paramagnet when the lattice geometry is a thick-plane or equivalently a thick-cylinder. These two distinct geometrical structures are comparable in terms of the existence of the edge states because: i/ both of them have boundaries on the upper and lower honeycomb layers. ii/ they have no boundaries in the interlayer space. The shape of the boundary may change but it is irrelevant as far as the edge states is concerned.  This is the first example of boundary conditions where edge states appear. The boundaries are marked in solid lines  for the upper layer and dashed red lines for the lower layer. To preserve  $\mathbb{Z}^{fv}_2$, the boundaries on the upper and on the lower layer should be identical. Spins, which are placed on the vertices of the triangular lattice (in black lines), are represented by blue arrows (these arrows are dashed in the lower layer). The domain walls ending at the boundary are in blue solid(dashed) lines on the upper (lower) layer. The direct lattice (hexagonal) is coloured in dark grey.  It is remarkable that no boundary is created in the interlayer space. This is the reason why edge states coming from $\mathbb{Z}^{fv}_2$ symmetry are not present. }
\end{figure*}

\noindent
Let us start by considering periodic boundary conditions in both directions. This corresponds to having a torus with an internal structure due to the interlayers, as we can see in Fig.\ref{fig:torus}. In (a), it is sketched a two dimensional hexagonal bilayer lattice with periodic boundary conditions. The picture for a square bilayer lattice is very similar. (b) shows the internal structure that a bilayer lattice has within the torus. This schematic figure is valid for both the square and the hexagonal bilayer lattices. As it can be seen from Fig.\ref{fig:torus},  a bilayer lattice with periodic boundary conditions in both directions has no geometrical boundaries to host the edge modes. As a result, there are no boundaries in the system and it is impossible to construct edge states wave functions. If a transversal cut is done on the torus, the system turns out to be a thick-cylinder. Periodic boundary conditions are set just in one direction but there are two geometrical edges in the other direction.  As it is shown in Fig.\ref{fig: Cylinder} in red solid lines, there are two boundaries on each cut of the thick-cylinder: one in the upper layer and another one in the lower layer. 
Thus, we have edge states on the upper and lower layer for the bNTP model because the domain walls superposition is alternating in sign on the layers. The edge states remain the same if the cylinder is cut into a thick-plane and are shown in Fig.\ref{fig: boundaries_bDS}. The edge modes can be parametrized by boundary spin configurations. For each configuration, $|\beta\rangle_{\mathbf{r_{pi}}}$ of spins $\{\uparrow,\downarrow\}$, the corresponding wave function is  defined as:
\begin{align}\label{eq:edges_ds}
|\text{edges}\rangle_{\mathbf{r_{pi}}}=(-1)^{\mathbf{DW}_{\mathbf{r_{pi}}}}|\beta\rangle_{\mathbf{r_{pi}}},
\end{align}
where $\mathbf{DW}_{\mathbf{r_{pi}}}$, i.e., the total number of domain walls in the system. As it is mentioned above, the domain walls that end on the boundary are closed up by assuming a {\it ghost} spin in the exterior of the system.
It is remarkable that these edge states preserve $\mathbb{Z}^{fv}_2$ spin symmetry. In order to preserve $\mathbb{Z}^{fv}_2$ flavour symmetry, the boundary conditions imposed should be symmetric with respect to exchange of upper and lower layers. This implies that the boundary condition for the upper layer should be the same as the ones for the lower layer.

\noindent
 Noticeably, we can create boundaries on the interlayer space and obtain edge states as long as $\mathbb{Z}^{fv}_2$ is preserved. To this end, we introduce boundary conditions in the interlayer space. These boundary conditions are shown in Fig.\ref{fig: edges} in red solid lines. These edge are made by cutting the system through links in the dual lattice that join $\mathbf{r_{1/2}}$ vertices . These type of vertices are related with degenerate plaquettes in the direct lattice. Thus, the domain walls corresponding to $\mathbf{r_{1/2}}$ vertices are degenerate plaquettes shown in solid blue lines.  Fig.\ref{fig: edges} shows the structure of the edge states for the bTP model, due to $\mathbb{Z}^{fv}_2$ flavour symmetry. The wave function explicit expression for each spin configuration $|\beta\rangle_{\mathbf{r_{1/2}}}$ is the following:
\begin{align}\label{eq:edges_bilayer}
|\text{edges}\rangle_{\mathbf{r_{1/2}}}=K^{\mathbf{DW}_{\mathbf{r_{1/2}}}}|\beta\rangle_{\mathbf{r_{1/2}}},
\end{align}
where $\mathbf{DW}_{\mathbf{r_{1/2}}}$. The edge modes exist when $K=-1$.

\noindent

The edge states that appear on the interlayer space are also present in the bNTP. The structure of the factor of the wave function corresponding to ${\mathbf{r_{1/2}}}$ vertices    is exactly the same. Consequently, the bNTP has two kinds of edge states: on the one hand, edge states on the layers, protected by spin-flip symmetry. On the other hand, gapless modes in the interlayer space due to spin-flavour symmetry. 

\section{Conclusions and Outlook}\label{conclusions}

\begin{figure}
\includegraphics[width=\linewidth]{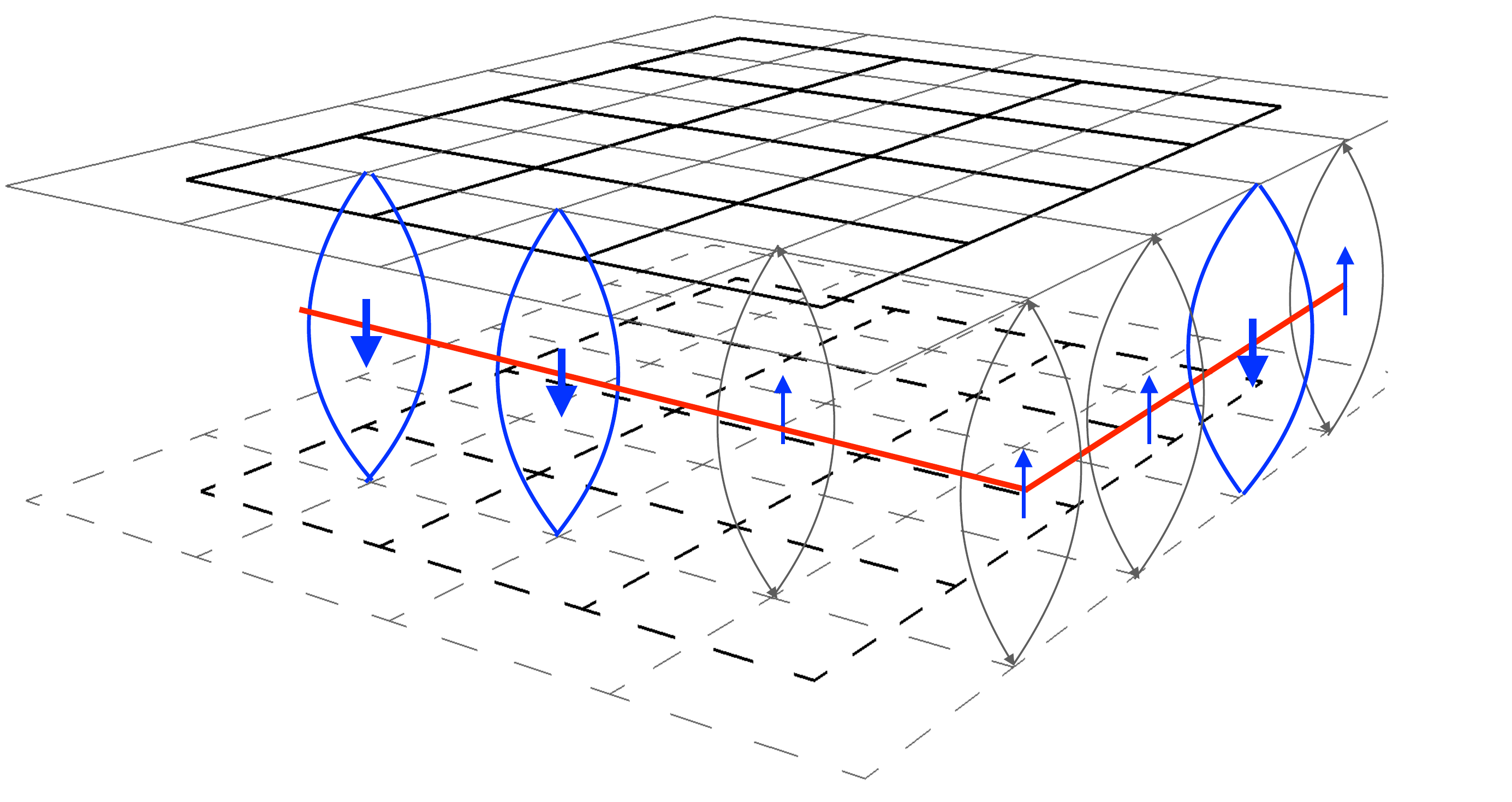}
\caption{\label{fig: edges} The edge states hosted in the bTP model due to the $\mathbb{Z}^{fv}_2$ flavour symmetry, are pictured in the figure. Blue arrows represent the spins, located on the vertices of the dual lattice (black solid lines on the upper layer and dashed black line on the lower one). A blue arrow should be placed in every vertex on the dual lattice but only few are shown for clarity. The spins  $\downarrow$ are represented by  bigger arrows as they give rise to particular domain walls.  The domain walls (in blue solid lines) are isolated because they are degenerate plaquettes in the direct lattice. The interlayer boundary is shown in red solid line and the direct lattice is coloured in dark grey. The explicit expression for the wave function is in Eq.\eqref{eq:edges_bilayer}. }
\end{figure}
\noindent
One of the most important and recent developments in the field of quantum topological phases of matter 
is the notion of symmetry-enriched topological order (SETO). This comes out as the result of 
the unification of the notion of an intrinsic topological order (TO) with the notion of global symmetry. SET phases have far reaching consequences like the fractionalization of quantum numbers describing the topological
charges in an ordinary TO, and the existence of symmetry defects that may have their own topological charges and induce
permutations of anyons, just to mention some of the leading consequences discovered thus far. Our work has precisely focused
on the construction of fractionalization mechanism that may produce more instances of SET phases.

\noindent
We may understand these topological fractionalization mechanisms in a qualitative way, as 
a new version of the method proposed by Haldane \cite{Haldane_88} to construct quantum Hall effects without external magnetic fields.
Now this model is understood as a type of topological insulator. In his celebrated model, Haldane deals with spinless fermions hopping on a single layer of an hexagonal lattice. Hopping fermions between next-to-nearest neighbour sites are designed 
to couple them to quenched magnetic fluxes forming a pattern of background fields. These patterns do not affect the fact that 
ground states remain lowest energy states, but open the possibility of producing topological edge states protected by a global
symmetry. Another simple example of this fermionic phases with background magnetic fluxes, but this time with non-zero net flux,
is the Creutz ladder \cite{creutz1,creutz2,creutz3,creutz4}. Notice that in the string-flux mechanism \cite{Hermele}, the physical picture is that of a set of electrical
charges hopping around background magnetic fluxes similarly to the Haldane model or other topological insulators.
Thus, the fractionalized method treated in our work can be thought of as spin versions of the fermionic mechanism used 
in topological insulators for fermions. It is remarkable the fact that we can apply the same physical concept in  spin systems and in  fermionic systems. More abstractly, the background fields are responsible for introducing Berry phases characterizing the 
topological features of the ground states in different models.
This idea may illuminate the way in which further new fractionalization methods can be devised for models with more general
quantum degrees of freedom and more general situations, like in the presence of thermal fluctuations \cite{DMTI,2dUhlmann,arovas,2dmaterials,self_correcting}.

\noindent
From the point of view of condensed matter physics, the bilayer DS model constructed in this work represents a 
new state of matter with fixed topological order (local gauge group) and fixed internal symmetry (global group).

\noindent
The rich structure of edge states for the Non-Trivial Paramagnet  is constructed explicitly in Fig.\ref{fig: boundaries_bDS} (combined with the frontal structure in Fig.\ref{fig: edges}). We also provide the edge states for its companion model the Trivial Paramagnet. While the Non-Trivial Paramagnet may host two different types of edge states (for both the $\mathbb{Z}^{fs}_2$ and $\mathbb{Z}^{fv}_2$ global symmetries), the Trivial Paramagnet model only presents one type of them, namely, those protected by $\mathbb{Z}^{fv}_2$.

\noindent
This work on the construction of the bilayer DS model may lead to new and interesting extensions. 
A natural generalization is to extend the model to host multilevel spins (or qudits). In this case, the intrinsic
topological order would have a local gauge group $\mathcal{G}=\mathbb{Z}_D, D\geq 2$. To this end, one
needs to deal with a multilayer lattice of honeycombs and incorporate a Cayley graph structure \cite{Hermele}.
Another interesting issue is to substitute the global on-site symmetries by spatial symmetries that appear very
often in lattice systems, namely, translational symmetry or rotational symmetry \cite{Essin_Hermele,Song_Hermele}.
Finally, a more demanding generalization is to incorporate symmetry-enriching topological effects beyond fractionalization
of topological charges like the inclusion of symmetry defects \cite{spto3,teo,teo2,bombin1,qi1,qi2,kong1,pramod2,tarantino}.

\section*{ACKNOWLEDGMENTS}
\noindent
We acknowledge  financial support from the Spanish MINECO grants FIS2012-33152, FIS2015-67411, and
the CAM research consortium QUITEMAD+, Grant
No. S2013/ICE-2801. The research of M.A.M.-D. has been supported in part by the U.S. Army Research Office through Grant No. W911N F-14-1-0103.

\appendix

\section{Zero-flux rule invariant subspace and hermiticity of the DS hamiltonian}\label{app:zero_flux_rule}
\noindent
This appendix is devoted to give the details on the zero-flux invariant subspace in the DS model. As it is mentioned in Section \ref{ssec:double_semion}, plaquette operators do not commute in the whole Hilbert space. To obtain an exactly solvable model it is necessary to deal exclusively with states in a certain subspace. 
\begin{figure} 
\includegraphics[width=0.5\linewidth]{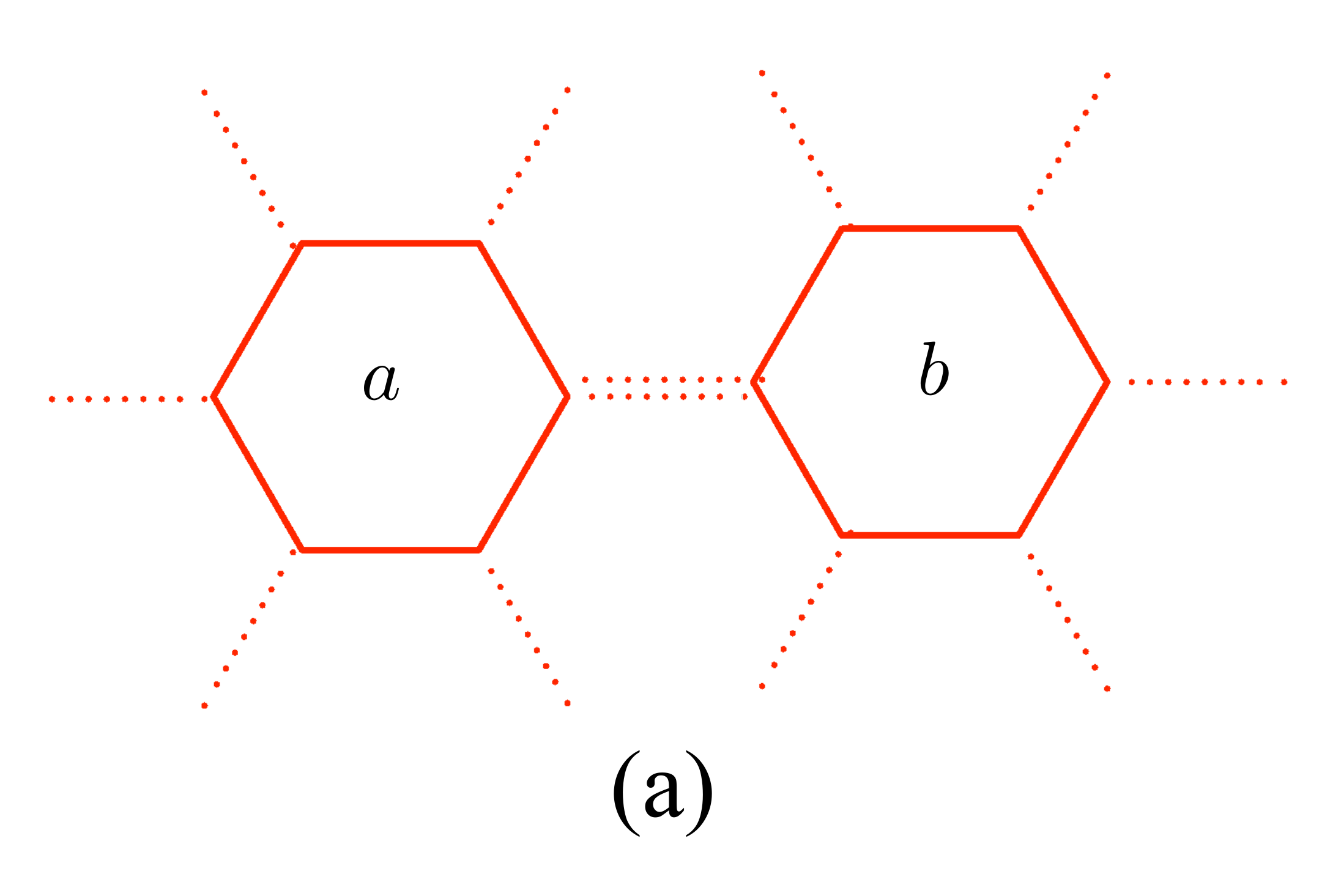}
\includegraphics[width=0.35\linewidth]{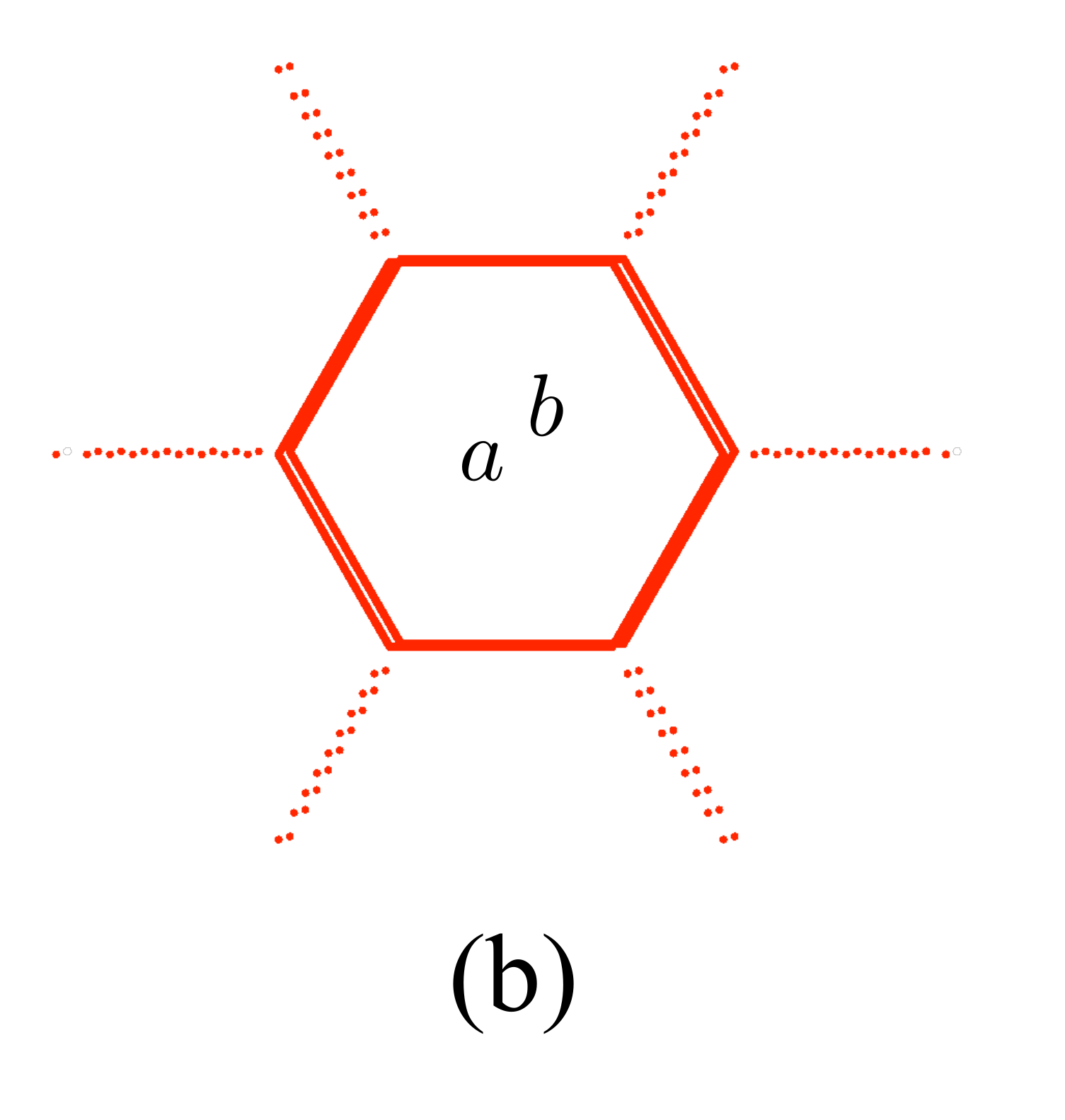}
\includegraphics[width=0.5\linewidth]{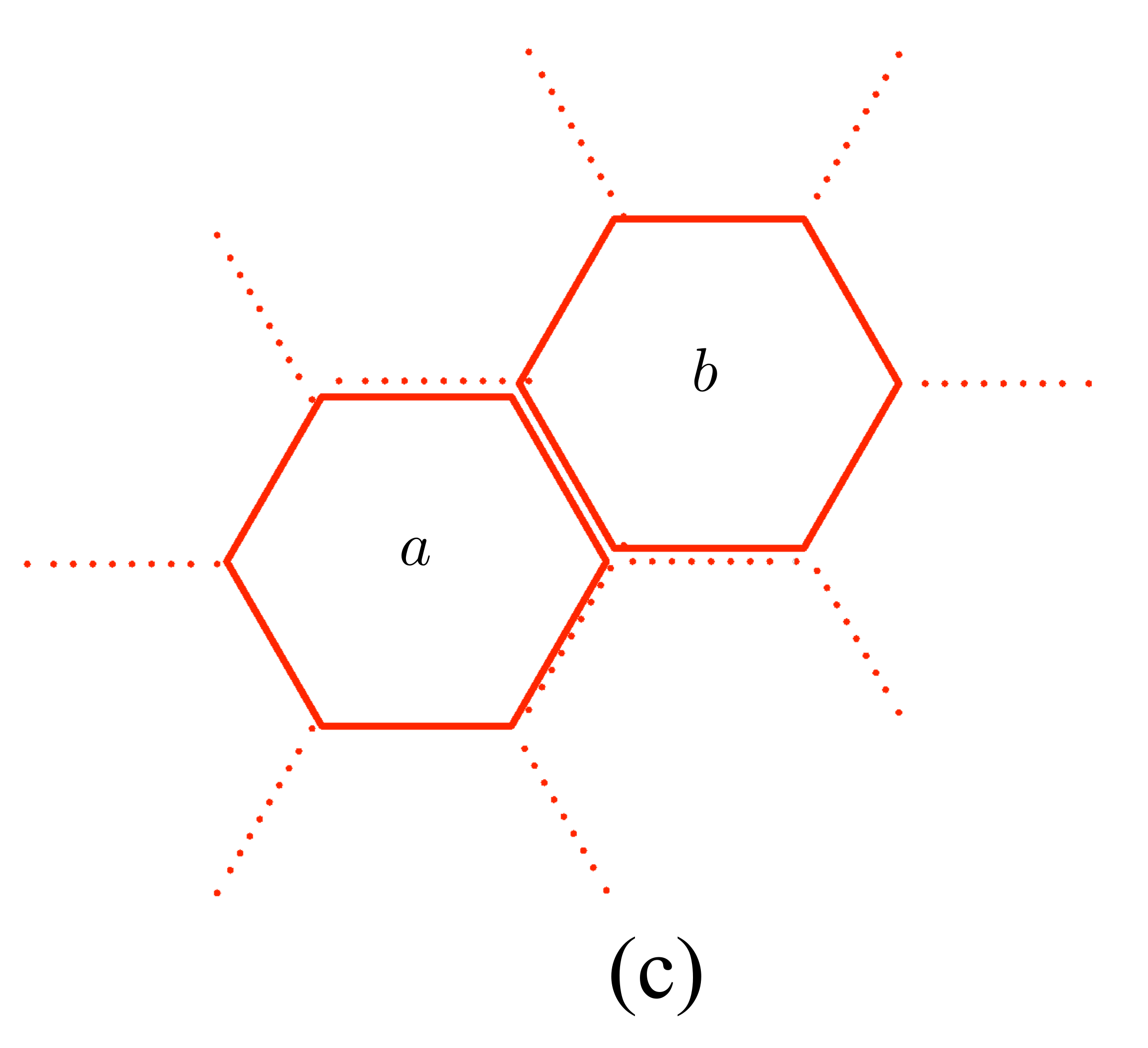}
\caption{\label{fig: position plaquettes} Two neighbouring plaquette operators in the DS model are shown in the figure. Solid lines represent the links where $\sigma^x_l$ is acting and dotted lines represent $S_l$ operator: $S_l=\textrm{i}^{(1-\sigma^z_l)/2}$ (a) In this position the two plaquette operators only share a link. (b) Two plaquette operators can be one on top of the other. Then, they share every link and act on these spins in the same manner. (c) This figure shows plaquette operators $a$ and $b$, which act on five spins with $\sigma^x_l$ and $S_l$. Due to the special commutation rule among these operators (Eq.(\ref{eq:commutator_x_s})), the commutator is only equal to zero on the invariant subspace (see Eq.(\ref{eq:commutator_c})).}
\end{figure}
Firstly, let us recall the definition of vertex and plaquette operators presented in Section \ref{ssec:double_semion}:
\begin{align}\label{eq: recall_ds}
&A_v=\prod_{i\in{s(v)}}\sigma^z_i, \hspace{0.9cm} B_p=-\prod_{i\in \partial p}\sigma^x_i\prod_{j\in s(p)}S_j,\\[3mm]
&\text{where}\hspace{0.7cm}
S_j=\textrm{i}^{(1-\sigma^z_j)/2}.\nonumber
\end{align}
We introduce now the $S_j$ operator to describe the plaquette operators in order to simplify the following calculations. It is easier to operate with  well-defined operators than computing phases, as we did in Section \ref{ssec:double_semion}. Hence we formulate the plaquette operators as a product of $\sigma^x$ and $S$ operators. In the $\sigma^z$ basis the $S_j$ operator can be written as:
\begin{align}
S_j= 
\begin{pmatrix}
1 & 0 \\ 0 & \textrm{i}
\end{pmatrix}.
\end{align}
$S_j$  commutes with $\sigma^z_l$. On the contrary, the commutation rule with $\sigma^x_l$ is:
\begin{align} \label{eq:commutator_x_s}
\sigma^x S=(+i\sigma^z) S \sigma^x .
\end{align}
Due to the factor $+i\sigma^z$, semions appears in the system. However it is precisely this factor that makes two neighbouring plaquettes not commute.
In order to show this fact, we first calculate the commutator of two plaquettes sharing some spins. The definition of a commutator through this appendix is the same used in Sec. \ref{ssec:double_semion}: $[B_{p_1},B_{p_2}]:=B_{p_1}B_{p_2} - B_{p_2}B_{p_1}$. There are three different relative positions where two plaquette operators share some spins. These three possible positions are shown in Fig.\ref{fig: position plaquettes}. In the following we analyse each case.

\noindent
The first situation to consider is shown in Fig.\ref{fig: position plaquettes} (a). The two plaquette operators share only a qubit and both of them act on the common spin with $S_l$ . As $S_j$ commutes with itself, it follows that plaquette operators situated in this position trivially commute. Fig.\ref{fig: position plaquettes} (b) illustrates the two plaquette operators one on top of the other. Then they share all spins and operators act twice on each spin. Since every operator commutes with itself, the two plaquette operators naturally commute.

\noindent
The most interesting situation is represented in  Fig.\ref{fig: position plaquettes} (c), where the plaquette operators share five qubits. Using the commutation rule among $\sigma^x$ and $S$ operators, shown in Eq. (\ref{eq:commutator_x_s}), the commutator  gives:   
 \begin{align}\label{eq:commutator_c}
 [B_{p_a},B_{p_b}]_{\mathrm{(c)}}=&(1-(-i) \sigma^z_1(+i)\sigma^z_2(-i) \sigma^z_3(+i)\sigma^z_4)\times\\
& \times \sigma^x_1S_1S_2\sigma^x_2\sigma^x_3S_3S_4\sigma^x_4(\sigma^z_5)^2=\nonumber\\
 =&(1- A_{v_1} A_{v_2})\sigma^x_1S_1S_2\sigma^x_2\sigma^x_3S_3S_4\sigma^x_4.\nonumber
 \end{align}
\begin{figure}
\includegraphics[width=0.95\linewidth]{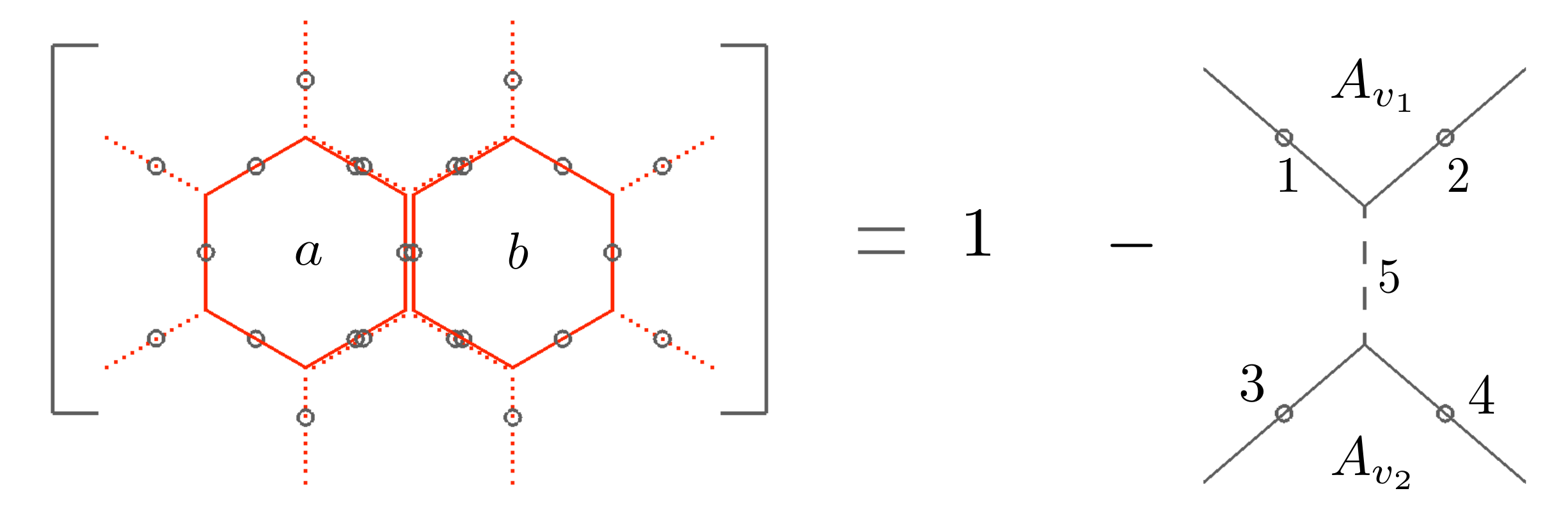}
\caption{\label{fig: commutator} The figure shows the resulting commutator of two plaquette operators in Fig.\ref{fig: position plaquettes} (c) for the DS model.  A  non-trivial result is obtained. It can be factorised in two vertex operators which are equal to one on the zero-flux rule subspace. Within the brackets, solid red  lines represent links where $\sigma^x_l$  acts, while dotted red lines are links where $S_l$ is acting. Grey circles represent spins on the links. Concerning the result of the commutator, grey solid lines represent links where $\sigma_l^z$  acts. The dashed grey line represents a vertical link where no operator is acting due to cancelations. The left-hand side is multiplied by a factor of $\sigma^x_l$ and $S_l$ written in Eq.(\ref{eq:commutator_c}).}
\end{figure}
\begin{figure}
\includegraphics[width=0.95\linewidth]{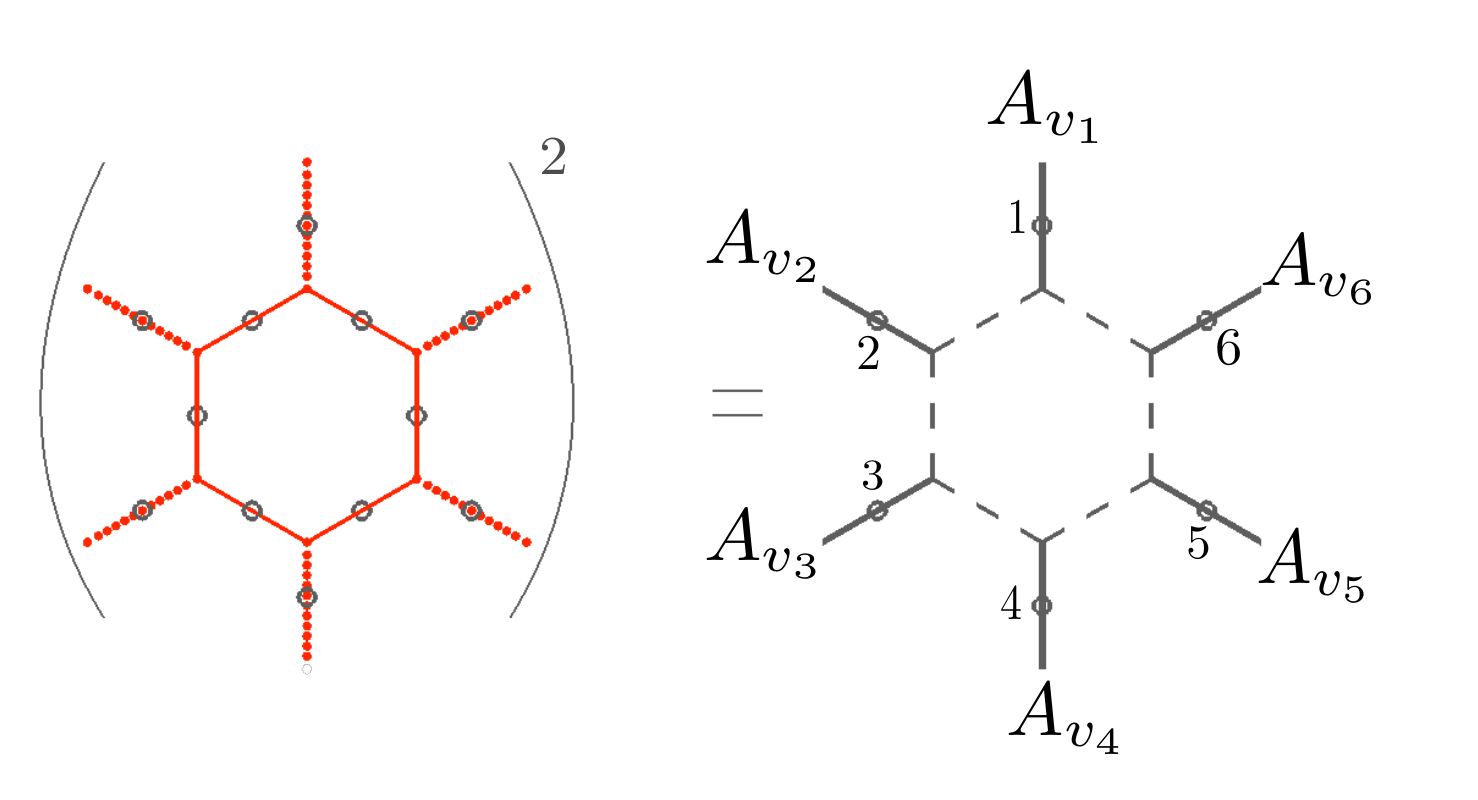}
\caption{\label{fig: square_plaquette} Pictorial view for the square of a plaquette operator in the DS model, in Eq.(\ref{eq:square_plaquette}). Solid red  lines represent links where $\sigma^x_l$ are acting. Dotted red lines are links where $S_l$ acts. Grey circles represents spins and grey solid lines represent links where $\sigma_l^z$  acts. Dashed grey line represent links where no operator acts due to cancelations. }
\end{figure}

\noindent
This result is sketched schematically in Fig.\ref{fig: commutator}. In brackets, representing the commutator, we plot the two plaquette operators showing spins on the links  as grey circles. Moreover, we graphically show the result of this commutator. The result can be expressed  as a product of two vertex operators. We plot as solid grey lines the links where $\sigma^z$ is acting. The link of spin $5$ is plotted with dashed grey lines. This means that no operator is acting on it. This is due to the fact that the two plaquette operators $a$ and $b$ act on spin $5$ with $\sigma^x$ and $(\sigma^x)^2=1$. As the result resembles graphically a scattering diagram, we call this product of two vertices \textit{scattering operator}. It is important to emphasize  that the resulting operator can be factorized into vertex operators. Concretely, it can be written as a product of two vertex operators. Therefore the plaquette operators do not commute in this particular case. In order to solve this problem and build an exactly solvable model, an invariant subspace where the scattering operator is equal to one (thus, the commutator is zero) is defined:
 \begin{align}\label{app:zero-flux-rule}
 &A_v=\prod_{i\in{s(v)}}\sigma^z_i =+1. 
 \end{align}
The above equation is the same as Eq.(\ref{zero-flux-rule}) in Sec.\ref{preliminars} where the zero-flux subspace is first mentioned. The product runs over the three links irradiating from a vertex, as it is shown in Fig.\ref{fig:ds_model}. 
 Acting on the ground state, this condition is always fulfilled. Therefore, the plaquette operators are well-defined on the ground state. When we create vertex excitation, we have to make sure that our string operators create an excitation but still commute with the plaquette operators along their path. This is the reason why vertex string operators are endowed with extra phases, as we have seen in Eq.(\ref{eq: string-operators}). However, there is a more complicated way to formulate the DS model where plaquette operators commute in the whole Hilbert space \cite{Fiona_Burnell}.
 
\noindent
Another remark should be done regarding plaquette operators in the DS model.  A plaquette operator should square to one. This property is used to show that the ground state in Eq.(\ref{eq: gs}) fulfill the lower energy condition, as it is mentioned in Sec.\ref{ssec:double_semion}. However this is not the case for the entire Hilbert space. Once again we need to restrict our calculation to the zero flux subspace to obtain the desired results. To show that, we specify the square of a DS plaquette operator:
\begin{align}\label{eq:square_plaquette}
(B_p)^2= \sigma^z_1\sigma^z_2\sigma^z_3\sigma^z_5\sigma^z_6.
\end{align}
The $\sigma^z$ operators appear becacuase $S^2_l=\sigma^z_l$. This is illustrated in Fig.\ref{fig: square_plaquette} and due to the form of the resulting operator, we call it a \textit{crown operator}. This result can be factorized into a product of vertex operators:
\begin{align}
(B_p)^2=\prod_{i=1}^6 A_{v_i}.\nonumber
\end{align}
The above equation means that the result of that product is $1$ within the zero flux rule subspace.

\noindent
Upon restricting the DS model to an invariant subspace,  we also show that the hamiltonian is hermitian. To do so we recall again the DS hamiltonian:
\begin{align}
&H=-\sum_v A_v -\sum_{p}B_p.
\intertext{where}
A_v=\prod_{i\in{s(v)}}&\sigma^z_i , \text{and }\hspace{1cm} B_p=-\prod_{i\in \partial p}\sigma^x_i\prod_{j\in s(p)}S_j.\nonumber
\end{align}
In contrast with the Kitaev model, the plaquette operators for the DS model present extra links associated with imaginary phases. These six extra links in $s(p)$ are irradiating from the hexagonal plaquette. These imaginary phases attached to the external legs of the plaquette operator contribute to the hamiltonian as a product of $S_l$ operator: $\prod_{j\in s(p)}S_j$. Although this product may lead to complex phases, the only two possible results on the subspace of zero flux rule are $1$ or $-1$. Consequently, the hamiltonian becomes real and hermitian on this subspace. Therefore, the zero flux rule plays an important role to get a hermitian hamiltonian. 

\noindent
The mathematical condition in Eq.(\ref{app:zero-flux-rule}) has its equivalent property in electrodynamics: the Gauss law. In this example, the zero flux rule is a Gauss law on a discrete space.

\section{Commutativity of Vertex and Plaquette Operators in the Bilayer DS Model}\label{app:commutativity}
\noindent
The model we present in Section \ref{model} is an exactly solvable model with topological features. This appendix is devoted to demonstrate that the bilayer Double Semion model is indeed an exactly solvable model so that every operator in the hamiltonian in Eq.(\ref{eq: hamiltonian}) commute with each other. 

\noindent
Firstly, we calculate the commutator among a vertex operator and each type of plaquette operator. Then we move on to calculate the commutator of plaquette operators among each other.

\begin{figure}
\includegraphics[width=0.3\linewidth]{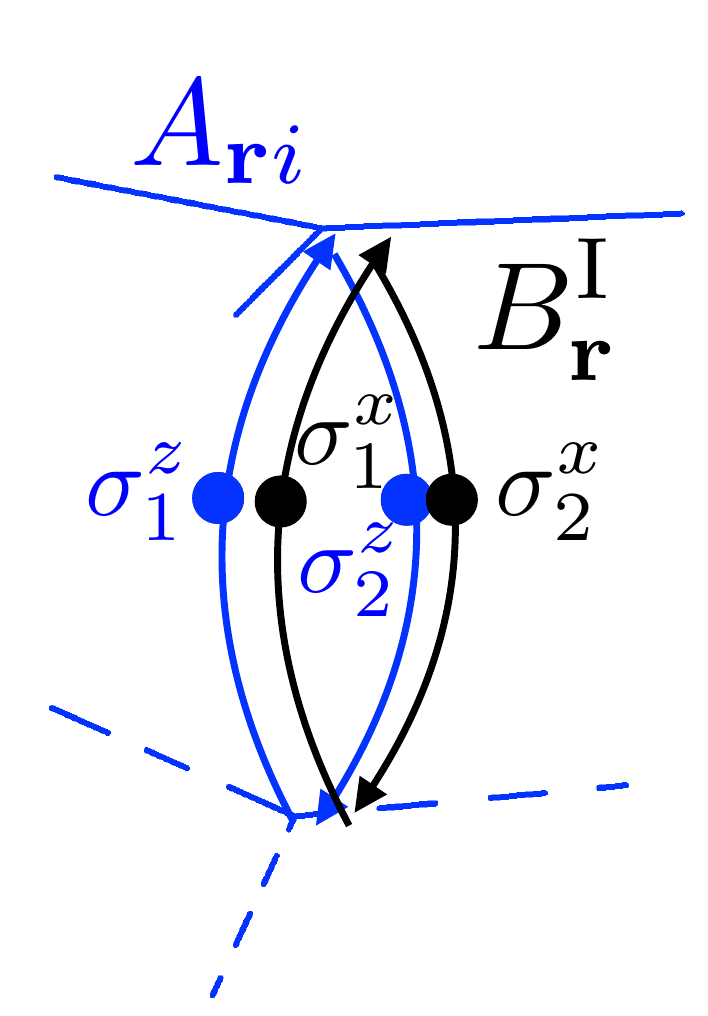}
\caption{\label{fig: commutator_A_B_I} In this figure it is shown a vertex operator (blue solid lines) and a plaquette operator of type I (black solid lines). The common spins are plotted twice as blue and black circles to show that the vertex and plaquette operators act on the same spin but with different operators.}
\end{figure}

\begin{figure}
\includegraphics[width=0.9\linewidth]{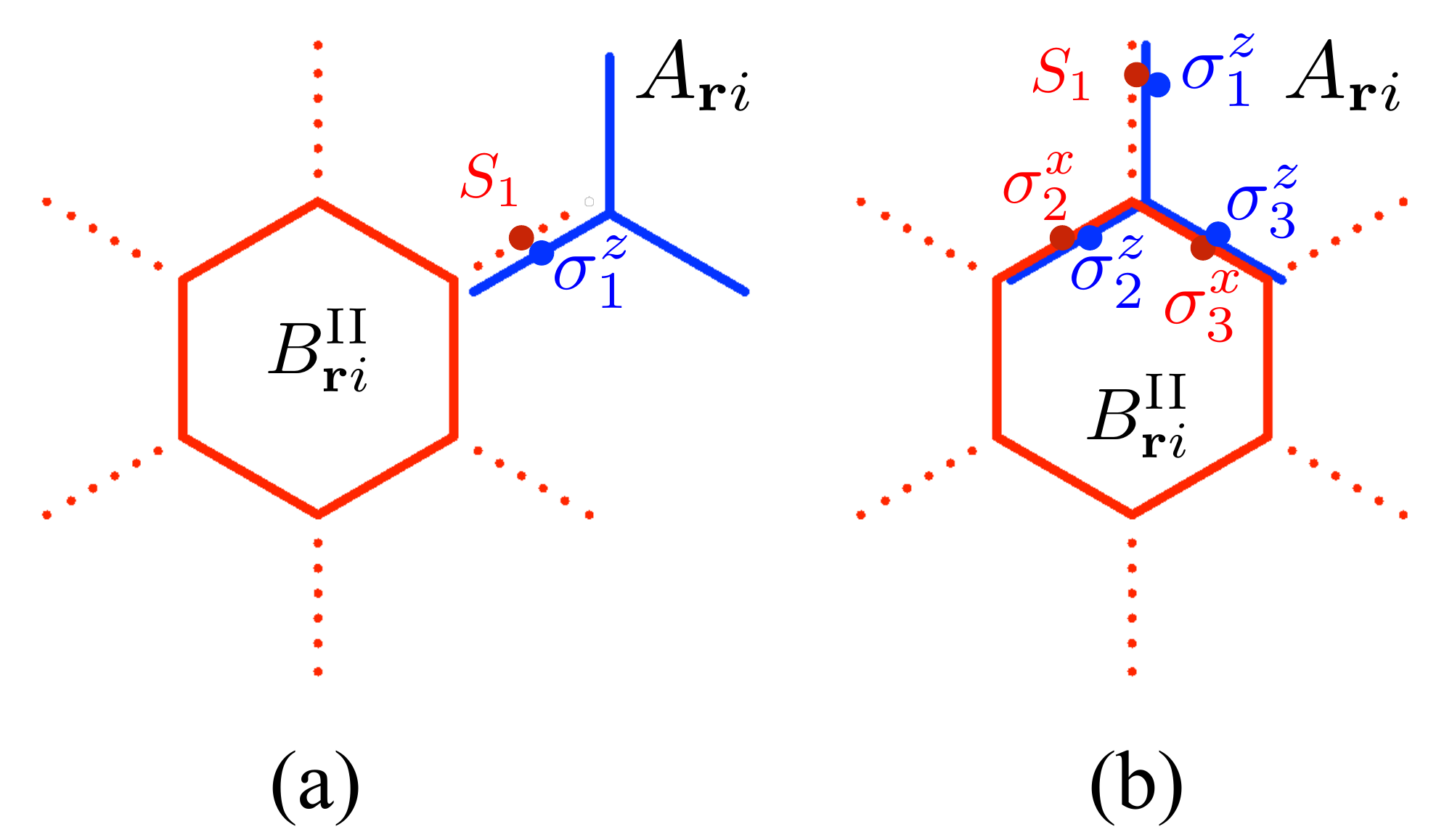}
\caption{\label{fig: position_vertex_plaquette_II} We show in this figure the two possible related positions between a vertex (blue solid lines) and a plaquette operator of type II. The links where $\sigma^x_l$ is acting are plotted in red solid lines, whereas $S_l$ are shown in red dotted lines. Following the notation introduced in Fig.\ref{fig: commutator_A_B_I}, the shared spins are plotted as red and blue circles associated with different operators depending on the colour. (a) and (b) represent the two operators sharing one or three spins respectively. }
\end{figure}

\noindent
Vertex operator and plaquette operator of type $\textrm{I}$ share two spins. The spins shared are shown in Fig.\ref{fig: commutator_A_B_I} as circles where $\sigma_i$ is also specified. The commutator is readily obtained taking into account that $\sigma^z_l$ and $\sigma^x_l$ anticommute. Therefore the result of the commutator is:
\begin{align}
[A_{\mathbf{r}i}, B^{\textrm{I}}_\mathbf{r}]=(1- (-1)^2)\sigma^z_1\sigma^x_1\sigma^z_2\sigma^x_2=0.
\end{align}
\newline
Next we calculate the commutator of $A_{\mathbf{r}i}$ and $B^{\textrm{II}}_{\mathbf{r}i}$.   These two operators can be in two different positions with respect to each other, as Fig.\ref{fig: position_vertex_plaquette_II} shows. 
\begin{figure}
\includegraphics[width=0.9\linewidth]{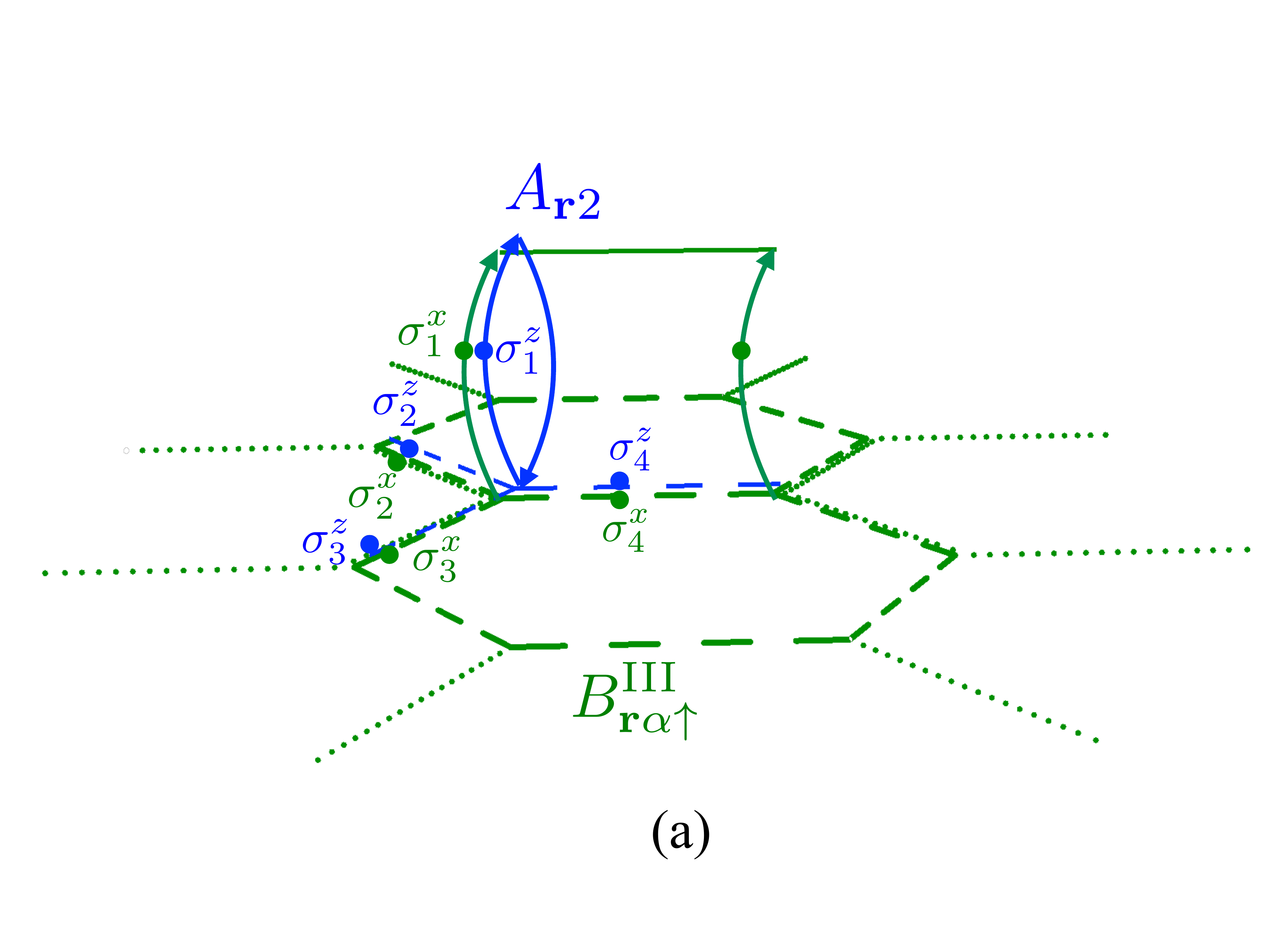}
\includegraphics[width=0.9\linewidth]{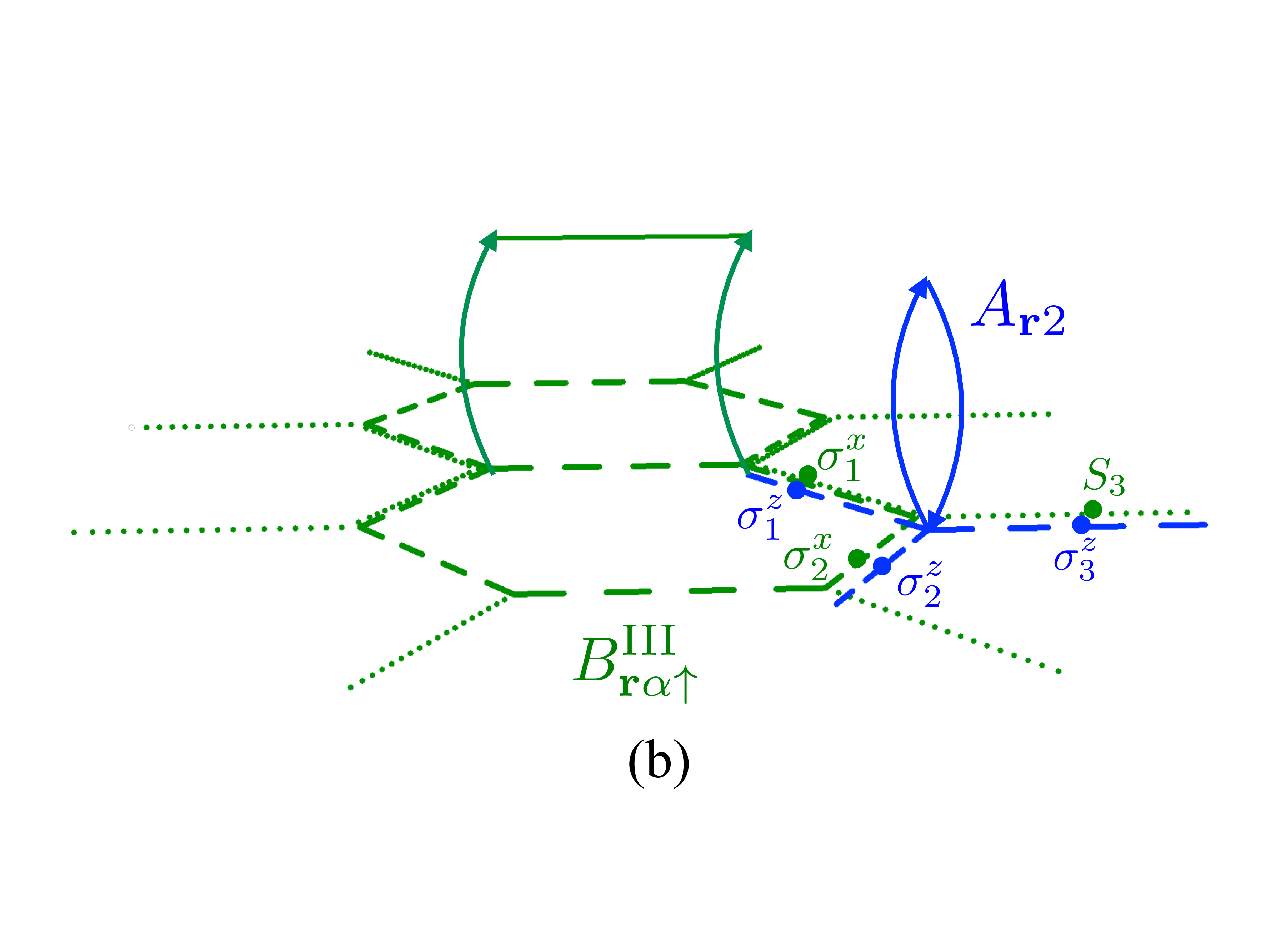}
\includegraphics[width=0.9\linewidth]{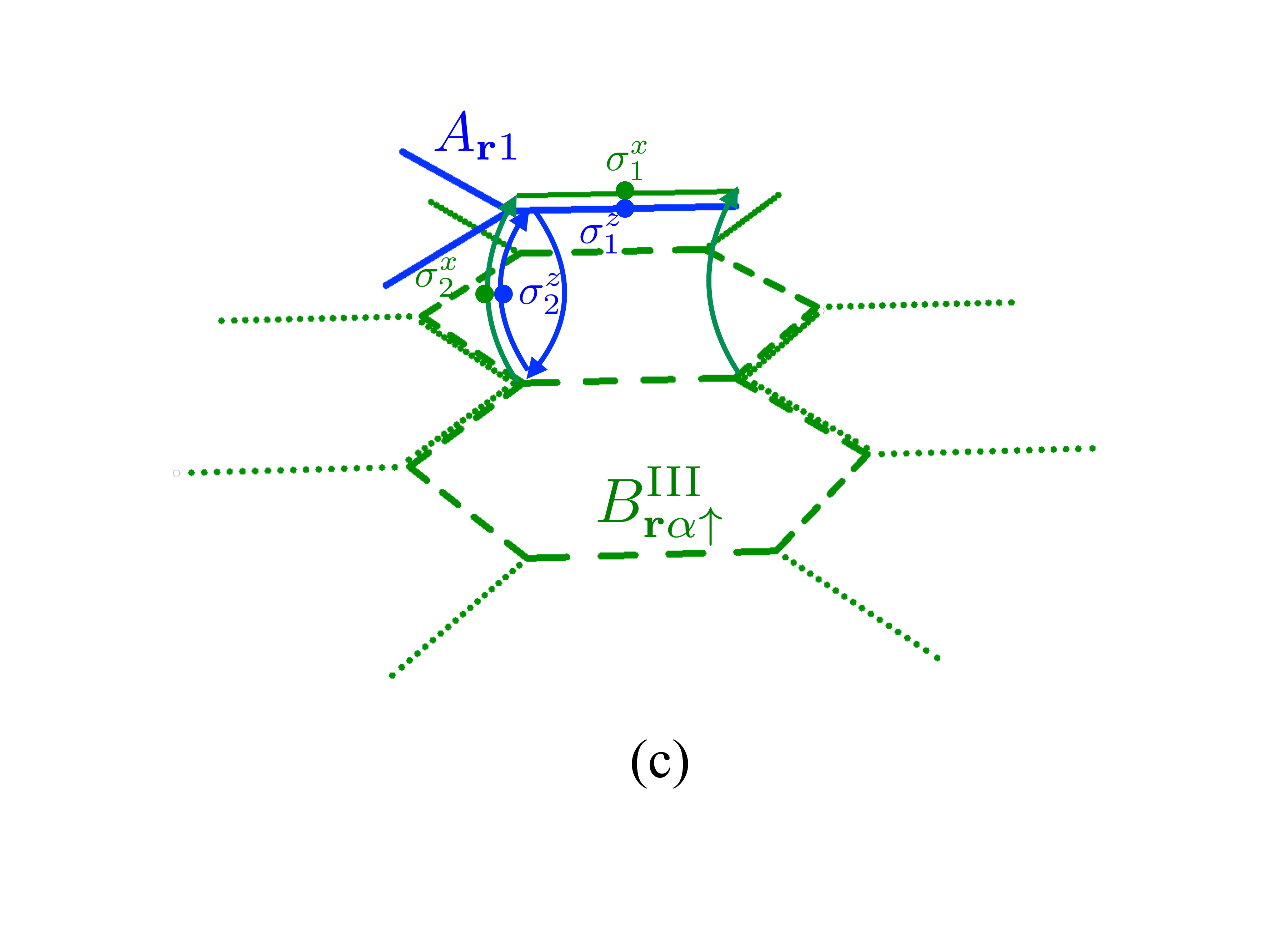}
\caption{\label{fig: position_vertex_plaquette_III} This figure shows the three possible positions among a vertex and a plaquette operator of type III. The vertex operator and the spins where it acts are shown in blue. The plaquette operator of type $\textrm{III}$ together with $\sigma^x$ and $S$ are illustrated in green. (a), (b) and (c)  sketch the possibilities considered in Eq. (\ref{app:a_bIII_a}), (\ref{app:a_bIII_b}) and (\ref{app:a_bIII_c}).}
\end{figure}
If we study the position shown in Fig.\ref{fig: position_vertex_plaquette_II} (a) and we remember that $\sigma^z_l$ and $S_l$ operators commute, we easily obtain :
\begin{align}
[A_{\mathbf{r}i}, B^{\textrm{II}}_{\mathbf{r}i}]_a= (1-1)S_1\sigma^z_1=0.
\end{align}

\noindent
In position (b) of Fig.\ref{fig: position_vertex_plaquette_II}, the two operators $A_{\mathbf{r}i}$ and $ B^{\textrm{II}}_{\mathbf{r}i}$ share three spins. Calculating the commutator between these two operators  we get:
 \begin{align}\label{eq:A_BII}
[A_{\mathbf{r}i}, B^{\textrm{II}}_{\mathbf{r}i}]_b= \left(1- (-1)^2\right)\sigma^z_1\sigma^z_2\sigma^z_3S_1\sigma^x_2\sigma^x_3=0.
\end{align}
The factors $(-1)$ in the above formula come from the fact that $\sigma^x$ and $\sigma^z$ anticommute. As it happens twice, the commutator gives a zero.

\noindent
Now, we move on to calculate the commutator among vertex operator and plaquette operator of type III. The Fig.\ref{fig: position_vertex_plaquette_III} illustrates the three possible positions where these two operators can be. The commutator for each case is obtained in the following. In Fig.\ref{fig: position_vertex_plaquette_III} (a) the two operators share four spins. We only need to recall the anticommutation rule for $\sigma^x_l$ and $\sigma^z_l$ to work out the result: 
\begin{align}\label{app:a_bIII_a}
[A_{\mathbf{r}2}, B^{\textrm{III}}_{\mathbf{r}\alpha\uparrow}]_a= (1-(-1)^4)\sigma^z_1\sigma^z_2\sigma^z_3\sigma^z_4\sigma^x_1\sigma^x_2\sigma^x_3\sigma^x_4=0.
\end{align} 

\noindent
The next case is shown in Fig.\ref{fig: position_vertex_plaquette_III} (b). The operators share only two spins and using the fact that $\sigma^z_l$ anticommutes with $\sigma^x_l$ and $S_j$ commutes with $\sigma^z_j$ we get:
\begin{align}\label{app:a_bIII_b}
[A_{\mathbf{r}i}, B^{\textrm{III}}_{\mathbf{r}\alpha\uparrow}]_b=(1-(-1)^2)\sigma^z_1\sigma^z_2\sigma^z_3\sigma^x_1\sigma^x_2S_3=0.
\end{align}
This  calculation is very similar to the commutator that appears in Eq.(\ref{eq:A_BII}).
In Fig.\ref{fig: position_vertex_plaquette_III} (c), a vertex operator on the upper layer is considered. Although the cover part of the $B^{\textrm{III}}_{\mathbf{r}\alpha\uparrow}$ remains in the lower layer, it shares two spins with the chosen vertex operator. The commutator is calculated in the following:
\begin{equation}\label{app:a_bIII_c}
[A_{\mathbf{r}1}, B^{\textrm{III}}_{\mathbf{r}\alpha\uparrow}]_c=(1-(-1)^2)\sigma^z_1\sigma^z_2\sigma^x_1\sigma^x_2=0.
\end{equation}

\noindent
 This finishes all possible combinations of vertex and plaquette operators. Now we focus on the commutator of plaquette operators among themselves. We start with $B^{\textrm{I}}_{\mathbf{r}}$ and $B^{\textrm{II}}_{\mathbf{r}i}$. However, these operators do not have any spin in common. The operator  $B^{\textrm{I}}_{\mathbf{r}}$ only acts on the interlayer space while $B^{\textrm{II}}_{\mathbf{r}i}$ is defined  on the layers. Thus, they trivially commute.

\begin{figure}
\includegraphics[width=0.8\linewidth]{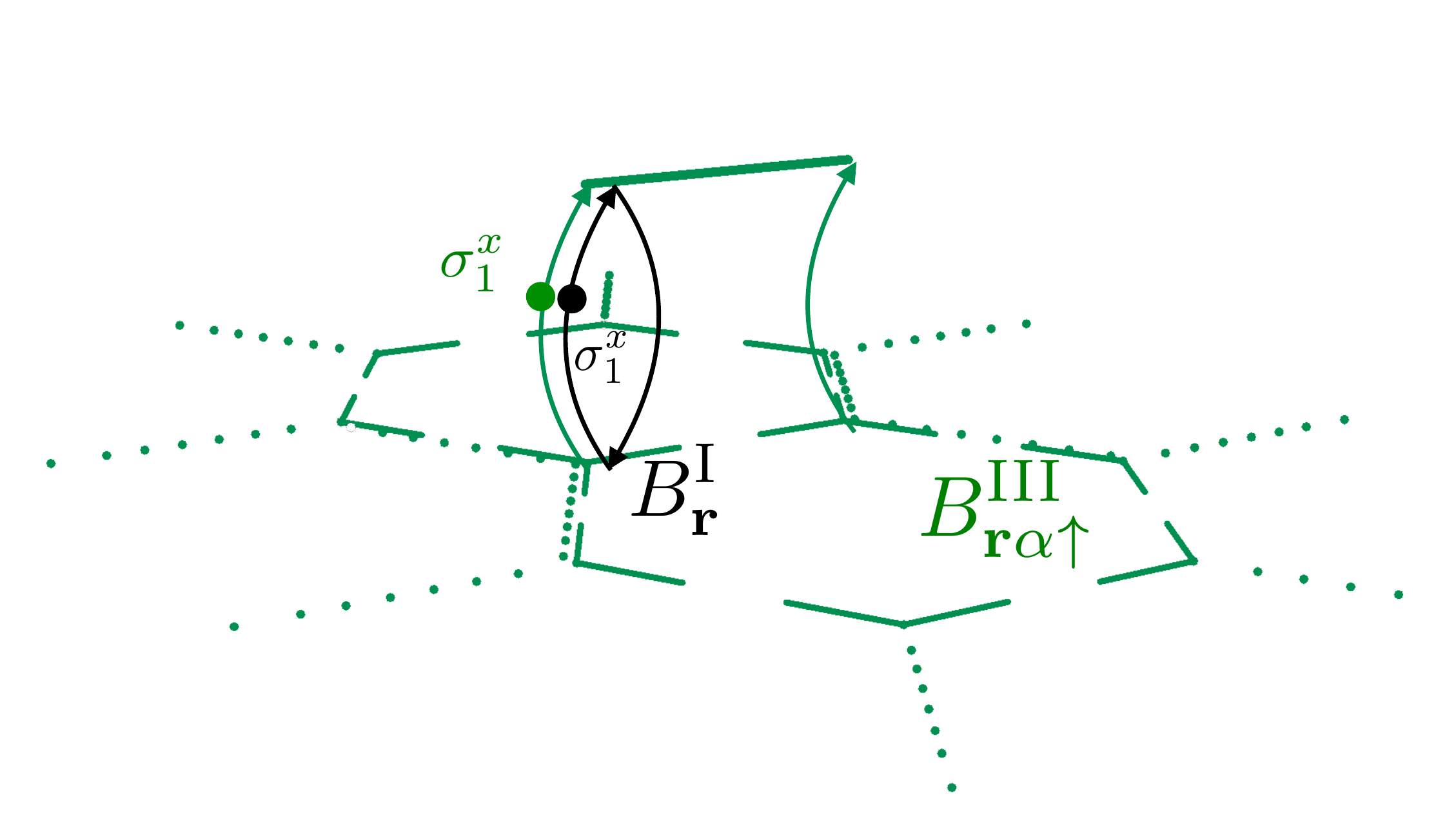}
\caption{\label{fig: commutator_B_I_B_III} The figure shows an example of two plaquette operators, one of type I and another of type II with a spin in common. The shared spin is plotted twice, each circle corresponding to  $B^{\textrm{I}}_{\mathbf{r}}$ (in black) and $B^{\textrm{III}}_{\mathbf{r}\alpha\uparrow}$ (in green)}
\end{figure}

\noindent
More interesting is the commutator between $B^{\textrm{I}}_{\mathbf{r}}$ and $B^{\textrm{III}}_{\mathbf{r}\alpha\uparrow}$, shown in Fig. (\ref{fig: commutator_B_I_B_III}).
Now the two operators are plaquette operators. The only spin they have in common is located in the interlayer links. All plaquette operators act with $\sigma^x_l$ on the interlayer layer. Therefore they trivially commute.

\begin{figure*}
\includegraphics[width=0.8\linewidth]{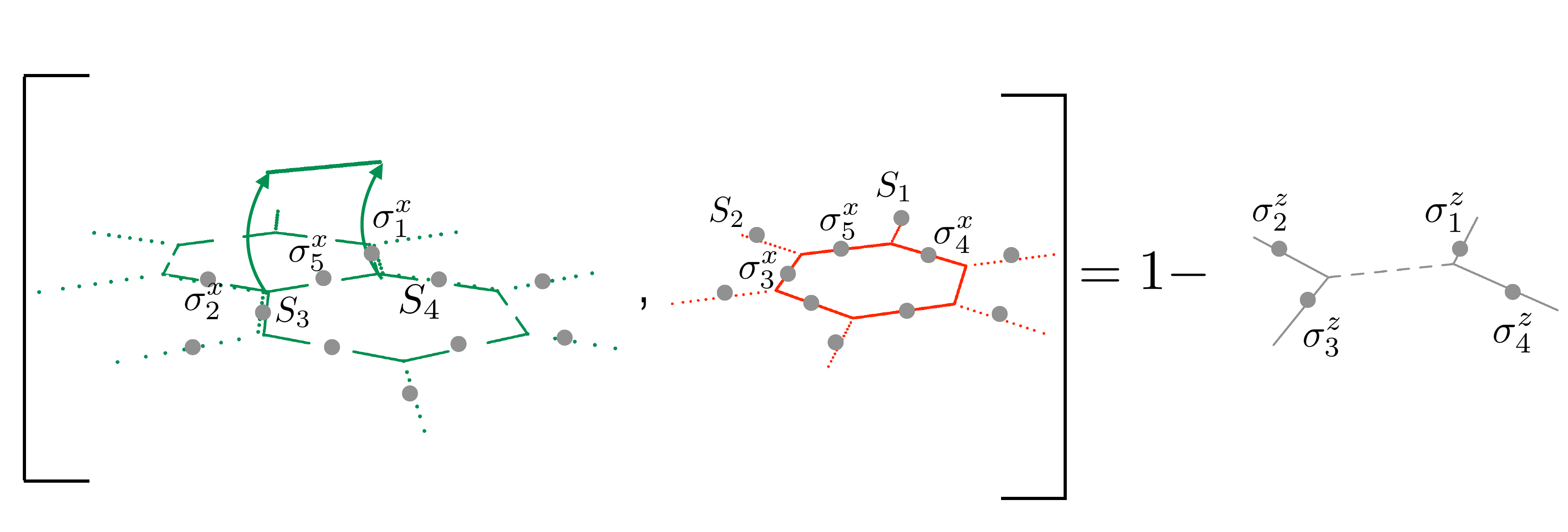}
\caption{\label{fig: commutator_B_III_B_II} The last commutator considered is pictured schematically in this figure. The brackets represent the commutator and the result is shown in grey.  The spins that $B^{\textrm{III}}_{\mathbf{r}\alpha\uparrow}$ (green) and $B^{\textrm{II}}_{\mathbf{r}i}$ (red) share are represented by grey circles. The relevant $\sigma^x_l$ and $S_l$ operators to calculate the commutator are shown next to the spins where they act.   }
\end{figure*}

\noindent
Finally, we solve the commutator of $B^{\textrm{II}}_{\mathbf{r}i}$ and $B^{\textrm{III}}_{\mathbf{r}\alpha\uparrow}$.  These two operators have many possible position to be located. However, the calculation for most of them resembles the previous cases. The only non trivial position is the one where they share twelve spins. This is possible when a plaquette of type $\textrm{II}$ is on top of the cover part of a plaquette of type $\textrm{III}$.  Although they share twelve spins only five (represented in Fig. \ref{fig: commutator_B_III_B_II} with a $\sigma^x$ or $S$ operator next to them) are important to work out the result of the commutator: 
\begin{align}
[B^{\textrm{III}}_{\mathbf{r}\alpha\uparrow},B^{\textrm{II}}_{\mathbf{r}i}, ]=(1-\sigma^z_1\sigma^z_2\sigma^z_3\sigma^z_4)\sigma^x_1\sigma^x_2S_3\sigma^x_4S_1S_2\sigma^x_3S_4.
\end{align}
This commutator is non zero in general. Notice that if the product $\sigma^z_1\sigma^z_2\sigma^z_3\sigma^z_4$ equals to $+1$ then the commutator cancels. The mentioned product can be factorized in vertex operators. Thus, it is 1 on the zero flux subspace and the operators $B^{\textrm{II}}_{\mathbf{r}i}$ and $B^{\textrm{III}}_{\mathbf{r}\alpha\uparrow}$ also commute.

\noindent
After these extensive calculations through all the commutators among vertex and plaquette operators, we have demonstrated that the model proposed in this work, the bilayer Double Semion, is a new exactly solvable example of symmetry enriched topological order.

\section{The geometrical obstruction to formulate the DS model on a square lattice}\label{app:square_lattice}
\begin{figure}
\includegraphics[width=0.6\linewidth]{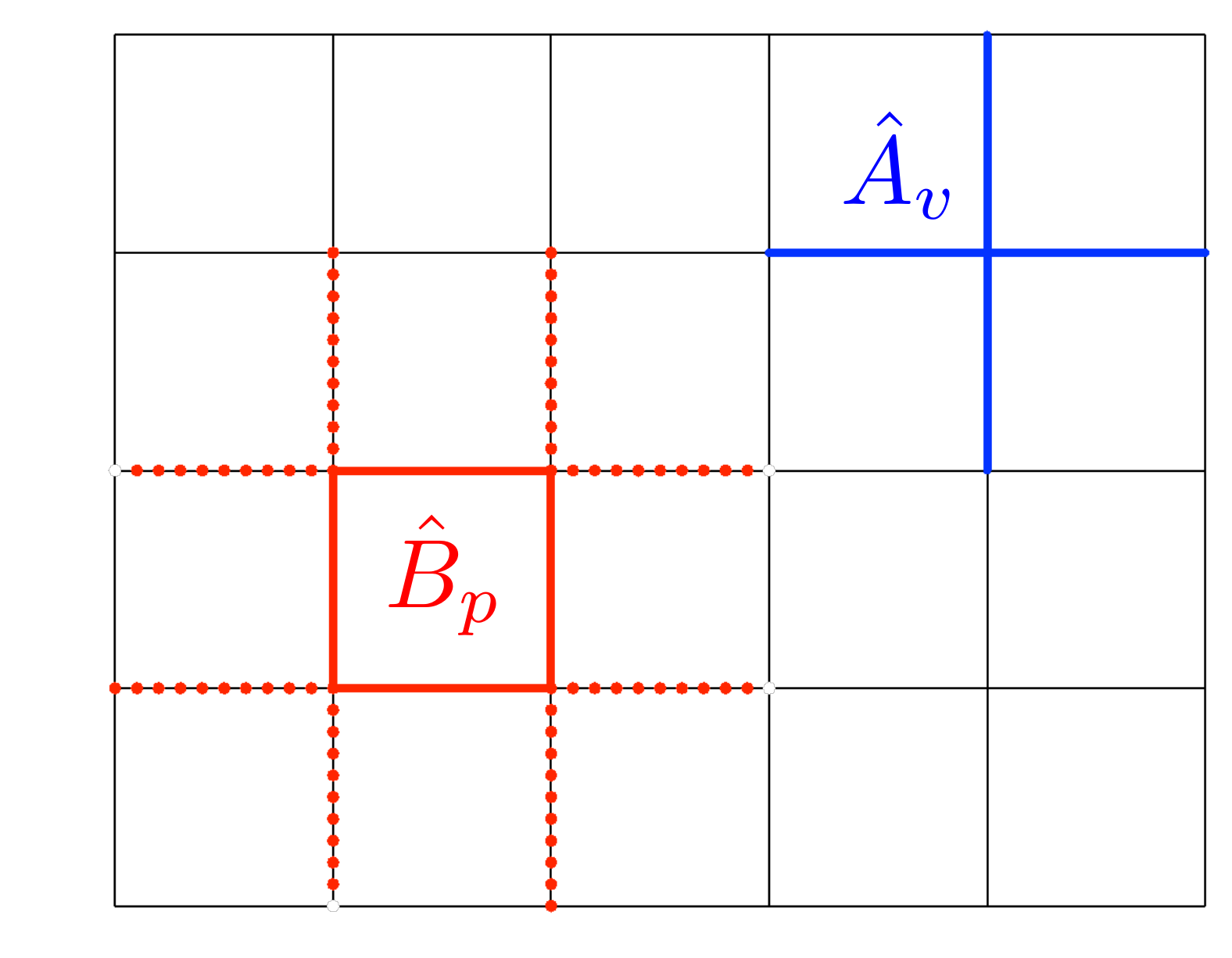}
\caption{\label{fig:square_lattice} DS model on a square lattice: a vertex operator (in blue) and a plaquette operator (in red) are shown. Links on solid blue (red) lines represents $\sigma^z_l$ ($\sigma^x_l$). The operator $S_l$ is indicated by dotted red lines. }
\end{figure}

\noindent
Let us explain the fundamental obstruction to construct the DS model on a square lattice. First, we introduce the natural way to 
 implement the DS model on a square lattice,  and then we shall give the intrinsic geometrical reason that prevents one from succeeding in a task like this. We consider a square lattice and  a spin {1/2} on each link of the lattice. The hamiltonian is the DS hamiltonian, constructed as a sum of vertex and plaquette operators. We recall the expression in Eq.\eqref{DS_hamiltonian}:
\begin{align}
H=-\sum_v A_v +\sum_{p}B_p.
\end{align}
The vertex and plaquette operators are defined as usual:
\begin{align}
 &A_v=\prod_{i\in{s(v)}}\sigma^z_i, \hspace{1cm }B_p=\prod_{i\in \partial p}\sigma^x_i\prod_{j\in s(p)}S_j,\\[3mm]
&\text{where}\hspace{0.7cm}
S_j=i^{(1-\sigma^z_j)/2},\nonumber
\end{align}
and  $s(v)$ are the four links entering a vertex $v$,  the symbol $\partial p$ is the set of links forming the boundary of a square plaquette and the product over $s(p)$ runs over the eight links irradiating from plaquette $p$. A picture of a vertex and a plaquette operator is shown in Fig.\ref{fig:square_lattice}. 

\begin{figure}
\includegraphics[width=\linewidth]{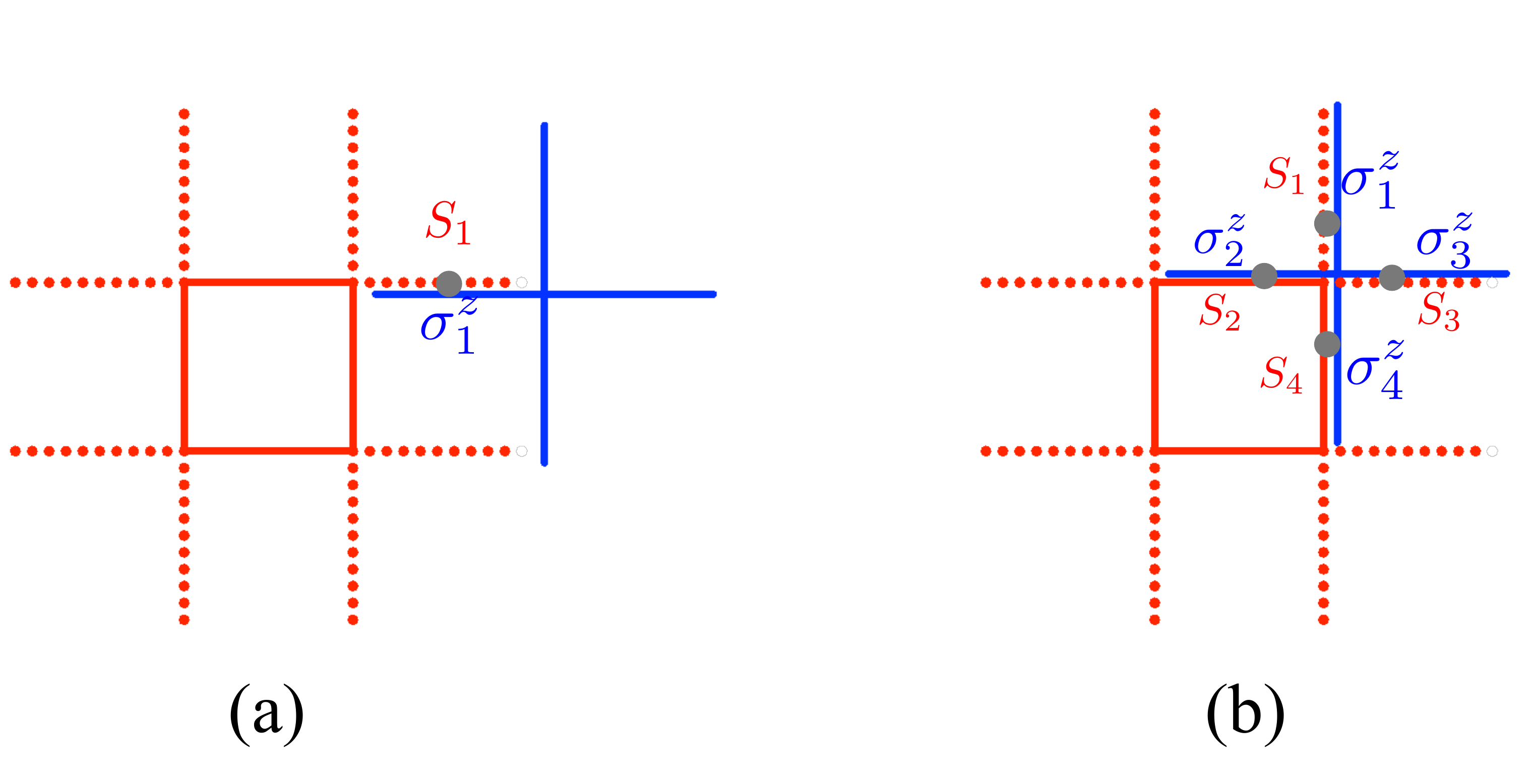}
\caption{\label{fig:position_A_B_square}  This  figure illustrates the two possible relative positions between a plaquette operator (red) and a vertex operator (blue) in the DS model. In both figures (a) (where there is one spin shared) and (b) (where there are four spins shared)  common spins are represented by grey circles. The operators that act on these spins are shown in the figure in the corresponding color: blue for vertex operator and red for plaquette operator.}
\end{figure}

\noindent
We demand that the DS model on a square lattice be well-defined and constitutes an exactly solvable model. In order to check whether
this holds true, we  must calculate the commutators among the operators in the hamiltonian. 
Vertex operators always commute among themselves since  they are built from $\sigma^z$ operators alone. More intriguing is the commutation between vertex and plaquette operators. There are two distinct positions where they share different number of spins as it is illustrated in Fig.\ref{fig:position_A_B_square}. 
We calculate in the following the commutators for both positions. First, we consider the situation when the operators share only one spin (shown in Fig.\ref{fig:position_A_B_square} (a)):
\begin{align}
[\hat{A}_v, \hat{B}_p]_{(a)}=(1-1)\sigma^z_1S_1=0.
\end{align}
This result is obtained taking into account that $\sigma^z_l$ and $S_l$ commute. 
Considering the position shown in Fig.\ref{fig:position_A_B_square} (b), the commutator is calculated:
\begin{align}\label{appen:A_B_b}
[\hat{A}_v, \hat{B}_p]_{(b)}=(1-(-1)^2)\sigma^z_1\sigma^z_2\sigma^z_3\sigma^z_4\sigma^x_1\sigma^x_2S_3S_4=0.
\end{align}
 To obtain the above result, we consider that the operators $\sigma^z_l$ and $\sigma^x_l$ anticommute. Next, we  calculate the commutator between two plaquette operators.  We recall from Appendix \ref{app:zero_flux_rule} that plaquette operators in the DS model do not commute in the whole Hilbert space. The model is well-defined on an invariant subspace called zero flux  subspace defined by:
\begin{align}\label{app:zero_square}
\hat{A}_v=+1.
\end{align}
This condition is equivalent to Eq.(\ref{app:zero-flux-rule}) but on a different lattice.   Eq.(\ref{app:zero-flux-rule}) is a product of three operators acting on three different spins whereas Eq.\eqref{app:zero_square} implies a restriction over four spins.
As a consequence, any product of $\sigma^z_l$ that can be factorized in vertex operators equals to one.
\begin{figure}
\includegraphics[width=0.8\linewidth]{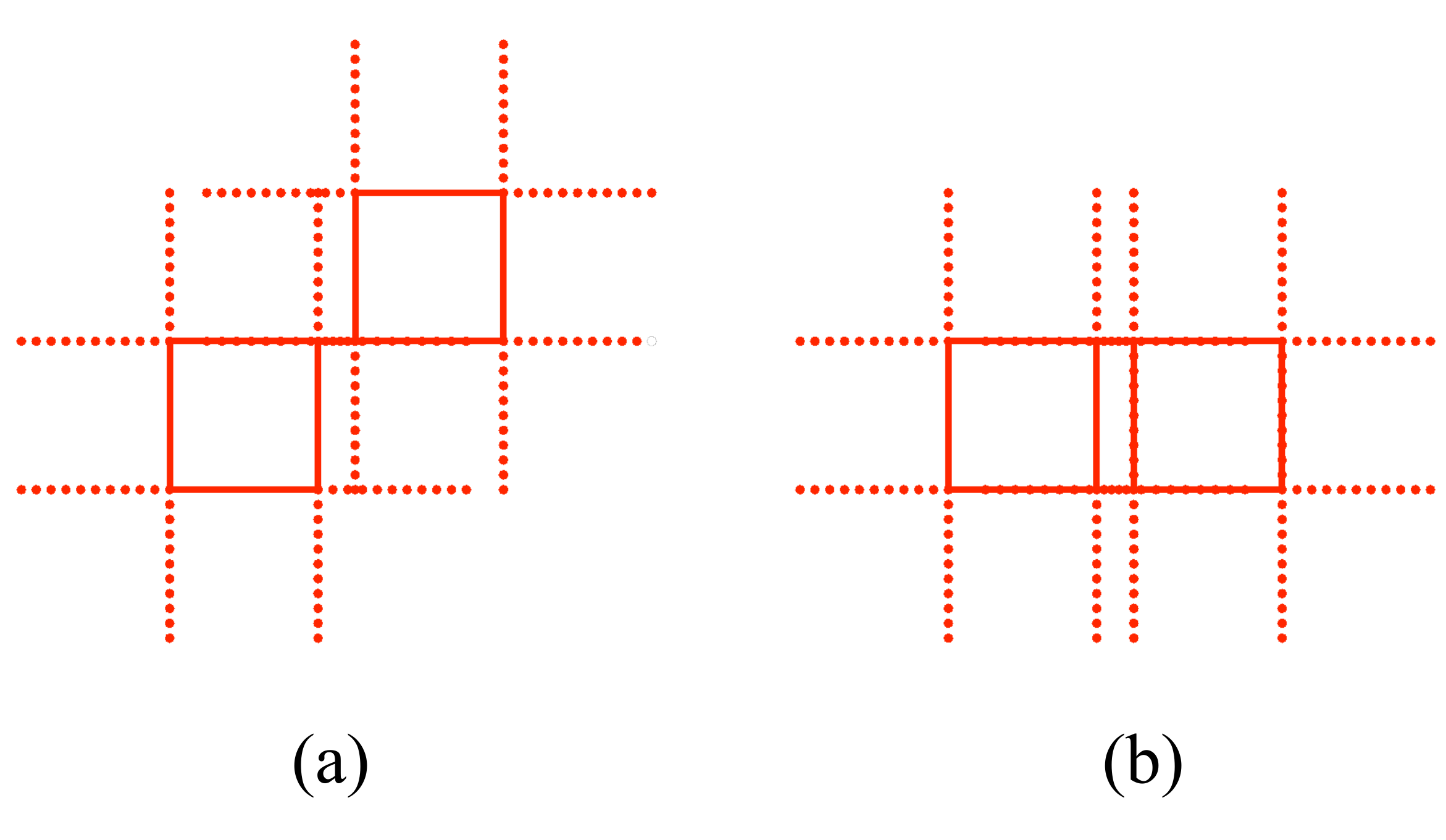}
\caption{\label{fig:position_B_B_square} The two possible relative positions for plaquette operators in the DS model on a square lattice
are shown. Solid red lines represent the action of $\sigma^z_l$ on spins and dotted red lines, the action of $S_l$. (a) corresponds to the case where the two plaquette operators share four spins. In (b), seven spins are shared by the two plaquette operators.  }
\end{figure}

\noindent
Notice though that two different relative positions among plaquette operators are possible in this lattice. These cases are shown in 
Fig.\ref{fig:position_B_B_square}.

\begin{figure}
\includegraphics[width=\linewidth]{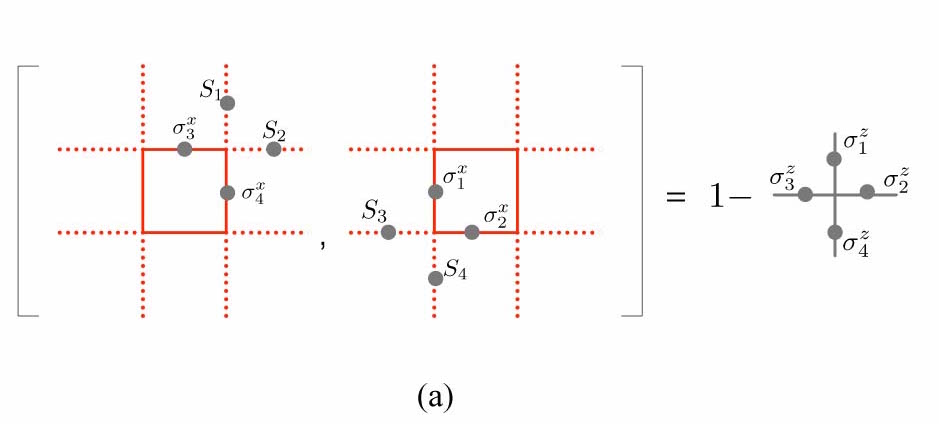}
\includegraphics[width=\linewidth]{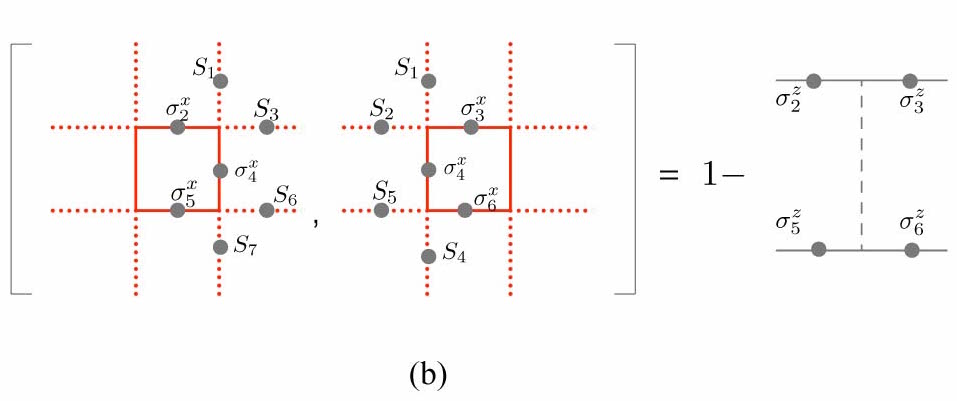}
\caption{\label{fig:commutator_B_B_1} Pictorial representation of the commutators of two plaquette operators for the DS model on a square lattice. The  shared spins are represented by grey circles.  Both (a) and (b) figures sketch the result of the corresponding commutators. The left hand side in (a) is multiplied by $S_1\sigma^x_1S_2\sigma^x_2\sigma^x_3S_3\sigma^x_4S_4$, a factor that appears in Eq.(\ref{appen:BB_square_1}). The corresponding expression for (b) is in Eq.(\ref{appen:BB_square_2}) and the left hand side in the picture is multiplied by $\sigma^x_2S_2S_3\sigma^x_3\sigma^x_5S_5S_6\sigma^x_6$.   }
\end{figure}

 In the first position, the plaquette operators share four spins. The commutator is:
 \begin{align}\label{appen:BB_square_1}
 [\hat{B}_{p1}, \hat{B}_{p2}]_a&=(1- (+i)\sigma^z_1(+i)\sigma^z_2(-i)\sigma^z_3 (-i)\sigma^z_4)\times\\
 &\times S_1\sigma^x_1S_2\sigma^x_2\sigma^x_3S_3\sigma^x_4S_4=\nonumber\\
 &=(1- \sigma^z_1\sigma^z_2\sigma^z_3\sigma^z_4)S_1\sigma^x_1S_2\sigma^x_2\sigma^x_3S_3\sigma^x_4S_4.\nonumber
 \end{align}
 The result is  graphically represented in  Fig.\ref{fig:commutator_B_B_1} (a) . The product $\sigma^z_1\sigma^z_2\sigma^z_3\sigma^z_4$ is indeed a vertex operator. Therefore, the result is $+1$ in the zero-flux rule invariant subspace and the model works fine thus far.

\noindent
The commutator for Fig.\ref{fig:commutator_B_B_1} (b) is:
  \begin{align}\label{appen:BB_square_2}
 [\hat{B}_p, \hat{B}_p]_b&=(1-(-i)\sigma^z_2(+i)\sigma^z_3(-i)\sigma^z_5 (+i)\sigma^z_6)\times\\
 &\times S_1\sigma^x_2S_3\sigma^x_4\sigma^x_5S_6S_7S_1S_2\sigma^x_3\sigma^x_4S_5\sigma^x_6S_7=\nonumber\\
 &=(1-\sigma^z_2\sigma^z_3\sigma^z_5\sigma^z_6 )\sigma^x_2S_2S_3\sigma^x_3\sigma^x_5S_5S_6\sigma^x_6\nonumber.
 \end{align}

\noindent
This commutator does not equal to zero on the invariant subspace. The result can not be factorized into a product of vertex operators. Therefore, the plaquette operators do not commute and the DS model is not well-defined on the square lattice.
 
We can conclude that the DS model can not be implemented on a lattice with coordination number equal to four.  
This fact creates a geometrical obstruction to place the DS model on a square lattice. In the bilayer DS model that is introduced in Sec.\ref{model}, the DS model is only implemented on the hexagonal layers. To build a bilayer system, it is necessary to include a square geometry joining the two layers. As the square geometry is incompatible with the DS model, it is not possible to extend the DS model to the whole system. This is the reason why    a Kitaev-like model is included on the links joining the layers in the bilayer DS model. 



\begin{thebibliography}{99}
\bibitem{change_notation} Notice that we have exchanged the notation of plaquette operators of type II and type III with respect to the notation 
in the original reference \cite{Hermele} since it is more convenient for comparison with the bilayer DS model.
\end{thebibliography}
\end{document}